\documentclass[journal=jpccck,manuscript=article]{achemso}

\usepackage{chemformula} 
\usepackage[T1]{fontenc} 

\usepackage[bb=boondox]{mathalfa}
\def\e{\begin{equation}}
\def\f{\end{equation}}
\def\_#1{{\bf #1}}

\def\.{\cdot}

\usepackage{comment}
\definecolor{myblue}{rgb}{0,0.44705, 0.7411}
\definecolor{bordo}{rgb}{0.63529, 0.07843,0.18431}
\definecolor{Mgreen}{rgb}{0.30196,0.68627, 0.290196}
\definecolor{Mred}{rgb}{0.89412, 0.10196,0.1098}

\usepackage{graphicx}
\usepackage{adjustbox}
\usepackage[caption=false]{subfig}
\usepackage{hyperref}

\makeatletter
\newcommand*{\rom}[1]{\expandafter\@slowromancap\romannumeral #1@}
\makeatother
\usepackage{dcolumn}
\usepackage{bm}

\usepackage{atbegshi}
\AtBeginDocument{\AtBeginShipoutNext{\AtBeginShipoutDiscard}\addtocounter{page}{-1}}

\author{Mariia Poleva}
\email{mariia.poleva@kit.edu}
 \affiliation{
Institute of Theoretical Solid State Physics,
Karlsruhe Institute of Technology (KIT),
Kaiserstr. 12, 76131 Karlsruhe, Germany
}

\author{Benedikt Zerulla}%
\affiliation{%
Institute of Nanotechnology,
Karlsruhe Institute of Technology (KIT),
Kaiserstr. 12, 76131 Karlsruhe, Germany
}%

\author{Christof Holzer}
\affiliation{
Institute of Theoretical Solid State Physics,
Karlsruhe Institute of Technology (KIT),
Kaiserstr. 12, 76131 Karlsruhe, Germany
}

\author{Vlasta Bonačić-Kouteck\'y}
\affiliation{
Center of Excellence for Science and Technology-Integration of Mediterranean
Region (STIM) at Interdisciplinary Center for Advanced Sciences and Technology
(ICAST), University of Split, Poljička cesta 35, 21000 Split, Croatia
}
\alsoaffiliation{
Chemistry Department, Humboldt University of Berlin, Brook-Taylor-Strasse 2,
12489 Berlin, Germany
}

\author{Anna Pniakowska}
\affiliation{
Institut lumière matière, UMR5306, Université Claude Bernard Lyon1-CNRS, Univ. Lyon 69622, Villeurbanne cedex, France
}
\alsoaffiliation{
Institute of Advanced Materials, Wroclaw University of Science and Technology, Wybrzeże Wyspiańskiego 27, 50-370 Wroclaw, Poland
}

\author{Joanna Olesiak-Banska}
\affiliation{
Institute of Advanced Materials, Wroclaw University of Science and Technology, Wybrzeże Wyspiańskiego 27, 50-370 Wroclaw, Poland
}

\author{Rodolphe Antoine}
\affiliation{
Institut lumière matière, UMR5306, Université Claude Bernard Lyon1-CNRS, Univ. Lyon 69622, Villeurbanne cedex, France
}

\author{Ivan Fernandez-Corbaton}
\affiliation{%
Institute of Nanotechnology,
Karlsruhe Institute of Technology (KIT),
Kaiserstr. 12, 76131 Karlsruhe, Germany
}%

\author{Carsten Rockstuhl}
\email{carsten.rockstuhl@kit.edu}
\affiliation{
Institute of Theoretical Solid State Physics,
Karlsruhe Institute of Technology (KIT),
Kaiserstr. 12, 76131 Karlsruhe, Germany
}%
\alsoaffiliation{%
Institute of Nanotechnology,
Karlsruhe Institute of Technology (KIT),
Kaiserstr. 12, 76131 Karlsruhe, Germany
}%

\author{Marjan Krsti\'c}
\email{marjan.krstic@kit.edu}
\affiliation{
Institute of Theoretical Solid State Physics,
Karlsruhe Institute of Technology (KIT),
Kaiserstr. 12, 76131 Karlsruhe, Germany
}%

\title{Predicting the optical properties of organometallic nanoparticles with a scale--bridging method: The importance of the embedding}

\abbreviations{IR,NMR,UV}

\begin{document}


\begin{abstract}
It remains a prime question of how to describe the optical properties of large molecular clusters accurately. Quantum chemical methods capture essential electronic details but are infeasible for entire clusters, while optical simulations handle cluster-scale effects but miss crucial quantum effects. To overcome such limitations, we apply here a multi--scale modeling approach, combining precise quantum chemistry calculations with Maxwell scattering simulations, to study the linear and nonlinear optical response of finite--size supramolecular gold-cysteine nanoparticles dispersed in water. In this approach, every molecular unit that forms the cluster is represented by a polarizability and a hyperpolarizability, and the overall response is obtained from solving an optical multiple scattering problem. We particularly demonstrate how important it is to accurately consider the environment of the individual molecular units when computing their polarizability and hyperpolarizability. In our quantum chemical simulations, we do so at the level of a static partial charge field that represents the presence of other molecular units. Without correctly considering these effects of the embedding, predictions would deviate from experimental observations even qualitatively. Our findings pave the way for more accurate predictions of the optical response of complex molecular systems, which is crucial for advancing applications in nanophotonics, biosensing, and molecular optoelectronics.      
\end{abstract}

\section{\label{sec:level1}Introduction}

Hybrid metal--organic nanomaterials constitute a large material class with applications in fields such as energy conversion \cite{https://doi.org/10.1002/ejic.201800829, https://doi.org/10.1002/aoc.1613}, chemicals production and storage \cite{THOMAS20044110, PHILIPPOT20031019, ZHANG201865}, biology \cite{https://doi.org/10.1002/cbic.201200159}, information technologies and telecommunication \cite{Xie2020}, medical treatment \cite{doi:10.1021/jm100020w, C2DT12460B, Fernández_2008, MAHADEVPATIL2023144110, doi:10.1021/acs.chemrev.5b00148}, sensing \cite{WANG2015139, C2CP00050D}, and many more \cite{https://doi.org/10.1002/anie.200905678, ZHANG2021213652}. One particularly interesting subclass of such materials is those containing gold atoms. Historically, gold was particularly suitable and interesting for experiments due to its unique combination of opacity and transparency, its ability to develop color in both reflected and transmitted light, its ability to maintain integrity while being divided, and its potential to alter the colors of light by varying the size of its particles, especially when combined with other molecules \cite{Faraday1857}. Gold metal--organic nanomaterials can exist in the form of nanoparticles. The size of these nanoparticles is such that they contain only several gold atoms \cite{LAVENN2014234, doi:10.1021/ar200331z, doi:10.1021/ja042218h, C5NR07810E, doi:10.1021/jz3007436, doi:10.1021/ja900386s, C5RA22741K, doi:10.1021/ja062584w, doi:10.1021/ja0483589} up to few hundreds or thousands of gold atoms \cite{doi:10.1021/jacs.9b10770, doi:10.1021/ja800341v, doi:10.1021/nn305856t}. The arrangement of the gold atoms in such nanoparticles typically takes the form of core-shell structures. There, gold atoms build a metallic core stabilized and protected by organic molecules \cite{C3DT51180D, GONZALEZDERIVERA201213, doi:10.1021/om400927g, C5NR07810E, doi:10.1021/jz3007436, doi:10.1021/acs.iecr.6b02925, B9NR00160C}. Certain magic sizes of such nanoparticles exhibit exquisite stability and intriguing optical features \cite{doi:10.1021/cr030698+, doi:10.1021/jz3007436, doi:10.1021/cm500139t, doi:10.1021/nl301988v, doi:10.1021/acs.jpcc.5b08341, doi:10.1021/ja800341v, C4NR03782K, C5NR08122J, doi:10.1021/acs.chemrev.5b00703}. However, core-shell nanoparticles are not the only geometrical shape explored. Many alternative geometries for gold metal--organic nanomaterials are accessible by carefully choosing the organic building blocks \cite{doi:10.1021/la049415e, RUSSIERANTOINE2016455}. There are observations of spiral-like or rod-shaped nanoparticles, or in the form of nanowires and sheets \cite{doi:10.1021/acs.chemrev.5b00148, RUSSIERANTOINE2016455}. All those different shapes have in common that they are synthesized by supramolecular assembling from precursors, usually in liquid environments. 

Gold metal--organic nanomaterials are extensively studied due to their exceptional ability to facilitate and manipulate light--matter interactions. On the one hand, gold nanoclusters are known to have pronounced linear and nonlinear optical properties at infrared (IR) \cite{doi:10.1021/jp993691y, C6NR02251K, C4NR01130A} and visible (Vis.) frequencies \cite{doi:10.1021/ar400295d, doi:10.1021/jp993691y, doi:10.1021/ja900386s, C6NR02251K}. On the other hand, the organic molecules are usually optically active in the visible and ultraviolet (UV). Together, these hybrids are quite interesting to tailor light matter-interaction and the optical response for different applications \cite{Faraday1857, RUSSIERANTOINE2016455, doi:10.1021/ar200331z, doi:10.1021/jp993691y, doi:10.1021/ja900386s, C6NR02251K}. Furthermore, the inertness of gold and biocompatibility of selected organic molecules used to stabilize the gold atoms makes them attractive for applications \emph{in vitro} and \emph{in vivo} \cite{C5RA11321K, C0NR00458H, C4NR01130A}. 

The extensive design space of gold metal--organic nanomaterials necessitates robust theoretical and computational tools to predict and optimize their chemical and optical properties for applications. Structural properties are commonly studied on the level of classical molecular mechanics, or quantum chemistry in the case of smaller molecular systems of up to a few hundred atoms. Optical properties are studied using quantum chemistry methods. To such end, density functional theory (DFT) is typically chosen among the available options, including coupled cluster or configuration interaction methods. The computational complexity of simulating the optical response, especially the nonlinear response, such as the second--harmonic generation or the two--photon absorption, limits the size to basic molecular structures containing less than a hundred atoms in total \cite{Sanader, Sanader2}. Typically, to further reduce the computational efforts and  speed up the calculations, extended surroundings, including material molecules or solvents important for the light--matter interactions, are neglected. These simplifications lead to predictions that are challenging to compare to experimental observations and do not allow for further extrapolation of optical properties to larger system sizes. 

To address many of these simplifications, we employ our recently developed workflows for scale--bridging simulations to study the linear and nonlinear properties of a specific gold metal--organic nanomaterial \cite{Fernandez-Corbaton:2020, Zerulla2024Feb}. In the current manuscript, we focus theoretically on the gold metal--organic nanomaterial in the form of thin sheets. These materials are considered to be infinitely extended in two dimensions but finite in the third one. Our workflow combines accurate quantum chemistry calculations with Maxwell scattering theory based on the transition matrix (T--matrix) and hyper--transition matrix (hyper--T--matrix) formalism to study the linear \cite{Zerulla2022May,https://doi.org/10.1002/adfm.202301093} and nonlinear \cite{Zerulla2024Feb, https://doi.org/10.1002/adom.202400150} response of structured molecular materials. Our multi--scale method enables the prediction of experimentally observable quantities, while accounting simultaneously for optical multiple scattering effects and the chemical details of the molecules. Although Maxwell's approach has been widely used to study nonlinear effects in semiconductors \cite{carletti2021reconfigurable, carletti2024intrinsic}, as well as in all-dielectric, plasmonic, and hybrid structures \cite{makarov2017efficient, smirnova2016multipolar, zheng2023advances, rahmani2018nonlinear}, the combination of quantum chemistry and Maxwell scattering has not yet been applied to study the nonlinear properties of solvent-dispersed metal-organic nanoparticles.

Our bottom--up approach starts with the quantum chemical description of the properties of finite--sized systems made from identical molecular units arranged on a periodic grid. Improvements incorporated into TURBOMOLE electronic structure software \cite{TURBOMOLE2022, TM_TODAY_TOMORROW} allow us to describe spatial domains containing up to 1,000 atoms explicitly on quantum chemical grounds. Molecules outside that quantum domain are accommodated either explicitly as partial charges or implicitly by describing the solvent environment through polarizable continuum models. Here, we combine both approaches to describe the embeddings. The former reflects the presence of molecular units outside the quantum domain through the polarization of slightly charged atoms. The latter reflects solvent effects, and we consider here a Conductor-like screening model (COSMO) \cite{klamtCOSMONewApproach1993}. The detailed consideration of the environment offers an improved description of each molecular unit within the sheets when transitioning to the optical simulations. These optical simulations are performed at the level of a full--wave Maxwell simulation, and each molecular unit is represented by its own T--matrix (expressing the linear properties) and the hyper--T--matrix (expressing the non--linear properties). With that approach, sheets with a thickness of several tens of nanometers can be accurately described.

More precisely, here we focus on one specific gold metal--organic nanomaterial: nanoparticles made of gold--L-cysteine layered sheets \cite{SOPTEI20158,Fakhouri2019,Ni2024}. These nanoparticles have a large spatial extent in two dimensions and are rather thin in the third. Hence, we treat them as infinitely extended sheets with a finite thickness. The layered structural motif resembles the $\beta$ structures in biochemistry as depicted in supplementary figure S2. Therefore, these nanoparticles are called Au-cysteine $\beta$--sheet nanoparticles. Experimentally, these nanoparticles are prepared in a self-assembled fashion in an aqueous surrounding from gold-thiolate (-Au-SR-) single atom strands and R = L-cysteines, coordinated by weaker H-bonds and van-der-Waals interactions between the neighboring organic molecules into planar structures (Fig. S2). Finally, multiple noncovalently bounded stacks of several 2D-structural motifs are self-arranged into a finite--size supramolecular nanoparticle dispersed in water. 

We study the linear and nonlinear optical response from such clusters and compare theoretical predictions with experimental results. We clearly show that, without considering the details of the finiteness of the molecular units and their surroundings given by the molecular units that form the entire cluster, even basic quantities, such as the correct sign for the dispersion of the SH signal as a function of the wavelength, are not predicted precisely.  

This article is structured in the following way. In Sec.~II we present the molecular material and its computational model. Section~III provides a brief introduction to the workflow and the methodology used. The results are presented in Sec.~IV. In the first sub--section, we focus on multi--scale simulations of the linear and nonlinear response based on a Maxwell scattering approach. In a second sub--section, we focus purely on quantum chemistry simulations. The key result of our work is showing the importance of considering a layer--dependent molecular response that precisely models the impact of the surroundings in the quantum chemical calculations. Finally, we summarize our results in Sec.~V and provide an outlook on how to extend our approach to even more precise theoretical modeling in the future.

\section{Molecular material and models}

\begin{figure*}[ht!]
\centering
\includegraphics[width=1.0\textwidth]{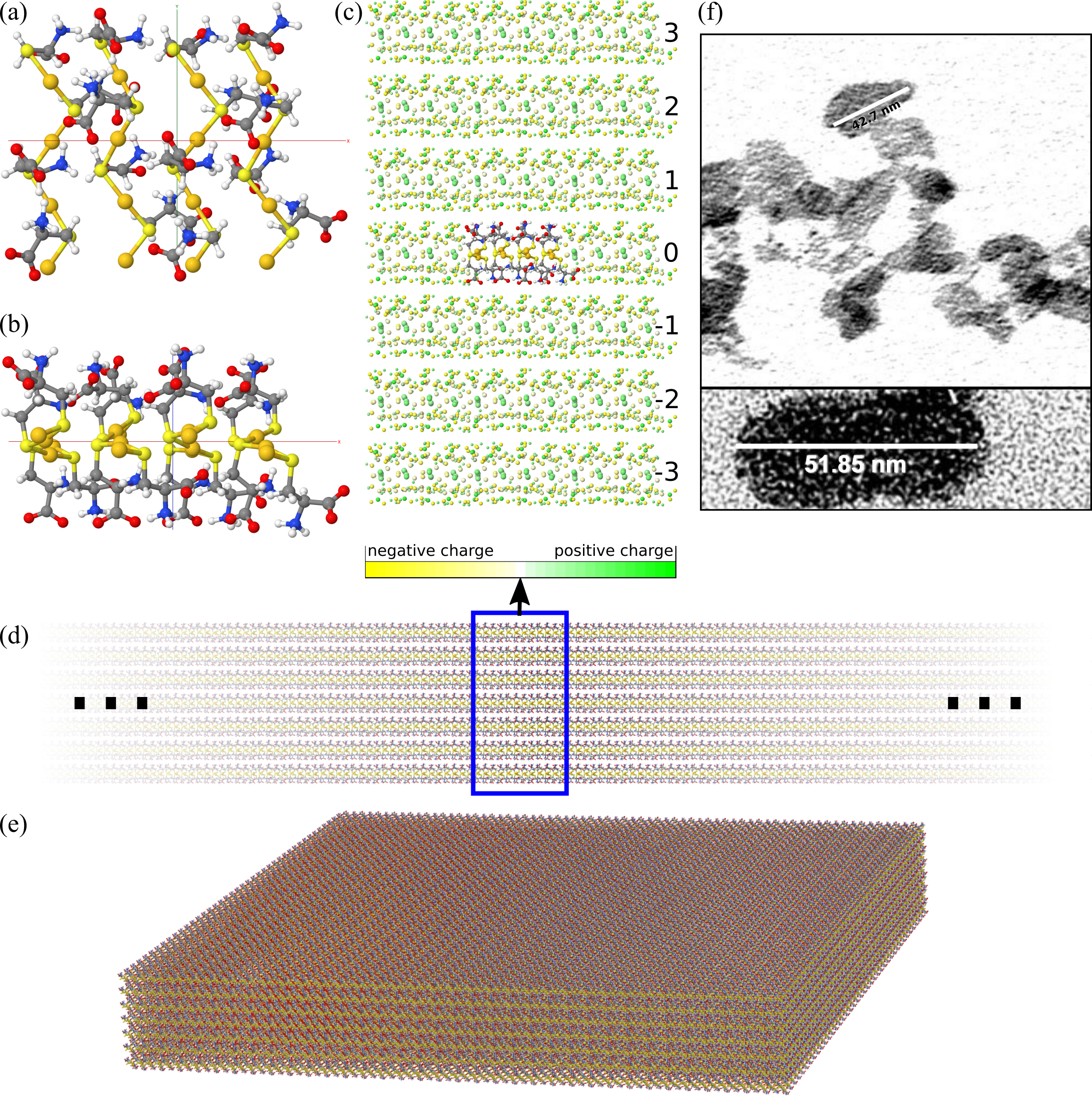}\caption{\textbf{(a)} A top--view of a periodic cell of DFT optimized molecular model of Au--cysteine nanoparticle without water molecules for simulations with implicit solvent surrounding. \textbf{(b)} A side--view of the same periodic cell from (a). \textbf{(c)} Embedding of the periodic molecular cell into the field of NBO partial charges. The quantum region (periodic cell) was positioned in the middle of each layer of the partial charges field from bottom to top of the molecular film (models "-3", ..., "3").  \textbf{(d)} A side--view of a 7 layers (9.1 nm thick) thick model of Au-cysteine thin film infinite in x-- and y--direction. \textbf{(e)} A simplistic molecular model of the finite--size supramolecular Au-cysteine nanoparticle built by 35 x 41 x 7 unit cells in x--, y-- and z--direction with total size of the nanoparticle 59.94 nm, 60.81 nm and 9.1 nm in x--, y-- and z--direction, respectively. \textbf{(f)} Tunneling-electron microscope images of examples of Au-Cys-$\beta$-sheet nanoparticles. }
\label{fig:MolModels}
\end{figure*}

We focus on one particular hybrid gold--cysteine nanoparticle. The structure can be realized experimentally, and it offers unique linear and nonlinear optical properties \cite{SOPTEI20158,Fakhouri2019,Ni2024}. These nanoparticles have been synthesized from solvated gold atoms and cysteine amino--acid and have an average size of $\sim$60 nm (Fig.~\ref{fig:MolModels} (f) and supplementary figures S4 and S5) in lateral dimensions with a thickness of $\sim$8--9~nm \cite{SOPTEI20158,Fakhouri2019,Ni2024}. Such a thickness corresponds to seven non--covalently coordinated layers, each being 1.3 nm thick \cite{SOPTEI20158,Fakhouri2019,Ni2024}.   

These nanoparticles exhibit significant linear absorption and pronounced non--resonant second-harmonic generation. Since the cysteine molecule is chiral, these nanoparticles also exhibit different interactions with left- and right-circularly polarized light, giving rise to a notable linear ECD signal in the UV--Vis. part of the electromagnetic (EM) spectrum. The nanoparticles also exhibit a nonlinear hyper-Rayleigh optical activity in the IR part of the EM spectrum \cite{Fakhouri2019}.

The size of these nanoparticles poses a challenge to study them theoretically and to unravel the complex relations between structural and optical properties. Although simplified small molecular models have been employed to study these and similar gold-organic nanoparticles and materials \cite{Fakhouri2019, RUSSIERANTOINE2016455,Ni2024}, it was not possible to address the intriguing light--matter interactions on the larger scales of the whole nanoparticle or thin film due to its sheer size. Thus, we apply here a unique multi--scale approach. This approach considers a precise quantum chemical description of the molecular building blocks of these nanoparticles. Our approach will capture the effects that modify the optical response at the level of an individual molecular unit that make up the nanoparticle. We combine this with a Maxwell scattering approach through T--matrices and hyper--T--matrices. The Maxwell solver accounts for all the multi--scattering and absorption details in the fundamental and SH frequencies and allows us to study much larger objects than just individual molecular units. The length scales accessible by such a Maxwell solver are inaccessible with quantum chemistry. In Fig.~\ref{fig:MolModels}, we present different molecular models used in this study to unravel the origin of interesting optical properties at the level of the nanoparticle.

We start by obtaining the geometrical structure of such materials. The small-- and wide--angle x--ray scattering (SWAXS) studies reveal that these Au--cysteine nanoparticles have a periodic crystalline structure \cite{SOPTEI20158, Fakhouri2019, Ni2024}. Therefore, 3D periodic DFT electronic structure calculations were used to minimize the energy of the geometry of the 3D periodic building block of a material surrounded by explicit water molecules (Fig. S1). The optimized periodic building block within each layer comprises 16 gold atoms, 16 cysteines, and 34 H\textsubscript{2}O surrounding molecules. The structure is presented in Figs.~\ref{fig:MolModels} (a) and (b) in a top-- and side--view, respectively. Contrary to the previously reported molecular structure of these nanoparticles obtained by molecular dynamic force-field geometry optimization \cite{SOPTEI20158}, the periodic DFT simulations revealed that each sheet of these nanoparticles is made of aligned single--atom thick Au-S-Au nanowires stabilized by interactions between cysteine molecules from the neighboring nanowires as depicted in Figs.~\ref{fig:MolModels} (a) and (b). 

In the calculation of the linear and nonlinear optical properties, we considered the optimized periodic cell and removed all H\textsubscript{2}O molecules to build a finite size molecular model in Fig.~\ref{fig:MolModels} (a) and (b) in TURBOMOLE. The influence of the water molecules was treated implicitly using the Conductor-like Screening Model (COSMO) \cite{klamtCOSMONewApproach1993} with parameters describing water. This allowed us to reduce the quantum region by 102 atoms in total and speed up calculations of the first hyperpolarizabilities by a factor of $\sim$1.9 while still obtaining results in excellent agreement with experimental measurements. We stress that the quantum region is this fraction of the structure that is considered quantum mechanically all the time. The details of the environment will change the properties of this quantum region depending on the exact spatial location within the nanoparticle. Therefore, how to construct the environment of each quantum region is important. 

Experimentally, this nanomaterial is realized as a sub--100 nm finite--size supramolecular nanoparticles. With that, it is very thin in one dimension and much larger in the two other dimensions. The nanoparticles are dissolved in water as a surrounding. To better capture the effects of the finiteness of the nanoparticle and improve our theoretical predictions, we considered two models. 

First, we consider the nanoparticle as a thin--film with the same thickness of $\sim$9 nm as the nanoparticle, but infinite in $x$-- and $y$--direction. The film is then considered to consist of seven discrete molecular layers. In this model, which we call the {\textbf{embedded model}}, we study the properties of the molecular unit in each layer and consider it as the quantum region. The molecular properties in each layer will be different because the molecular unit in each layer experiences a different surrounding. That surrounding is considered at the level of partial changes derived from the molecular unit in the other layers. Besides the molecular surroundings in direct proximity, we consider solvent effects. Therefore, this model is composed of 7 layers ("-3" to "3", Figs.~\ref{fig:MolModels} (c,d)). Figure~\ref{fig:MolModels} (c) highlights the case where the central layer is considered the quantum region, and the other layers are considered at the level of partial charges. We assume that top (layer "3") and bottom (layer "-3") surface dominate the optical properties of the nanoparticle. They account for almost 77\% of the total nanoparticle surface in our basic rectangular approximation of the nanoparticle depicted in Fig.~\ref{fig:MolModels} (e). 

Second, to directly compare theoretical predictions with the experimental measurements, we consider a rectangular finite--size supramolecular nanoparticle composed of 35 unit cells in $x$--direction, 41 unit cells in $y$--direction, and 7 unit cells in $z$--direction. This results in 10,045 unit cells with the $\sim$60 nm length in lateral dimensions and $\sim$9 nm in the third dimension, corresponding to the mean of the particle size distribution from experiments done at the same materials \cite{SOPTEI20158, Fakhouri2019, Ni2024}. Each of the seven layers is built of 1,435 periodic cells in the form of a sheet of the material that is non--covalently coordinated with neighboring sheets above and below. 

However, the spatial details of this nanoparticle are not considered in the optical simulation. We opt to model infinite periodic films rather than finite structures for our optical simulations. To understand the optical response from an actual finite--sized supramolecular nanoparticle, we average the response from the film over all possible incidence angles from one half--space.
This approach is the only one that is computationally possible while still yielding reliable and representative results of the system's behavior. By leveraging the periodic boundary conditions, we reduce the complexity of the simulations without compromising the accuracy of the obtained results.

To emphasize the importance of taking the partial charges into account, we also perform both TD--DFT and optical simulations for the \textbf{non--embedded model}. In this model, we consider in the TD--DFT simulations a single unit cell dispersed in water without accounting for the further molecular surroundings. 

\section{Methodology and Workflow}

This section describes the methodology used in this work and outlines the workflow. The section consists of two sub--sections. In the first sub--section, we describe the optical simulations. Here, the nanoparticle, in essence, is considered as a stack of layers consisting of a periodic arrangement of molecular units. Each molecular unit
is described by a T--matrix and a hyper--T--matrix. In the second subsection, we describe the quantum--chemical approach to calculate these two quantities using TD--DFT.  

\subsection{T--matrix and hyper--T--matrix formalism}

The crucial object in our approach to predict the optical response of molecular nanophotonic materials is the transition matrix, or, in short, the T--matrix. Once the T--matrix of a molecule is known, we can efficiently simulate the response of films composed of layers of periodic arrangements of molecular units. Stacking multiple layers is an effective approach for handling thin films composed of discrete layers. The possibility of constructing the T--matrix for molecular materials based on precise quantum chemistry calculations allows us to bridge scales. 

The entire methodology relies on an expansion of all fields involved into vector spherical harmonics in a specific basis. The T--matrix of a scatterer links the amplitude coefficients of the incident field with those of the scattered field in a matrix--vector--product \cite{Waterman, Tsang2000-BasicTheoryofElect, Rahimzadegan}. The T--matrix method is usually applied when describing the linear response of a scatterer or molecule. However, the formalism has recently been extended to accommodate nonlinear optical effects in molecules. The hyper--T--matrix then encodes such nonlinear effects, and it is obtained from the molecular hyperpolarizabilities \cite{Zerulla2024Feb, https://doi.org/10.1002/adom.202400150}.

When considering the linear response of a 2D periodic array of identical scatterers, we need to solve \cite{Zerulla2022May}
\begin{equation}\label{eq_c_a_lattice}
\boldsymbol{c_0}^{\omega}=\left(\mathbb{1}-\mathrm{T}\left(\omega, \omega\right)\sum_{\boldsymbol{R} \neq 0} \mathbf{C}^{(3)}(-\boldsymbol{R}) \mathrm{e}^{\mathrm{i} \boldsymbol{k}_{\mid \mid}^{\omega}\boldsymbol{R}}\right)^{-1} \mathrm{~T}\left(\omega, \omega\right) \boldsymbol{a_0}^{\omega}\,.
\end{equation}
This equation establishes a connection between the amplitudes expanding the scattered field $\boldsymbol{c_0}^{\omega}$ from the particle at the lattice origin and the amplitude coefficients expanding the incident field $\boldsymbol{a_0}^{\omega}$. Here, $\omega$ represents the frequency of incident and scattered fields.  $\mathrm{T}\left(\omega, \omega\right)$ is a T--matrix of a single building block, and it could be derived using different approaches, {\it e.g.}, experimental measurements, finite element calculations, or the TD--DFT approach discussed below. In this work, the latter is applied since we wish to handle molecules. Note that Eq.~\ref{eq_c_a_lattice} is expressed in a basis of well--defined parity, {\it i.e.}, TM and TE modes. In our case, the TM and TE modes are represented as vector spherical waves $\mathbf{N}\left(k(\omega),\boldsymbol{r}\right)$ and $\mathbf{M}\left(k(\omega),\boldsymbol{r}\right)$, respectively, where $\boldsymbol{r}$ represents the point at which the fields are measured, $k\left(\omega\right)$ is the wavenumber. Moreover, $\mathbf{C}^{(3)}(-\boldsymbol{R})$ is a matrix expressing the translation coefficients for vector spherical waves, $\boldsymbol{R}$ is a lattice point in the 2D array and $\boldsymbol{k}_{\mid \mid}^{\omega}$ is the tangential component of the wave vector of the incident plane wave. More information on T--matrix formalism can be found in the Supporting Information.

The physical meaning of the equation is rather straight. The optical response of the scatterer in the lattice is given as a product of a T--matrix with the amplitude coefficients expanding the incident field. The T--matrix is renormalized due to the interaction with all the scatterers forming the lattice, and this lattice interaction appears in the denominator.   

In this work, we consider thin films that consist of multiple equidistant layers of 2D periodic lattices. To predict the response of the multilayered structure, one should solve Eq.~\ref{eq_c_a_lattice} for each layer individually, and subsequently stack them on top of each other employing the Q-matrices from equations (6)-(9) in Ref.~\cite{Beutel2021Jun}, which solve the system rigorously.

The nonlinear response of a system is typically characterized using the nonlinear bulk susceptibility, which often requires approximations or even relies on phenomenological approaches for estimation. However, in this work, we rely on a recently developed approach involving the hyper--T--matrix formalism \cite{Zerulla2024Feb, https://doi.org/10.1002/adom.202400150}. In a nutshell, the approach predicts the nonlinear multipolar response of a system based on its hyperpolarizabilities and multipolar decomposition of incident plane waves. The hyperpolarizabilities are calculated using TD--DFT. 

In slightly more detail, the treatment of the nonlinear response consists of different steps. First, we define a matrix that connects the expansion coefficients of the incident waves (the number of which is determined by the nonlinear effect considered) with the scattered field at the sum of the frequencies considered in the illumination. This is precisely what the hyper--T--matrix represents. Our focus will be on the nonlinear dipolar response of the second harmonic generation (SHG). Using the hyper--T--matrix, we connect the dipolar expansion coefficients of the scattered field $c_{1m}^{\Omega}$ at the second harmonic (SH) frequency $\Omega = 2\omega$ to the expansion coefficients $a_{1r}^{\omega}$ and $a_{2s}^{\omega}$ of the incident waves at the fundamental frequency $\omega$.
The hyper--T--matrix is then defined as, 
\begin{equation}\label{eq_THyper}
c_{1m}^{\Omega}=\sum_{r,s}\mathrm{T}_{mrs}^{\mathrm{Hyper}}\left(-\Omega;\omega,\omega\right)a_{1r}^{\omega}a_{2s}^{\omega}\,.\end{equation}
Building on the previous work in nonlinear multi--scale simulations \cite{Zerulla2024Feb, https://doi.org/10.1002/adom.202400150}, we introduce a new formula for deriving the hyper--T--matrix from first hyperpolarizabilities,
\begin{align}\label{hyper-T}
    T^{\text{Hyper}}_{mrs}\left(-\Omega\right) = \frac{c_{\mathrm{h}}\left(\Omega\right) Z_{\mathrm{h}}\left(\Omega\right) \left(k_{\mathrm{h}}\left(\Omega\right) \right)^3}{2 \cdot\left(6 \pi\right)^{\frac{3}{2}}} \times \notag \\
    \times \sum_{ijk} {A}_{mi} \beta_{ijk}\left(-\Omega \right) [\textbf{A}^{-1}]_{jr} [\textbf{A}^{-1}]_{ks},
\end{align}
where the matrix \textbf{A} transforms the vector from Cartesian to spherical coordinates (\cite{Fernandez-Corbaton:2020}, Eq.~(S8)),
$\boldsymbol{\beta}_{ijk}\left(-\Omega \right)$ is the first hyperpolarizability in the Cartesian basis describing the SHG process, $c_{\mathrm{h}}\left(\Omega\right) =1 / \sqrt{\varepsilon_{\mathrm{h}}\left(\Omega\right) \mu_{\mathrm{h}}\left(\Omega\right)}$
is the speed of light in the surrounding medium, $Z_{\mathrm{h}}\left(\Omega\right) =\sqrt{\mu_{\mathrm{h}}\left(\Omega\right) \varepsilon_{\mathrm{h}}\left(\Omega\right) }$ its wave impedance, and $k_{\mathrm{h}}\left(\Omega\right) =\Omega \sqrt{\varepsilon_{\mathrm{h}}\left(\Omega\right) \mu_{\mathrm{h}}\left(\Omega\right)}$ its wave number \cite{Zerulla2024Feb}. The $\varepsilon_{\mathrm{h}}\left(\Omega\right)$ is the permittivity and $\mu_{\mathrm{h}}\left(\Omega\right)$ is the permeability of the surrounding medium. The expression in Eq.~\ref{hyper-T} results in faster calculations than using the equivalent expression in \cite[Eq.~(7)]{Zerulla2024Feb}.

Given the scattered second-harmonic (SH) dipole moment of the single unit cell at the origin of the layer, $c^{\Omega}_{1m}$ (Eq.~\ref{eq_THyper}), we calculate the SH response of the entire periodic layer using Eq.~(8) from Ref.~\cite{Zerulla2024Feb}.  This equation resembles Eq.~\ref{eq_c_a_lattice}, but differs in two ways: the frequency $\omega$ is replaced by the SH frequency $\Omega$, and the T--matrix term outside the brackets is absent on the right-hand side. In this case, $c^{\Omega}_{1m}$ act as effective incident field coefficients, so multiplication by the T--matrix is unnecessary.

We employ the established formalism to calculate the SHG electric field arising from the multilayered structure in Fig.~\ref{fig:MolModels} (d) illuminated by two incident plane waves. This involves formulating the following steps within the optical part of the procedure: 
\begin{itemize}
  \item Using Eq.~\ref{eq_c_a_lattice}, we calculate the linear optical response for each layer of the multilayered structure.
  \item For each layer, we sum up the incident field with the field reflected and transmitted through all the other layers in the structure to get the expression for the total linear field incident on this layer.
  \item Once we know the total incident field at the fundamental frequency, we calculate the nonlinear dipolar response of the unit cell at the origin in terms of the SHG electric field using the hyper--T--matrix in Eq.~\ref{eq_THyper}.
  \item Then we calculate the SHG scattering response of the entire layer using Eq.~(8) from Ref.~\cite{Zerulla2024Feb}.
  \item Finally, the SHG field from each layer propagates through the structure while being reflected and transmitted at every interface. The contributions from all layers are then summed at the top to yield the overall SHG field leaving the structure.
\end{itemize}
All of the steps presented above are conducted using the publicly available in--house developed software \textit{treams} \cite{Beutel2024Apr}.

\subsection{Time--dependent Density Functional Theory}

To predict the linear and nonlinear optical response from the nanoparticles, we need to know the T--matrices and the hyper--T--matrices expressing the properties of the molecule. They are obtained using time--dependent density functional theory (TD--DFT).

The description starts with the consideration of the optimized periodic cell of Au--cysteine molecular material. Such unit cell was used as a finite--size model to calculate the ground state electron density as described in SI. Upon converging the ground state electron density, we employed TD--DFT to calculate damped dynamic first hyperpolarizability tensors for non--embedded and embedded models. We also calculated the linear properties of the molecules in the form of damped dynamic polarizabilities at the fundamental and second--harmonic wavelengths. The calculated polarizabilities and first hyperpolarizabilities from quantum chemistry calculations were used to construct T--matrices and hyper--T--matrices for the scattering simulations of the thin Au-cysteine nanofilm.

Furthermore, we calculated the first hyperpolarizabilities in the molecular frame ${\beta}_{ijk}$ to calculate the first hyperpolarizabilities measured using hyper--Rayleigh scattering technique ${\beta}_{HRS}$ in the laboratory frame for direct comparison. The relations connecting molecular first hyperpolarizabilities with those obtained experimentally are \cite{Antoine}:

\begin{equation}\label{eq_beta_HRS}
 \Bigl \langle {\beta}_{HRS}^{2} \bigr \rangle = \Bigl \langle {\beta}_{XZZ}^{2} \bigr \rangle + \Bigl \langle {\beta}_{ZZZ}^{2} \bigr \rangle
\end{equation}

\begin{equation}\label{eq_beta_XZZ}
\begin{split}
 \Bigl \langle {\beta}_{XZZ}^{2} \bigr \rangle = \frac{1}{35}\sum_{i}{\beta}_{iii}^{2} -\frac{2}{105}\sum_{i\neq j}{\beta}_{iii}^{2}{\beta}_{ijj}^{2} +\frac{11}{105}\sum_{i\neq j}{\beta}_{iij}^{2} \\ - \frac{2}{105}\sum_{ijk, cyclic}\left({\beta}_{iij}^{2}{\beta}_{jkk}^{2} + \frac{8}{35}{\beta}_{ijk}^{2}\right)
 \end{split}
\end{equation}

\begin{equation}\label{eq_beta_ZZZ}
\begin{split}
 \Bigl \langle {\beta}_{ZZZ}^{2} \bigr \rangle = \frac{1}{7}\sum_{i}{\beta}_{iii}^{2} +\frac{6}{35}\sum_{i\neq j}{\beta}_{iii}^{2}{\beta}_{ijj}^{2} +\frac{9}{35}\sum_{i\neq j}{\beta}_{iij}^{2} \\ + \frac{6}{35}\sum_{ijk, cyclic}\left({\beta}_{iij}^{2}{\beta}_{jkk}^{2} + \frac{12}{35}{\beta}_{ijk}^{2}\right)
 \end{split}
\end{equation}

\section{Results}
This section presents our simulation results and compares them with experimental data from the literature \cite{Fakhouri2019}. We demonstrate that accurately predicting the dispersion of the second-harmonic (SH) signal requires a precise treatment of the molecular details of the embedding. Here, dispersion refers to the variation of SH signal intensity with wavelength over an experimentally accessible range. Neglecting the embedding yields a prediction of an anomalous dispersion, whereas incorporating it correctly leads to a normal dispersion, in full agreement with experimental observations. Normal dispersion here means a decrease in the SHG signal with increasing wavelength. An anomalous dispersion would mean an increase in the SHG signal with increasing wavelength. 
This different quantitative behavior underlines the importance of correctly considering the embedding in such descriptions. 

In the first sub--section, we illustrate these effects through a combined quantum chemical and optical simulation approach. The second sub--section further substantiates our findings by showing that the fundamental mechanisms already emerge from quantum-chemical calculations.

\subsection{Linear and nonlinear optical response of the gold--cysteine thin film from the multi--scale scattering simulations}
In this subsection, we analyze the linear and nonlinear scattering from the gold--cysteine nanoparticles dispersed in water. We compare our results from the multi--scale approach with experimental data published in the literature \cite{Fakhouri2019}. It is important to note that while the experimental measurements were conducted on gold--cysteine nanoparticles randomly dispersed in water, our study considers thin films made from a periodic arrangement of molecular units. As a result, an exact replication of the experimental results is not expected. Nevertheless, our aim is to demonstrate that the overall behavior of the physical quantities is similar. Because the measurements were done with nanoparticles in solution that have a random orientation relative to the incident field, we rotationally averaged both linear and nonlinear quantities. The rotational averaging is performed by rotating an incident plane wave over equally distributed angles in the lower half-space and subsequently averaging the results. 

\begin{figure}[t!]
\centering
\includegraphics[width=1.0\columnwidth]{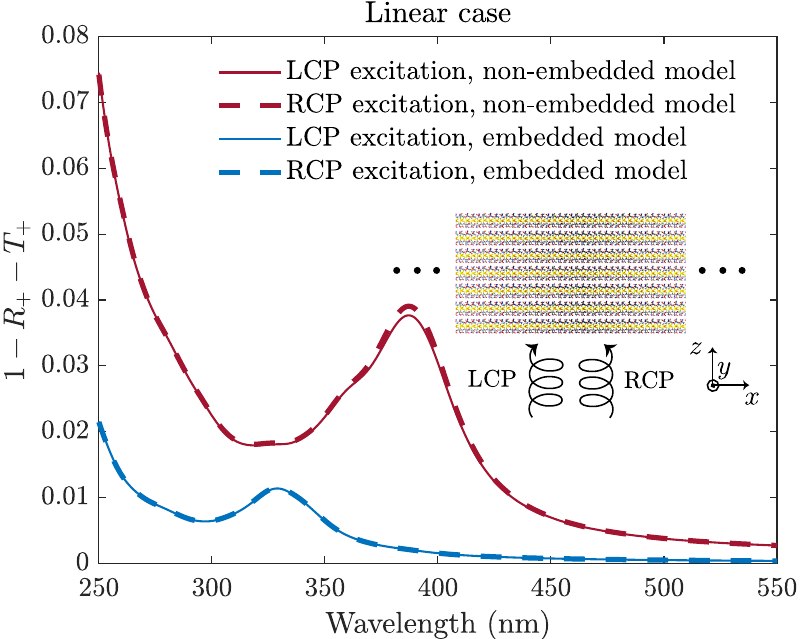}
\caption{Absorption spectrum of a thin film of gold--cysteine nanoparticles with a width of 9.1~nm (7~layers) upon illumination with a left-- and right--handed circularly polarized wave. The system under consideration is shown as an insert. For illustrative purposes, the incident light is shown in the case of normal incidence.} \label{fig_beta_sheets_lin_lcp_abs}
\end{figure}

The absorption spectrum of the left- and right-handed circular polarized waves is presented in Fig.~\ref{fig_beta_sheets_lin_lcp_abs}. The calculation considers an infinite film in the $(x, y)$-plane made from seven molecular layers. The total thickness is $9.1$~nm. The wave vector of the incident wave is rotationally averaged in the lower half-space. In this specific case, we selected the incident field to consist of circularly polarized waves. This choice enabled us to compute the circular dichroism (CD) spectrum, which we then compared against the experimental data. The spectrum of CD can be found in the Supporting Information. 

As discussed above, we study the response in two different configurations. In the first scenario, which we call the non--embedded model, all the unit cells have the same linear and nonlinear responses. The T--matrices and hyper--T--matrices are calculated without taking the embedding into account. The second case considered is the embedded model, where partial charges are taken into account in TD--DFT simulations, as discussed in Chapter \rom{2}. From a computational point of view, in this model, both linear and nonlinear responses of each unit cell depend on the layer. 

In Fig.~\ref{fig_beta_sheets_lin_lcp_abs}, we compare the absorption for the previously mentioned non--embedded and embedded models. In both cases, we simulated a material composed of seven infinite sheets stacked above each other. Both models show a reasonable agreement with the experimental data (see Fig. S6 and Fig.~2 (a) in Reference \cite{Fakhouri2019}). However, the spectral position of the absorption peak in the embedded model aligns much better with the experimental results. From Fig. \ref{fig_beta_sheets_lin_lcp_abs}, we see that in the non--embedded case, for both left- and right-handed circularly polarized incident light, the absorption peak is at $\approx 390$ nm. However, the experimentally measured peak value occurs at $\approx 350$ nm. For the embedded model, on the other hand, the absorption peak is positioned at $\approx 330$ nm, which is much closer to the experimental value.

\begin{figure*}[htp!]
\centering
\includegraphics[width=1.0\textwidth]{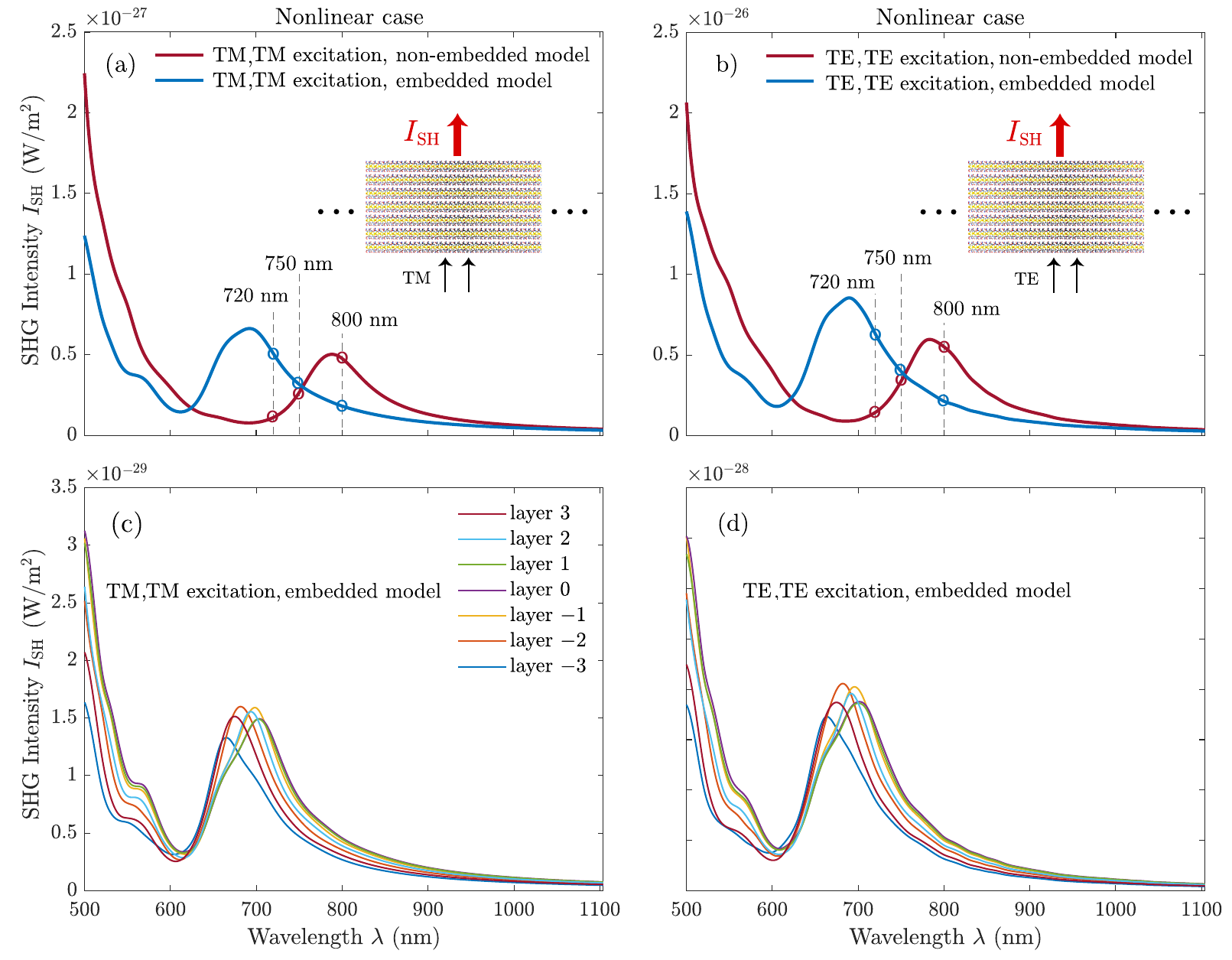}
\caption{Spectral dependence of the second harmonic (SH) signal from a gold–cysteine film under the illumination of two plane waves with (a) TM--polarization, (b) TE--polarization. The system under consideration is shown in the insets. For illustrative purposes, the incident light is shown in the case of normal incidence. SH intensity contributions from each of the seven layers within the embedded model under (c) TM, TM and, (d) TE, TE excitation.} \label{fig_beta_sheets_nonlin} 
\end{figure*}

We now turn our attention to the nonlinear properties of the system. In Fig.~\ref{fig_beta_sheets_nonlin}, the intensities of SHG plane waves propagating in the positive z--direction are depicted, resulting from the SHG scattering of two incident plane waves propagating in the positive $z$--direction with the amplitude of $1$ V/m upon interaction with a gold--cysteine film. In Fig.~\ref{fig_beta_sheets_nonlin} (a), both incident plane waves are TM--polarized, and, in Fig.~\ref{fig_beta_sheets_nonlin} (b), TE--polarized. At normal incidence, the TM--polarization corresponds to a polarization along the $x$--axis, and the TE--polarization corresponds to a
polarization along the $y$-axis. For oblique incidence, the fields are
rotated accordingly. The incidence angle is the same for both plane waves, and the wave vectors are rotationally averaged in the lower half-space. 

In the nonlinear case, we also compare the non--embedded and embedded models. The simulation for the non--embedded model, shown in red, employs identical T-matrices and hyper--T--matrices for all molecular building blocks, calculated without considering the embedding. However, this method predicts an incorrect trend for three values at three wavelengths indicated in Figs.~\ref{fig_beta_sheets_nonlin} (a)-(b) and in the table \ref{tbl:beta_HRS}. In Ref.~\cite{Fakhouri2019}, the measured values for these three wavelengths exhibit a normal dispersion with increasing wavelength, contrary to the simulations of the simplified molecular model shown in Figs.~\ref{fig_beta_sheets_nonlin} (a)-(b) as the red curves. However, when we examine the embedded model, we obtain results that exhibit the correct trend, aligning with experimental observations (blue curves). In Figs.~\ref{fig_beta_sheets_nonlin} (c)-(d), one can see the contributions of each of the seven layers to the total SH intensity for both the TM and TE excitations. All layers contribute more or less equally to the overall sum. 

From our optical simulations, we conclude that considering only the properties of the non--embedded unit cell and placing it into an infinite periodic structure using electromagnetic interactions does not accurately describe the system's behavior. To obtain more accurate results, it is necessary to adopt a more complex embedded model, where the surrounding is taken into account on the quantum chemistry level. Only with this approach can we obtain results that align  with the experimental data. In the following subsection, we demonstrate that this conclusion can already be drawn from the quantum chemistry simulations.

Finally, we note that although the results of the experiments are presented in terms of ${\beta}_{HRS}$, and we have calculated the SH intensities, these intensities are theoretically expected to be proportional to the square of ${\beta}_{HRS}$ \cite{Antoine}.

\subsection{Quantum chemistry calculations}

Indeed, the same spectral features observed on the level of the Maxwell simulations can already be seen in the analysis of the nanomaterial based on finite--size molecular models and quantum chemistry calculations. 

\begin{figure*}[tp!]
\centering
\subfloat{
\includegraphics[width=0.47\textwidth]{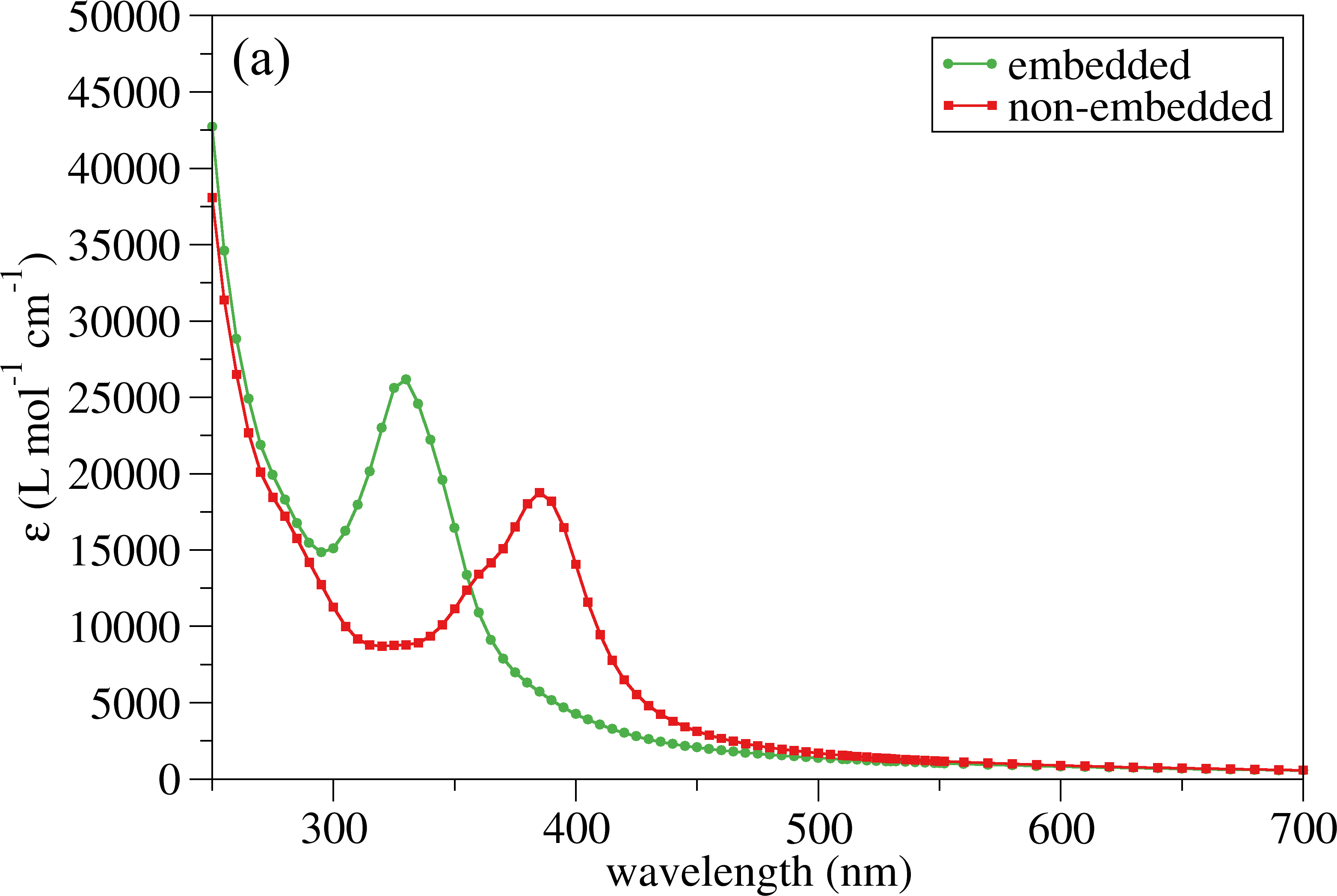}
 }\hspace{0.1cm}
\subfloat{
\includegraphics[width=0.47\textwidth]{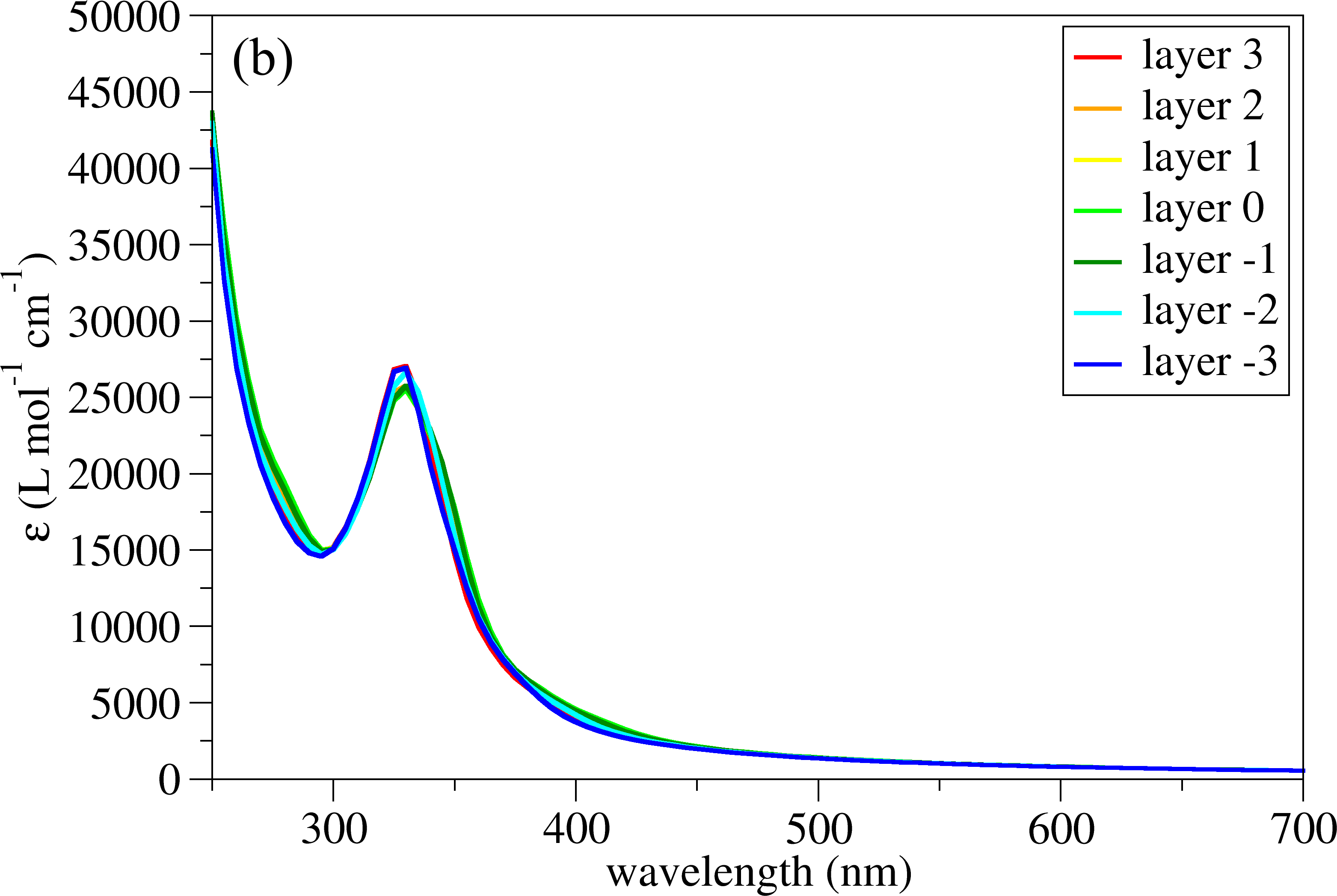}
}
\caption{\textbf{(a)} Comparison of the linear absorption predicted using two different models. On the one hand, we consider a non--embedded molecular model based on quantum chemistry calculations of one periodic cell of Au-cysteine material. This can be compared to the average of all 7 layers considered in the embedded model ("-3" to "3"). \textbf{(b)} Spectral dependency of the absorption in each of the seven layers considered in the embedded model.}
\label{fig:UV-Vis_embedding_bulk}
\end{figure*}

\begin{figure*}[tp!]
\centering
\subfloat{
\includegraphics[width=0.48\textwidth]{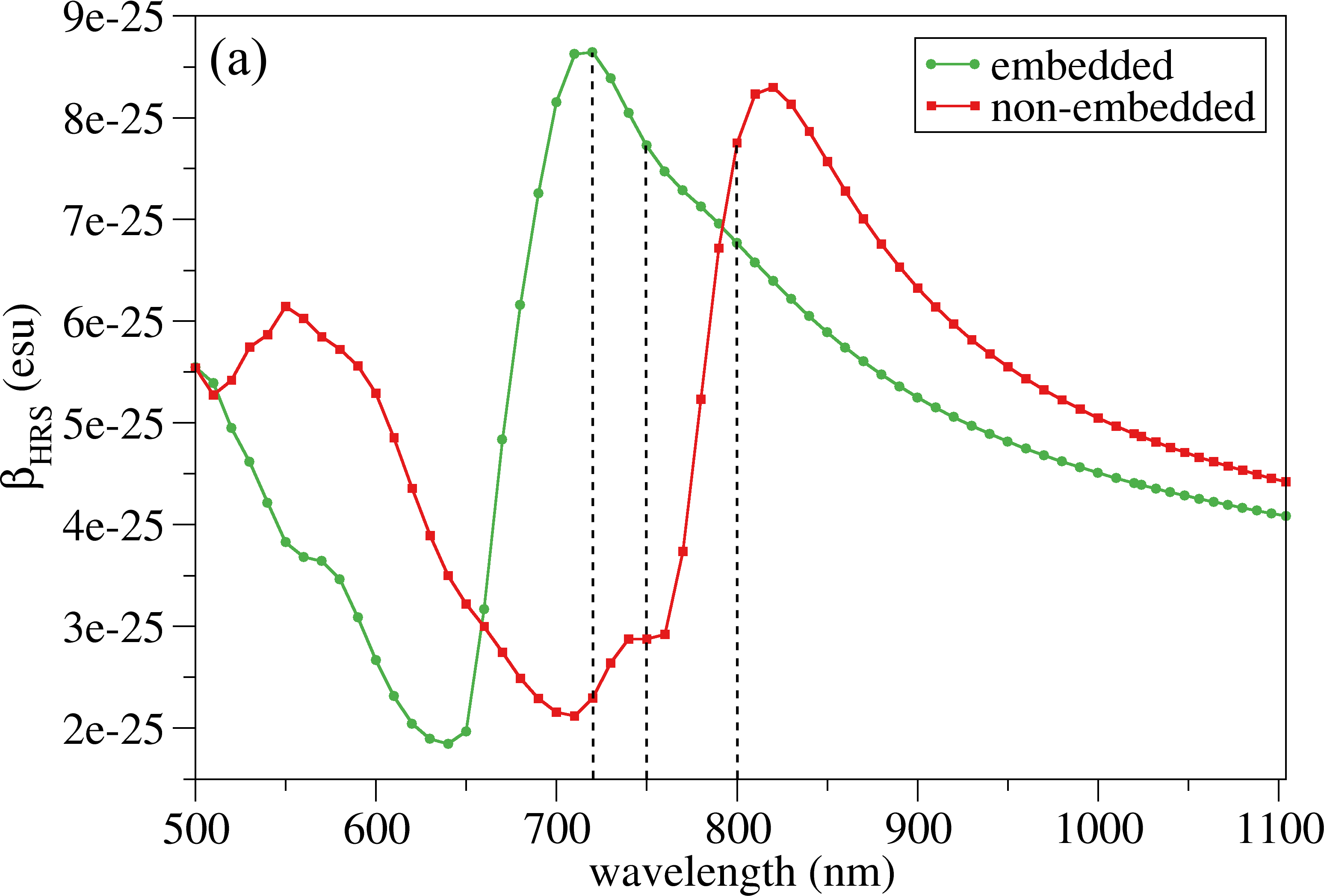}
 }
\subfloat{
\includegraphics[width=0.49\textwidth]{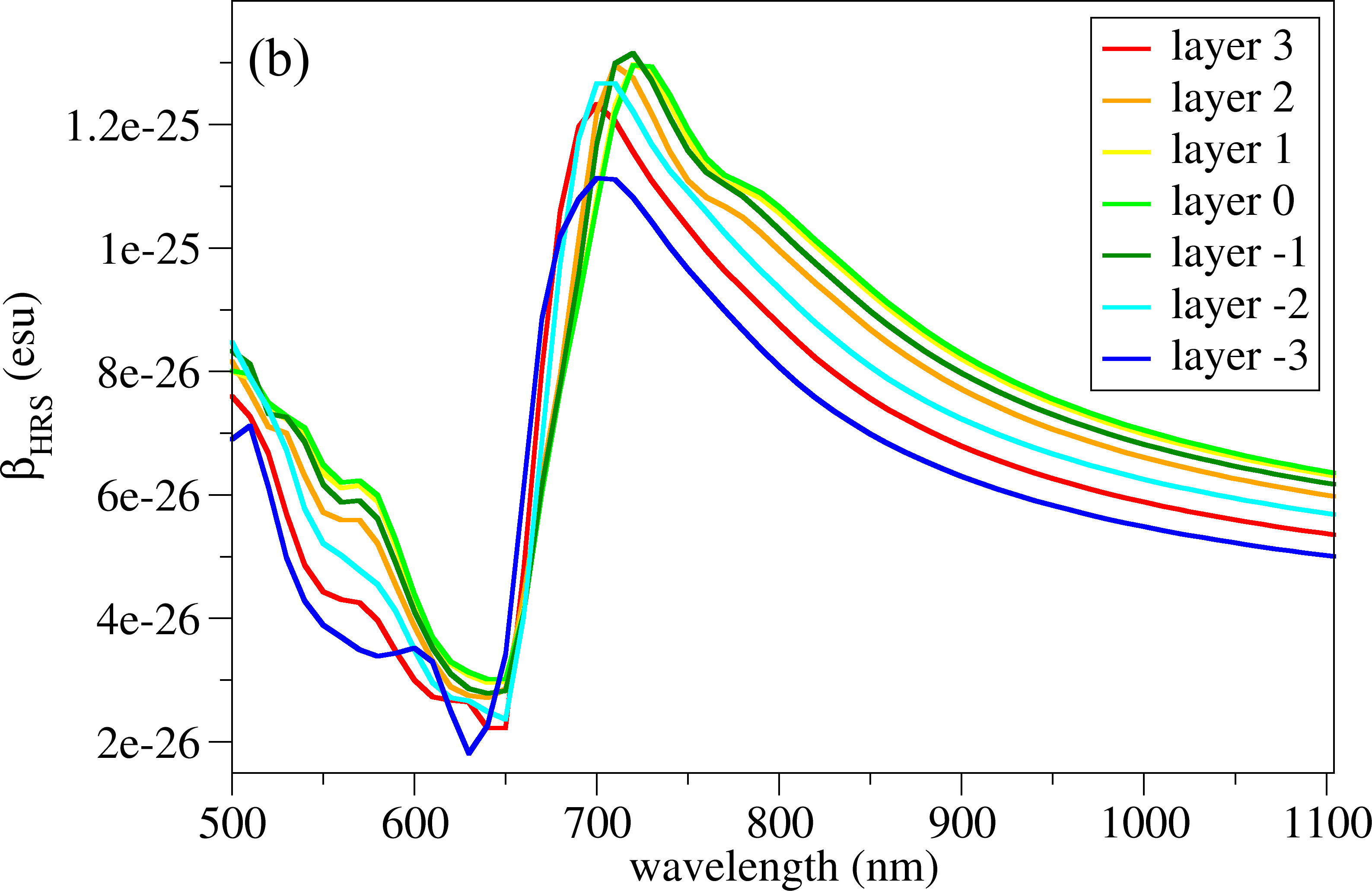}
}
\caption{\textbf{(a)} A comparison of the predicted hyper–Rayleigh scattering first hyperpolarizabilities ${\beta}_{HRS}$ using Eqs.~\eqref{eq_beta_HRS}  -- \eqref{eq_beta_ZZZ} for the simplified rectangular nanoparticle using the non--embedded model based on single periodic cell and the 7--layer model with periodic cell embedded within the corresponding partial charges of each layer as depicted in subfigure \textbf{(b)}. Elements of ${\beta}$ tensor are calculated using TD--DFT.}
\label{fig:SHG_embedding_bulk}
\end{figure*}

The comparison of the linear UV--Vis. absorption spectra of the single periodic cell of these hybrid nanomaterials with the averaged 7--layers model embedded in the field of partial charges to account for the influence of the surrounding nanomaterial is presented in Fig.~\ref{fig:UV-Vis_embedding_bulk} (a). The qualitative differences between the two models observed with the Maxwell solver can already be seen at the molecular level. Although both models express similar spectral features, the 7--layer model agrees better with the experimental observation of the absorption features for the nanoparticle where the first absorption band has a peak below 350 nm (Fig. S6) \cite{Fakhouri2019,SOPTEI20158,Ni2024}. In Fig.~\ref{fig:UV-Vis_embedding_bulk} (b) the individual UV--Vis. absorption spectra of all seven embedded layers are depicted. As can be concluded, a periodic cell shows very similar absorption features independent of its position within the thin film. 

When the linear electronic circular dichroism (ECD) of this material is considered, the same trend is observed. The non--embedded model fails to capture the observed features correctly. The averaged 7--layer embedded model ECD, on the other hand, shows a positive, although much weaker peak above 350~nm, in good agreement with experiments (Fig. S8) \cite{Fakhouri2019}, as well as negative signals below 350~nm (Fig.~S3 (a)). In contrast to the absorption analysis of each of the 7 layers embedded into the field of partial charges, the ECD of each layer model shows minimal position-dependent behavior, as can be seen in Fig.~S3 (b).

\setlength{\extrarowheight}{1pt}
\begin{table*}[ht!]
\centering
\caption{Hyper–Rayleigh scattering first hyperpolarizabilities ${\beta}_{HRS}$, predicted by TD--DFT simulations, and the SH  intensities $I_{SHG}$ of the scattered field, obtained from optical simulations for TE-incident waves presented for two molecular models: a non--embedded model based on a single periodic cell and an embedded model with a periodic cell embedded within a field of partial charges. These results are compared with experimentally obtained values from Ref.~\cite{Fakhouri2019}. The simple non--embedded model (\textcolor{bordo}{red}, \textcolor{Mred}{red}) fails to reproduce the trend of measured ${\beta}_{HRS}$ with increasing wavelength, whereas the more advanced embedded model (\textcolor{myblue}{blue}, \textcolor{Mgreen}{green}) successfully captures the experimental trend.}
\label{tbl:beta_HRS}
\begin{adjustbox}{width=1.0\textwidth}

\fontsize{10}{13}\selectfont
\setlength{\tabcolsep}{0.2em}
\begin{tabular}{|c||c|c||c|c||c|}
\hline\hline
\textbf{wavelength} & \textbf{non--embedded, Optics} &  \textbf{non--embedded, DFT} & \textbf{embedded, Optics} & \textbf{embedded, DFT} & \textbf{experiment} \cite{Fakhouri2019} \\
(nm) & $I_{SHG}$ ($\cdot$10\textsuperscript{-27} W/m$^2$)&${\beta}_{HRS}$ ($\cdot$10\textsuperscript{-25} esu) &  $I_{SHG}$ ($\cdot$10\textsuperscript{-27} W/m$^2$)&${\beta}_{HRS}$ ($\cdot$10\textsuperscript{-25} esu) & ${\beta}_{HRS}$ ($\cdot$10\textsuperscript{-25} esu) \\
 \hline
 720 & \textcolor{bordo}{1.44} &  \textcolor{Mred}{2.29} & \textcolor{myblue}{6.30} &\textcolor{Mgreen}{8.64} &  1.94   \\ 
 750 & \textcolor{bordo}{3.37} & \textcolor{Mred}{2.86} & \textcolor{myblue}{3.98} &\textcolor{Mgreen}{7.73} &  1.42   \\ 
 800 & \textcolor{bordo}{5.51} & \textcolor{Mred}{7.72} & \textcolor{myblue}{2.13} &\textcolor{Mgreen}{6.77} & 1.05    \\ 
 \hline\hline
\end{tabular}
\end{adjustbox}
\end{table*}

Now, we continue with the prediction and comparison of the nonlinear optical properties from the quantum chemical simulations. Specifically, we discuss the spectral dependency of ${\beta}_{HRS}$ that can be determined from the molecular ${\beta}_{ijk}$ complex tensors as described in Sec.~III. B. Figure~\ref{fig:SHG_embedding_bulk} compares the spectral dependent ${\beta}_{HRS}$ from 500~nm to 1104~nm. To obtain values close to the measured values \cite{Fakhouri2019}, we multiply the calculated values for the non--embedded model with 10,045 (number of unit cells in the simplified nanoparticle model) to reproduce the signal of the approximated rectangular nanoparticle as depicted in Fig.~\ref{fig:MolModels} (e). Similarly, we multiplied the predicted value from DFT by 1435 (number of unit cells in one layer of the nanoparticle) for each of the 7 layers to reach the same nanoparticle model size. Both nonlinear spectra for a single periodic cell in the implicit water surrounding, representing non--embedded material, and the embedded model of the Au-cysteine nanoparticle show qualitatively the same spectral dependency. However, the polarization of the surrounding material through static charges shifts the overall ${\beta}_{HRS}$ spectrum in the case of the embedded model to higher energies (lower wavelengths). 

The consequence of improving the quantum modeling with the inclusion of the surrounding through partial charges is obvious from Table~\ref{tbl:beta_HRS}, where it can be seen that the embedded model correctly captures the experimentally observed normal dispersion for ${\beta}_{HRS}$ with increasing wavelength \cite{Fakhouri2019}. On the contrary, the much simpler non--embedded model based on a single periodic cell of the material predicts the opposite and wrong trend. The change in the trends is due to a shift of the main absorption peak by approximately 65 nm between the two models. 

The ``position'' dependent ${\beta}_{HRS}$ of each of the seven layers is visualized in Fig.~\ref{fig:SHG_embedding_bulk} (b). It shows an interesting position dependence where layers close to the surface (``3" and ``-3") exhibit a slight shift of the signal towards higher energies compared to the central layer together with a change in intensity up to $\sim$20\%. Thus, each of the seven layers contributes differently to the overall observed ${\beta}_{HRS}$ signal.

Quantitatively, our calculations overestimate the measured values by a factor of $\sim$3-6. This can be attributed to the choice of the DFT functional and atomic orbital basis set used in quantum calculations, as well as our simplistic model of the rectangular nanoparticle. Nevertheless, the general conclusion can be made here, that for the correct prediction of the optical properties of large and complex novel molecular nanomaterials, such as these hybrid metal--organic materials, on the quantum level it is important to include the surrounding effects of the material to the explicitly considered quantum region through embedding procedures. Furthermore, an inclusion of an implicit description of the solvent effects is also advised.

\section{Conclusions}

In this work, we applied a multi--scale modeling framework to explore linear and nonlinear light--matter interactions of molecular materials to study the properties of sub--100 nm sized gold--cysteine nanoparticles. Numerically, the nanoparticles were treated in the form of thin films. We find that the experimental results are accurately predicted only if one uses embedded molecular models in which the surroundings are accounted for through partial charges. In such cases, the predicted spectra are in excellent agreement with previously reported measurements. In sharp contrast, assuming that the molecular units in the film can be described based on non--embedded isolated molecular models produces the wrong predictions. A minor influence of layer dependency is observed only for nonlinear optical properties. This can be explained by the surface effects contributing to the calculated first hyperpolarizabilities for molecular models of layers close to the surface of the film/nanoparticle.

Our advanced molecular modeling shows that, for improved prediction accuracy, the influence of the surrounding molecular material should not be neglected. We account for it by embedding the molecular periodic cell within the static partial charge field of the same cell to include polarization effects of the neighboring molecules. Our workflow for studies of nonlinear effects based on hyper--T--matrix formalism was used to further improve the optical response predictions of such large molecular systems beyond sizes accessible for quantum description. 

The presented scale--bridging approach is general, but can be further improved by homogenizing the nonlinear optical properties of the molecular materials. Such homogenization would allow us to treat arbitrarily shaped molecular systems. Alternatively, we could also refine the molecular description by including a position-dependent linear and nonlinear response of each molecular unit within the nanoparticle also within the $x$-- and $y$--direction. Future work can then remove the assumption of an infinitely extended system, by computing layer and lateral--position--dependent microscopic responses and solving the non-periodic multi-scattering problem. This would allow us to achieve a fully finite--size description of large nanoparticles in different surroundings on the level of a Maxwell theory. Currently, further research in those directions is ongoing. 

\section*{Data availability}
The outputs of quantum chemistry simulations to model these nanoparticles are deposited within NOMAD repository under the following DOI: \url{10.17172/NOMAD/2025.02.27-1}.
Other data produced within this study are available from the authors upon reasonable request. Our in--house developed software \textit{treams} is available from the following url: \url{https://github.com/tfp-photonics/treams}

\begin{acknowledgement}
M.P., M.K., and C.R. acknowledge support by the Deutsche Forschungsgemeinschaft (DFG, German Research Foundation) under Germany’s Excellence Strategy via the Excellence Cluster 3D Matter Made to Order (EXC-2082/1-390761711) and from the Carl Zeiss Foundation via the CZF-Focus@HEiKA Program. M.K., C.H., and C.R. acknowledge funding by the Volkswagen Foundation. I.F.C. and C.R. acknowledge support by the Helmholtz Association via the Helmholtz program “Materials Systems Engineering” (MSE). B.Z. and C.R. acknowledge support by the KIT through the “Virtual Materials Design” (VIRTMAT) project. R.A, A.P., and J.O.B. acknowledge PHC POLONIUM (Project: 49252PM, “Gold nanoclusters in chiral nanoassemblies - nonlinear optical properties”) for support. A.P. acknowledges ERASMUS+ program for funding Traineeship
for Academic Year 2021/2022 at Institut Lumière Matière, France. M.K. and C.R. acknowledge support by the state of Baden--Württemberg through bwHPC and the German Research Foundation (DFG) through grant no. INST 40/575-1 FUGG (JUSTUS 2 cluster) and the HoreKa supercomputer funded by the Ministry of Science, Research and the Arts Baden--Württemberg and by the Federal Ministry of Education and Research. 
\end{acknowledgement}


\bibliography{main}

\providecommand{\latin}[1]{#1}
\makeatletter
\providecommand{\doi}
  {\begingroup\let\do\@makeother\dospecials
  \catcode`\{=1 \catcode`\}=2 \doi@aux}
\providecommand{\doi@aux}[1]{\endgroup\texttt{#1}}
\makeatother
\providecommand*\mcitethebibliography{\thebibliography}
\csname @ifundefined\endcsname{endmcitethebibliography}  {\let\endmcitethebibliography\endthebibliography}{}
\begin{mcitethebibliography}{76}
\providecommand*\natexlab[1]{#1}
\providecommand*\mciteSetBstSublistMode[1]{}
\providecommand*\mciteSetBstMaxWidthForm[2]{}
\providecommand*\mciteBstWouldAddEndPuncttrue
  {\def\EndOfBibitem{\unskip.}}
\providecommand*\mciteBstWouldAddEndPunctfalse
  {\let\EndOfBibitem\relax}
\providecommand*\mciteSetBstMidEndSepPunct[3]{}
\providecommand*\mciteSetBstSublistLabelBeginEnd[3]{}
\providecommand*\EndOfBibitem{}
\mciteSetBstSublistMode{f}
\mciteSetBstMaxWidthForm{subitem}{(\alph{mcitesubitemcount})}
\mciteSetBstSublistLabelBeginEnd
  {\mcitemaxwidthsubitemform\space}
  {\relax}
  {\relax}

\bibitem[Bellini \latin{et~al.}(2018)Bellini, Bevilacqua, Marchionni, Miller, Filippi, Grützmacher, and Vizza]{https://doi.org/10.1002/ejic.201800829}
Bellini,~M.; Bevilacqua,~M.; Marchionni,~A.; Miller,~H.~A.; Filippi,~J.; Grützmacher,~H.; Vizza,~F. Energy Production and Storage Promoted by Organometallic Complexes. \emph{European Journal of Inorganic Chemistry} \textbf{2018}, \emph{2018}, 4393--4412\relax
\mciteBstWouldAddEndPuncttrue
\mciteSetBstMidEndSepPunct{\mcitedefaultmidpunct}
{\mcitedefaultendpunct}{\mcitedefaultseppunct}\relax
\EndOfBibitem
\bibitem[Bönnemann and Khelashvili(2010)Bönnemann, and Khelashvili]{https://doi.org/10.1002/aoc.1613}
Bönnemann,~H.; Khelashvili,~G. Efficient fuel cell catalysts emerging from organometallic chemistry. \emph{Applied Organometallic Chemistry} \textbf{2010}, \emph{24}, 257--268\relax
\mciteBstWouldAddEndPuncttrue
\mciteSetBstMidEndSepPunct{\mcitedefaultmidpunct}
{\mcitedefaultendpunct}{\mcitedefaultseppunct}\relax
\EndOfBibitem
\bibitem[Thomas and Raja(2004)Thomas, and Raja]{THOMAS20044110}
Thomas,~J.~M.; Raja,~R. Catalytic significance of organometallic compounds immobilized on mesoporous silica: economically and environmentally important examples. \emph{Journal of Organometallic Chemistry} \textbf{2004}, \emph{689}, 4110--4124, 40th Anniversary Issue - Dedicated to Professor Colin Eaborn\relax
\mciteBstWouldAddEndPuncttrue
\mciteSetBstMidEndSepPunct{\mcitedefaultmidpunct}
{\mcitedefaultendpunct}{\mcitedefaultseppunct}\relax
\EndOfBibitem
\bibitem[Philippot and Chaudret(2003)Philippot, and Chaudret]{PHILIPPOT20031019}
Philippot,~K.; Chaudret,~B. Organometallic approach to the synthesis and surface reactivity of noble metal nanoparticles. \emph{Comptes Rendus Chimie} \textbf{2003}, \emph{6}, 1019--1034\relax
\mciteBstWouldAddEndPuncttrue
\mciteSetBstMidEndSepPunct{\mcitedefaultmidpunct}
{\mcitedefaultendpunct}{\mcitedefaultseppunct}\relax
\EndOfBibitem
\bibitem[Zhang \latin{et~al.}(2018)Zhang, Li, Li, Zhong, Liao, and Li]{ZHANG201865}
Zhang,~H.; Li,~L.; Li,~Z.; Zhong,~W.; Liao,~H.; Li,~Z. Controllable synthesis of {SnO2}@carbon hollow sphere based on bi-functional metallo-organic molecule for high-performance anode in {Li}-ion batteries. \emph{Applied Surface Science} \textbf{2018}, \emph{442}, 65--70\relax
\mciteBstWouldAddEndPuncttrue
\mciteSetBstMidEndSepPunct{\mcitedefaultmidpunct}
{\mcitedefaultendpunct}{\mcitedefaultseppunct}\relax
\EndOfBibitem
\bibitem[Patra and Gasser(2012)Patra, and Gasser]{https://doi.org/10.1002/cbic.201200159}
Patra,~M.; Gasser,~G. Organometallic Compounds: An Opportunity for Chemical Biology? \emph{ChemBioChem} \textbf{2012}, \emph{13}, 1232--1252\relax
\mciteBstWouldAddEndPuncttrue
\mciteSetBstMidEndSepPunct{\mcitedefaultmidpunct}
{\mcitedefaultendpunct}{\mcitedefaultseppunct}\relax
\EndOfBibitem
\bibitem[Xie \latin{et~al.}(2020)Xie, Ouyang, Wang, Lee, and Fong]{Xie2020}
Xie,~Y.; Ouyang,~S.; Wang,~D.; Lee,~W.-Y.; Fong,~H.~H. Highly smooth and conductive silver film with metallo-organic decomposition ink for all-solution-processed flexible organic thin-film transistors. \emph{Journal of Materials Science} \textbf{2020}, \emph{55}, 15908--15918\relax
\mciteBstWouldAddEndPuncttrue
\mciteSetBstMidEndSepPunct{\mcitedefaultmidpunct}
{\mcitedefaultendpunct}{\mcitedefaultseppunct}\relax
\EndOfBibitem
\bibitem[Gasser \latin{et~al.}(2011)Gasser, Ott, and Metzler-Nolte]{doi:10.1021/jm100020w}
Gasser,~G.; Ott,~I.; Metzler-Nolte,~N. Organometallic Anticancer Compounds. \emph{Journal of Medicinal Chemistry} \textbf{2011}, \emph{54}, 3--25, PMID: 21077686\relax
\mciteBstWouldAddEndPuncttrue
\mciteSetBstMidEndSepPunct{\mcitedefaultmidpunct}
{\mcitedefaultendpunct}{\mcitedefaultseppunct}\relax
\EndOfBibitem
\bibitem[Patra \latin{et~al.}(2012)Patra, Gasser, and Metzler-Nolte]{C2DT12460B}
Patra,~M.; Gasser,~G.; Metzler-Nolte,~N. Small organometallic compounds as antibacterial agents. \emph{Dalton Trans.} \textbf{2012}, \emph{41}, 6350--6358\relax
\mciteBstWouldAddEndPuncttrue
\mciteSetBstMidEndSepPunct{\mcitedefaultmidpunct}
{\mcitedefaultendpunct}{\mcitedefaultseppunct}\relax
\EndOfBibitem
\bibitem[Fernández \latin{et~al.}(2008)Fernández, García-Barrasa, Laguna, de~Luzuriaga, Monge, and Torres]{Fernández_2008}
Fernández,~E.~J.; García-Barrasa,~J.; Laguna,~A.; de~Luzuriaga,~J. M.~L.; Monge,~M.; Torres,~C. The preparation of highly active antimicrobial silver nanoparticles by an organometallic approach. \emph{Nanotechnology} \textbf{2008}, \emph{19}, 185602\relax
\mciteBstWouldAddEndPuncttrue
\mciteSetBstMidEndSepPunct{\mcitedefaultmidpunct}
{\mcitedefaultendpunct}{\mcitedefaultseppunct}\relax
\EndOfBibitem
\bibitem[{Mahadev Patil} \latin{et~al.}(2023){Mahadev Patil}, Poddar, Parihar, Sen, Mohapatra, {Murty U}, and Pemmaraju]{MAHADEVPATIL2023144110}
{Mahadev Patil},~P.; Poddar,~N.; Parihar,~N.; Sen,~S.; Mohapatra,~P.; {Murty U},~S.; Pemmaraju,~D.~B. Optoresponsive Pheophorbide-Silver based organometallic nanomaterials for high efficacy multimodal theranostics in Melanoma. \emph{Chemical Engineering Journal} \textbf{2023}, \emph{470}, 144110\relax
\mciteBstWouldAddEndPuncttrue
\mciteSetBstMidEndSepPunct{\mcitedefaultmidpunct}
{\mcitedefaultendpunct}{\mcitedefaultseppunct}\relax
\EndOfBibitem
\bibitem[Chen \latin{et~al.}(2016)Chen, Roy, Yang, and Prasad]{doi:10.1021/acs.chemrev.5b00148}
Chen,~G.; Roy,~I.; Yang,~C.; Prasad,~P.~N. Nanochemistry and Nanomedicine for Nanoparticle-based Diagnostics and Therapy. \emph{Chemical Reviews} \textbf{2016}, \emph{116}, 2826--2885, PMID: 26799741\relax
\mciteBstWouldAddEndPuncttrue
\mciteSetBstMidEndSepPunct{\mcitedefaultmidpunct}
{\mcitedefaultendpunct}{\mcitedefaultseppunct}\relax
\EndOfBibitem
\bibitem[Wang \latin{et~al.}(2015)Wang, Liu, Tong, and Ha]{WANG2015139}
Wang,~J.; Liu,~H.-B.; Tong,~Z.; Ha,~C.-S. Fluorescent/luminescent detection of natural amino acids by organometallic systems. \emph{Coordination Chemistry Reviews} \textbf{2015}, \emph{303}, 139--184\relax
\mciteBstWouldAddEndPuncttrue
\mciteSetBstMidEndSepPunct{\mcitedefaultmidpunct}
{\mcitedefaultendpunct}{\mcitedefaultseppunct}\relax
\EndOfBibitem
\bibitem[Bonačić-Koutecký \latin{et~al.}(2012)Bonačić-Koutecký, Kulesza, Gell, Mitrić, Antoine, Bertorelle, Hamouda, Rayane, Broyer, Tabarin, and Dugourd]{C2CP00050D}
Bonačić-Koutecký,~V.; Kulesza,~A.; Gell,~L.; Mitrić,~R.; Antoine,~R.; Bertorelle,~F.; Hamouda,~R.; Rayane,~D.; Broyer,~M.; Tabarin,~T. \latin{et~al.}  Silver cluster–biomolecule hybrids: from basics towards sensors. \emph{Phys. Chem. Chem. Phys.} \textbf{2012}, \emph{14}, 9282--9290\relax
\mciteBstWouldAddEndPuncttrue
\mciteSetBstMidEndSepPunct{\mcitedefaultmidpunct}
{\mcitedefaultendpunct}{\mcitedefaultseppunct}\relax
\EndOfBibitem
\bibitem[Chen \latin{et~al.}(2009)Chen, Li, Lu, Chui, Ma, and Che]{https://doi.org/10.1002/anie.200905678}
Chen,~Y.; Li,~K.; Lu,~W.; Chui,~S. S.-Y.; Ma,~C.-W.; Che,~C.-M. Photoresponsive Supramolecular Organometallic Nanosheets Induced by {PtII}$\cdot\cdot\cdot${PtII} and {CH}$\cdot\cdot\cdot$$\pi$ Interactions. \emph{Angewandte Chemie International Edition} \textbf{2009}, \emph{48}, 9909--9913\relax
\mciteBstWouldAddEndPuncttrue
\mciteSetBstMidEndSepPunct{\mcitedefaultmidpunct}
{\mcitedefaultendpunct}{\mcitedefaultseppunct}\relax
\EndOfBibitem
\bibitem[Zhang \latin{et~al.}(2021)Zhang, Lin, Liu, Tan, Dai, and Xia]{ZHANG2021213652}
Zhang,~X.; Lin,~S.; Liu,~S.; Tan,~X.; Dai,~Y.; Xia,~F. Advances in organometallic/organic nanozymes and their applications. \emph{Coordination Chemistry Reviews} \textbf{2021}, \emph{429}, 213652\relax
\mciteBstWouldAddEndPuncttrue
\mciteSetBstMidEndSepPunct{\mcitedefaultmidpunct}
{\mcitedefaultendpunct}{\mcitedefaultseppunct}\relax
\EndOfBibitem
\bibitem[Faraday(1857)]{Faraday1857}
Faraday,~M. Experimental Relations of Gold (and other Metals) to Light. \emph{Philos. Trans.} \textbf{1857}, 145– 181\relax
\mciteBstWouldAddEndPuncttrue
\mciteSetBstMidEndSepPunct{\mcitedefaultmidpunct}
{\mcitedefaultendpunct}{\mcitedefaultseppunct}\relax
\EndOfBibitem
\bibitem[Lavenn \latin{et~al.}(2014)Lavenn, Albrieux, Tuel, and Demessence]{LAVENN2014234}
Lavenn,~C.; Albrieux,~F.; Tuel,~A.; Demessence,~A. Synthesis, characterization and optical properties of an amino-functionalized gold thiolate cluster: {Au10(SPh-pNH2)10}. \emph{Journal of Colloid and Interface Science} \textbf{2014}, \emph{418}, 234--239\relax
\mciteBstWouldAddEndPuncttrue
\mciteSetBstMidEndSepPunct{\mcitedefaultmidpunct}
{\mcitedefaultendpunct}{\mcitedefaultseppunct}\relax
\EndOfBibitem
\bibitem[Qian \latin{et~al.}(2012)Qian, Zhu, Wu, and Jin]{doi:10.1021/ar200331z}
Qian,~H.; Zhu,~M.; Wu,~Z.; Jin,~R. Quantum Sized Gold Nanoclusters with Atomic Precision. \emph{Accounts of Chemical Research} \textbf{2012}, \emph{45}, 1470--1479, PMID: 22720781\relax
\mciteBstWouldAddEndPuncttrue
\mciteSetBstMidEndSepPunct{\mcitedefaultmidpunct}
{\mcitedefaultendpunct}{\mcitedefaultseppunct}\relax
\EndOfBibitem
\bibitem[Negishi \latin{et~al.}(2005)Negishi, Nobusada, and Tsukuda]{doi:10.1021/ja042218h}
Negishi,~Y.; Nobusada,~K.; Tsukuda,~T. Glutathione-Protected Gold Clusters Revisited: Bridging the Gap between Gold(I)-Thiolate Complexes and Thiolate-Protected Gold Nanocrystals. \emph{Journal of the American Chemical Society} \textbf{2005}, \emph{127}, 5261--5270, PMID: 15810862\relax
\mciteBstWouldAddEndPuncttrue
\mciteSetBstMidEndSepPunct{\mcitedefaultmidpunct}
{\mcitedefaultendpunct}{\mcitedefaultseppunct}\relax
\EndOfBibitem
\bibitem[Xu \latin{et~al.}(2016)Xu, Li, Gao, and Zeng]{C5NR07810E}
Xu,~W.~W.; Li,~Y.; Gao,~Y.; Zeng,~X.~C. Medium-sized {Au40(SR)24} and {Au52(SR)32} nanoclusters with distinct gold-kernel structures and spectroscopic features. \emph{Nanoscale} \textbf{2016}, \emph{8}, 1299--1304\relax
\mciteBstWouldAddEndPuncttrue
\mciteSetBstMidEndSepPunct{\mcitedefaultmidpunct}
{\mcitedefaultendpunct}{\mcitedefaultseppunct}\relax
\EndOfBibitem
\bibitem[Ghosh \latin{et~al.}(2012)Ghosh, Udayabhaskararao, and Pradeep]{doi:10.1021/jz3007436}
Ghosh,~A.; Udayabhaskararao,~T.; Pradeep,~T. One-Step Route to Luminescent {Au18SG14} in the Condensed Phase and Its Closed Shell Molecular Ions in the Gas Phase. \emph{The Journal of Physical Chemistry Letters} \textbf{2012}, \emph{3}, 1997--2002\relax
\mciteBstWouldAddEndPuncttrue
\mciteSetBstMidEndSepPunct{\mcitedefaultmidpunct}
{\mcitedefaultendpunct}{\mcitedefaultseppunct}\relax
\EndOfBibitem
\bibitem[Wu \latin{et~al.}(2009)Wu, Gayathri, Gil, and Jin]{doi:10.1021/ja900386s}
Wu,~Z.; Gayathri,~C.; Gil,~R.~R.; Jin,~R. Probing the Structure and Charge State of Glutathione-Capped {Au25(SG)18} Clusters by NMR and Mass Spectrometry. \emph{Journal of the American Chemical Society} \textbf{2009}, \emph{131}, 6535--6542, PMID: 19379012\relax
\mciteBstWouldAddEndPuncttrue
\mciteSetBstMidEndSepPunct{\mcitedefaultmidpunct}
{\mcitedefaultendpunct}{\mcitedefaultseppunct}\relax
\EndOfBibitem
\bibitem[Liu \latin{et~al.}(2016)Liu, Tian, and Cheng]{C5RA22741K}
Liu,~Y.; Tian,~Z.; Cheng,~L. Size evolution and ligand effects on the structures and stability of {(AuL)n (L = Cl{,} SH{,} SCH3{,} PH2{,} P(CH3)2{,} n = 1–13)} clusters. \emph{RSC Adv.} \textbf{2016}, \emph{6}, 4705--4712\relax
\mciteBstWouldAddEndPuncttrue
\mciteSetBstMidEndSepPunct{\mcitedefaultmidpunct}
{\mcitedefaultendpunct}{\mcitedefaultseppunct}\relax
\EndOfBibitem
\bibitem[Gr{\"o}nbeck \latin{et~al.}(2006)Gr{\"o}nbeck, Walter, and Häkkinen]{doi:10.1021/ja062584w}
Gr{\"o}nbeck,~H.; Walter,~M.; Häkkinen,~H. Theoretical Characterization of Cyclic Thiolated Gold Clusters. \emph{Journal of the American Chemical Society} \textbf{2006}, \emph{128}, 10268--10275, PMID: 16881657\relax
\mciteBstWouldAddEndPuncttrue
\mciteSetBstMidEndSepPunct{\mcitedefaultmidpunct}
{\mcitedefaultendpunct}{\mcitedefaultseppunct}\relax
\EndOfBibitem
\bibitem[Negishi \latin{et~al.}(2004)Negishi, Takasugi, Sato, Yao, Kimura, and Tsukuda]{doi:10.1021/ja0483589}
Negishi,~Y.; Takasugi,~Y.; Sato,~S.; Yao,~H.; Kimura,~K.; Tsukuda,~T. Magic-Numbered $\text{Au}_n$ Clusters Protected by Glutathione Monolayers (n = 18, 21, 25, 28, 32, 39): Isolation and Spectroscopic Characterization. \emph{Journal of the American Chemical Society} \textbf{2004}, \emph{126}, 6518--6519, PMID: 15161256\relax
\mciteBstWouldAddEndPuncttrue
\mciteSetBstMidEndSepPunct{\mcitedefaultmidpunct}
{\mcitedefaultendpunct}{\mcitedefaultseppunct}\relax
\EndOfBibitem
\bibitem[Stauber \latin{et~al.}(2020)Stauber, Qian, Han, Rheingold, Král, Fujita, and Spokoyny]{doi:10.1021/jacs.9b10770}
Stauber,~J.~M.; Qian,~E.~A.; Han,~Y.; Rheingold,~A.~L.; Král,~P.; Fujita,~D.; Spokoyny,~A.~M. An Organometallic Strategy for Assembling Atomically Precise Hybrid Nanomaterials. \emph{Journal of the American Chemical Society} \textbf{2020}, \emph{142}, 327--334, PMID: 31782986\relax
\mciteBstWouldAddEndPuncttrue
\mciteSetBstMidEndSepPunct{\mcitedefaultmidpunct}
{\mcitedefaultendpunct}{\mcitedefaultseppunct}\relax
\EndOfBibitem
\bibitem[Ramakrishna \latin{et~al.}(2008)Ramakrishna, Varnavski, Kim, Lee, and Goodson]{doi:10.1021/ja800341v}
Ramakrishna,~G.; Varnavski,~O.; Kim,~J.; Lee,~D.; Goodson,~T. Quantum-Sized Gold Clusters as Efficient Two-Photon Absorbers. \emph{Journal of the American Chemical Society} \textbf{2008}, \emph{130}, 5032--5033, PMID: 18357982\relax
\mciteBstWouldAddEndPuncttrue
\mciteSetBstMidEndSepPunct{\mcitedefaultmidpunct}
{\mcitedefaultendpunct}{\mcitedefaultseppunct}\relax
\EndOfBibitem
\bibitem[Aldeek \latin{et~al.}(2013)Aldeek, Muhammed, Palui, Zhan, and Mattoussi]{doi:10.1021/nn305856t}
Aldeek,~F.; Muhammed,~M. A.~H.; Palui,~G.; Zhan,~N.; Mattoussi,~H. Growth of Highly Fluorescent Polyethylene Glycol- and Zwitterion-Functionalized Gold Nanoclusters. \emph{ACS Nano} \textbf{2013}, \emph{7}, 2509--2521, PMID: 23394608\relax
\mciteBstWouldAddEndPuncttrue
\mciteSetBstMidEndSepPunct{\mcitedefaultmidpunct}
{\mcitedefaultendpunct}{\mcitedefaultseppunct}\relax
\EndOfBibitem
\bibitem[Yan \latin{et~al.}(2013)Yan, Yuan, and Dyson]{C3DT51180D}
Yan,~N.; Yuan,~Y.; Dyson,~P.~J. Nanometallic chemistry: deciphering nanoparticle catalysis from the perspective of organometallic chemistry and homogeneous catalysis. \emph{Dalton Trans.} \textbf{2013}, \emph{42}, 13294--13304\relax
\mciteBstWouldAddEndPuncttrue
\mciteSetBstMidEndSepPunct{\mcitedefaultmidpunct}
{\mcitedefaultendpunct}{\mcitedefaultseppunct}\relax
\EndOfBibitem
\bibitem[{Gonzàlez de Rivera} \latin{et~al.}(2012){Gonzàlez de Rivera}, Angurell, Rossell, Seco, and Llorca]{GONZALEZDERIVERA201213}
{Gonzàlez de Rivera},~F.; Angurell,~I.; Rossell,~O.; Seco,~M.; Llorca,~J. Organometallic surface functionalization of gold nanoparticles. \emph{Journal of Organometallic Chemistry} \textbf{2012}, \emph{715}, 13--18\relax
\mciteBstWouldAddEndPuncttrue
\mciteSetBstMidEndSepPunct{\mcitedefaultmidpunct}
{\mcitedefaultendpunct}{\mcitedefaultseppunct}\relax
\EndOfBibitem
\bibitem[Orefuwa \latin{et~al.}(2014)Orefuwa, Ravanbakhsh, Neal, King, and Mohamed]{doi:10.1021/om400927g}
Orefuwa,~S.~A.; Ravanbakhsh,~M.; Neal,~S.~N.; King,~J.~B.; Mohamed,~A.~A. Robust Organometallic Gold Nanoparticles. \emph{Organometallics} \textbf{2014}, \emph{33}, 439--442\relax
\mciteBstWouldAddEndPuncttrue
\mciteSetBstMidEndSepPunct{\mcitedefaultmidpunct}
{\mcitedefaultendpunct}{\mcitedefaultseppunct}\relax
\EndOfBibitem
\bibitem[Knoppe \latin{et~al.}(2016)Knoppe, Zhang, Wan, Wang, Wang, and Verbiest]{doi:10.1021/acs.iecr.6b02925}
Knoppe,~S.; Zhang,~Q.-F.; Wan,~X.-K.; Wang,~Q.-M.; Wang,~L.-S.; Verbiest,~T. Second-Order Nonlinear Optical Scattering Properties of Phosphine-Protected {Au20} Clusters. \emph{Industrial \& Engineering Chemistry Research} \textbf{2016}, \emph{55}, 10500--10506\relax
\mciteBstWouldAddEndPuncttrue
\mciteSetBstMidEndSepPunct{\mcitedefaultmidpunct}
{\mcitedefaultendpunct}{\mcitedefaultseppunct}\relax
\EndOfBibitem
\bibitem[Jin(2010)]{B9NR00160C}
Jin,~R. Quantum sized, thiolate-protected gold nanoclusters. \emph{Nanoscale} \textbf{2010}, \emph{2}, 343--362\relax
\mciteBstWouldAddEndPuncttrue
\mciteSetBstMidEndSepPunct{\mcitedefaultmidpunct}
{\mcitedefaultendpunct}{\mcitedefaultseppunct}\relax
\EndOfBibitem
\bibitem[Daniel and Astruc(2004)Daniel, and Astruc]{doi:10.1021/cr030698+}
Daniel,~M.-C.; Astruc,~D. Gold Nanoparticles: Assembly, Supramolecular Chemistry, Quantum-Size-Related Properties, and Applications toward Biology, Catalysis, and Nanotechnology. \emph{Chemical Reviews} \textbf{2004}, \emph{104}, 293--346, PMID: 14719978\relax
\mciteBstWouldAddEndPuncttrue
\mciteSetBstMidEndSepPunct{\mcitedefaultmidpunct}
{\mcitedefaultendpunct}{\mcitedefaultseppunct}\relax
\EndOfBibitem
\bibitem[Zeng \latin{et~al.}(2014)Zeng, Chen, Li, and Jin]{doi:10.1021/cm500139t}
Zeng,~C.; Chen,~Y.; Li,~G.; Jin,~R. Magic Size {Au64(S-c-C6H11)32} Nanocluster Protected by Cyclohexanethiolate. \emph{Chemistry of Materials} \textbf{2014}, \emph{26}, 2635--2641\relax
\mciteBstWouldAddEndPuncttrue
\mciteSetBstMidEndSepPunct{\mcitedefaultmidpunct}
{\mcitedefaultendpunct}{\mcitedefaultseppunct}\relax
\EndOfBibitem
\bibitem[Philip \latin{et~al.}(2012)Philip, Chantharasupawong, Qian, Jin, and Thomas]{doi:10.1021/nl301988v}
Philip,~R.; Chantharasupawong,~P.; Qian,~H.; Jin,~R.; Thomas,~J. Evolution of Nonlinear Optical Properties: From Gold Atomic Clusters to Plasmonic Nanocrystals. \emph{Nano Letters} \textbf{2012}, \emph{12}, 4661--4667, PMID: 22845756\relax
\mciteBstWouldAddEndPuncttrue
\mciteSetBstMidEndSepPunct{\mcitedefaultmidpunct}
{\mcitedefaultendpunct}{\mcitedefaultseppunct}\relax
\EndOfBibitem
\bibitem[Knoppe \latin{et~al.}(2015)Knoppe, Häkkinen, and Verbiest]{doi:10.1021/acs.jpcc.5b08341}
Knoppe,~S.; Häkkinen,~H.; Verbiest,~T. Nonlinear Optical Properties of Thiolate-Protected Gold Clusters: A Theoretical Survey of the First Hyperpolarizabilities. \emph{The Journal of Physical Chemistry C} \textbf{2015}, \emph{119}, 27676--27682\relax
\mciteBstWouldAddEndPuncttrue
\mciteSetBstMidEndSepPunct{\mcitedefaultmidpunct}
{\mcitedefaultendpunct}{\mcitedefaultseppunct}\relax
\EndOfBibitem
\bibitem[Russier-Antoine \latin{et~al.}(2014)Russier-Antoine, Bertorelle, Vojkovic, Rayane, Salmon, Jonin, Dugourd, Antoine, and Brevet]{C4NR03782K}
Russier-Antoine,~I.; Bertorelle,~F.; Vojkovic,~M.; Rayane,~D.; Salmon,~E.; Jonin,~C.; Dugourd,~P.; Antoine,~R.; Brevet,~P.-F. Non-linear optical properties of gold quantum clusters. The smaller the better. \emph{Nanoscale} \textbf{2014}, \emph{6}, 13572--13578\relax
\mciteBstWouldAddEndPuncttrue
\mciteSetBstMidEndSepPunct{\mcitedefaultmidpunct}
{\mcitedefaultendpunct}{\mcitedefaultseppunct}\relax
\EndOfBibitem
\bibitem[Russier-Antoine \latin{et~al.}(2016)Russier-Antoine, Bertorelle, Hamouda, Rayane, Dugourd, Sanader, Bonačić-Koutecký, Brevet, and Antoine]{C5NR08122J}
Russier-Antoine,~I.; Bertorelle,~F.; Hamouda,~R.; Rayane,~D.; Dugourd,~P.; Sanader,~{\v{Z}}.; Bonačić-Koutecký,~V.; Brevet,~P.-F.; Antoine,~R. Tuning {A}g29 nanocluster light emission from red to blue with one and two-photon excitation. \emph{Nanoscale} \textbf{2016}, \emph{8}, 2892--2898\relax
\mciteBstWouldAddEndPuncttrue
\mciteSetBstMidEndSepPunct{\mcitedefaultmidpunct}
{\mcitedefaultendpunct}{\mcitedefaultseppunct}\relax
\EndOfBibitem
\bibitem[Jin \latin{et~al.}(2016)Jin, Zeng, Zhou, and Chen]{doi:10.1021/acs.chemrev.5b00703}
Jin,~R.; Zeng,~C.; Zhou,~M.; Chen,~Y. Atomically Precise Colloidal Metal Nanoclusters and Nanoparticles: Fundamentals and Opportunities. \emph{Chemical Reviews} \textbf{2016}, \emph{116}, 10346--10413, PMID: 27585252\relax
\mciteBstWouldAddEndPuncttrue
\mciteSetBstMidEndSepPunct{\mcitedefaultmidpunct}
{\mcitedefaultendpunct}{\mcitedefaultseppunct}\relax
\EndOfBibitem
\bibitem[Esumi \latin{et~al.}(2004)Esumi, Sarashina, and Yoshimura]{doi:10.1021/la049415e}
Esumi,~K.; Sarashina,~S.; Yoshimura,~T. Synthesis of Gold Nanoparticles from an Organometallic Compound in Supercritical Carbon Dioxide. \emph{Langmuir} \textbf{2004}, \emph{20}, 5189--5191, PMID: 15986650\relax
\mciteBstWouldAddEndPuncttrue
\mciteSetBstMidEndSepPunct{\mcitedefaultmidpunct}
{\mcitedefaultendpunct}{\mcitedefaultseppunct}\relax
\EndOfBibitem
\bibitem[Russier-Antoine \latin{et~al.}(2016)Russier-Antoine, Bertorelle, Kulesza, Soleilhac, Bensalah-Ledoux, Guy, Dugourd, Brevet, and Antoine]{RUSSIERANTOINE2016455}
Russier-Antoine,~I.; Bertorelle,~F.; Kulesza,~A.; Soleilhac,~A.; Bensalah-Ledoux,~A.; Guy,~S.; Dugourd,~P.; Brevet,~P.-F.; Antoine,~R. Chiral supramolecular gold-cysteine nanoparticles: Chiroptical and nonlinear optical properties. \emph{Progress in Natural Science: Materials International} \textbf{2016}, \emph{26}, 455--460, Special Issue for Nano Materials\relax
\mciteBstWouldAddEndPuncttrue
\mciteSetBstMidEndSepPunct{\mcitedefaultmidpunct}
{\mcitedefaultendpunct}{\mcitedefaultseppunct}\relax
\EndOfBibitem
\bibitem[Schaaff and Whetten(2000)Schaaff, and Whetten]{doi:10.1021/jp993691y}
Schaaff,~T.~G.; Whetten,~R.~L. Giant Gold-Glutathione Cluster Compounds: Intense Optical Activity in Metal-Based Transitions. \emph{The Journal of Physical Chemistry B} \textbf{2000}, \emph{104}, 2630--2641\relax
\mciteBstWouldAddEndPuncttrue
\mciteSetBstMidEndSepPunct{\mcitedefaultmidpunct}
{\mcitedefaultendpunct}{\mcitedefaultseppunct}\relax
\EndOfBibitem
\bibitem[Van~Steerteghem \latin{et~al.}(2016)Van~Steerteghem, Van~Cleuvenbergen, Deckers, Kumara, Dass, Häkkinen, Clays, Verbiest, and Knoppe]{C6NR02251K}
Van~Steerteghem,~N.; Van~Cleuvenbergen,~S.; Deckers,~S.; Kumara,~C.; Dass,~A.; Häkkinen,~H.; Clays,~K.; Verbiest,~T.; Knoppe,~S. Symmetry breaking in ligand-protected gold clusters probed by nonlinear optics. \emph{Nanoscale} \textbf{2016}, \emph{8}, 12123--12127\relax
\mciteBstWouldAddEndPuncttrue
\mciteSetBstMidEndSepPunct{\mcitedefaultmidpunct}
{\mcitedefaultendpunct}{\mcitedefaultseppunct}\relax
\EndOfBibitem
\bibitem[Guevel \latin{et~al.}(2014)Guevel, Tagit, Rodríguez, Trouillet, Pernia~Leal, and Hildebrandt]{C4NR01130A}
Guevel,~X.~L.; Tagit,~O.; Rodríguez,~C.~E.; Trouillet,~V.; Pernia~Leal,~M.; Hildebrandt,~N. Ligand effect on the size{,} valence state and red/near infrared photoluminescence of bidentate thiol gold nanoclusters. \emph{Nanoscale} \textbf{2014}, \emph{6}, 8091--8099\relax
\mciteBstWouldAddEndPuncttrue
\mciteSetBstMidEndSepPunct{\mcitedefaultmidpunct}
{\mcitedefaultendpunct}{\mcitedefaultseppunct}\relax
\EndOfBibitem
\bibitem[Knoppe and B{\"u}rgi(2014)Knoppe, and B{\"u}rgi]{doi:10.1021/ar400295d}
Knoppe,~S.; B{\"u}rgi,~T. Chirality in Thiolate-Protected Gold Clusters. \emph{Accounts of Chemical Research} \textbf{2014}, \emph{47}, 1318--1326, PMID: 24588279\relax
\mciteBstWouldAddEndPuncttrue
\mciteSetBstMidEndSepPunct{\mcitedefaultmidpunct}
{\mcitedefaultendpunct}{\mcitedefaultseppunct}\relax
\EndOfBibitem
\bibitem[Zhang \latin{et~al.}(2015)Zhang, Ye, Zhang, Li, Dong, Jiang, and Wang]{C5RA11321K}
Zhang,~W.; Ye,~J.; Zhang,~Y.; Li,~Q.; Dong,~X.; Jiang,~H.; Wang,~X. One-step facile synthesis of fluorescent gold nanoclusters for rapid bio-imaging of cancer cells and small animals. \emph{RSC Adv.} \textbf{2015}, \emph{5}, 63821--63826\relax
\mciteBstWouldAddEndPuncttrue
\mciteSetBstMidEndSepPunct{\mcitedefaultmidpunct}
{\mcitedefaultendpunct}{\mcitedefaultseppunct}\relax
\EndOfBibitem
\bibitem[Polavarapu \latin{et~al.}(2011)Polavarapu, Manna, and Xu]{C0NR00458H}
Polavarapu,~L.; Manna,~M.; Xu,~Q.-H. Biocompatible glutathione capped gold clusters as one- and two-photon excitation fluorescence contrast agents for live cells imaging. \emph{Nanoscale} \textbf{2011}, \emph{3}, 429--434\relax
\mciteBstWouldAddEndPuncttrue
\mciteSetBstMidEndSepPunct{\mcitedefaultmidpunct}
{\mcitedefaultendpunct}{\mcitedefaultseppunct}\relax
\EndOfBibitem
\bibitem[Russier-Antoine \latin{et~al.}(2017)Russier-Antoine, Bertorelle, Calin, Sanader, Krstić, Comby-Zerbino, Dugourd, Brevet, Bonačić-Koutecký, and Antoine]{Sanader}
Russier-Antoine,~I.; Bertorelle,~F.; Calin,~N.; Sanader,~{\v{Z}}.; Krstić,~M.; Comby-Zerbino,~C.; Dugourd,~P.; Brevet,~P.-F.; Bonačić-Koutecký,~V.; Antoine,~R. Ligand-core NLO-phores: a combined experimental and theoretical approach to the two-photon absorption and two-photon excited emission properties of small-ligated silver nanoclusters. \emph{Nanoscale} \textbf{2017}, \emph{9}, 1221--1228\relax
\mciteBstWouldAddEndPuncttrue
\mciteSetBstMidEndSepPunct{\mcitedefaultmidpunct}
{\mcitedefaultendpunct}{\mcitedefaultseppunct}\relax
\EndOfBibitem
\bibitem[Bertorelle \latin{et~al.}(2017)Bertorelle, Russier-Antoine, Calin, Comby-Zerbino, Bensalah-Ledoux, Guy, Dugourd, Brevet, Sanader, Krstić, Bonačić-Koutecký, and Antoine]{Sanader2}
Bertorelle,~F.; Russier-Antoine,~I.; Calin,~N.; Comby-Zerbino,~C.; Bensalah-Ledoux,~A.; Guy,~S.; Dugourd,~P.; Brevet,~P.-F.; Sanader,~{\v{Z}}.; Krstić,~M. \latin{et~al.}  {Au10(SG)10}: A Chiral Gold Catenane Nanocluster with Zero Confined Electrons. Optical Properties and First-Principles Theoretical Analysis. \emph{The Journal of Physical Chemistry Letters} \textbf{2017}, \emph{8}, 1979--1985\relax
\mciteBstWouldAddEndPuncttrue
\mciteSetBstMidEndSepPunct{\mcitedefaultmidpunct}
{\mcitedefaultendpunct}{\mcitedefaultseppunct}\relax
\EndOfBibitem
\bibitem[Fernandez-Corbaton \latin{et~al.}(2020)Fernandez-Corbaton, Beutel, Rockstuhl, Pausch, and Klopper]{Fernandez-Corbaton:2020}
Fernandez-Corbaton,~I.; Beutel,~D.; Rockstuhl,~C.; Pausch,~A.; Klopper,~W. Computation of electromagnetic properties of molecular ensembles. \emph{ChemPhysChem} \textbf{2020}, \emph{21}, 878--887\relax
\mciteBstWouldAddEndPuncttrue
\mciteSetBstMidEndSepPunct{\mcitedefaultmidpunct}
{\mcitedefaultendpunct}{\mcitedefaultseppunct}\relax
\EndOfBibitem
\bibitem[Zerulla \latin{et~al.}(2024)Zerulla, Beutel, Holzer, Fernandez-Corbaton, Rockstuhl, and Krstić]{Zerulla2024Feb}
Zerulla,~B.; Beutel,~D.; Holzer,~C.; Fernandez-Corbaton,~I.; Rockstuhl,~C.; Krstić,~M. {A Multi-Scale Approach to Simulate the Nonlinear Optical Response of Molecular Nanomaterials}. \emph{Adv. Mater.} \textbf{2024}, \emph{36}, 2311405\relax
\mciteBstWouldAddEndPuncttrue
\mciteSetBstMidEndSepPunct{\mcitedefaultmidpunct}
{\mcitedefaultendpunct}{\mcitedefaultseppunct}\relax
\EndOfBibitem
\bibitem[Zerulla \latin{et~al.}(2022)Zerulla, Krsti{\ifmmode\acute{c}\else\'{c}\fi}, \latin{et~al.} others]{Zerulla2022May}
Zerulla,~B.; Krsti{\ifmmode\acute{c}\else\'{c}\fi},~M.; others {A Multi-Scale Approach for Modeling the Optical Response of Molecular Materials Inside Cavities}. \emph{Adv. Mater.} \textbf{2022}, \emph{34}, 2200350\relax
\mciteBstWouldAddEndPuncttrue
\mciteSetBstMidEndSepPunct{\mcitedefaultmidpunct}
{\mcitedefaultendpunct}{\mcitedefaultseppunct}\relax
\EndOfBibitem
\bibitem[Zerulla \latin{et~al.}(2023)Zerulla, Li, Beutel, Oßwald, Holzer, Bürck, Bräse, Wöll, Fernandez-Corbaton, Heinke, Rockstuhl, and Krstić]{https://doi.org/10.1002/adfm.202301093}
Zerulla,~B.; Li,~C.; Beutel,~D.; Oßwald,~S.; Holzer,~C.; Bürck,~J.; Bräse,~S.; Wöll,~C.; Fernandez-Corbaton,~I.; Heinke,~L. \latin{et~al.}  Exploring Functional Photonic Devices made from a Chiral Metal–Organic Framework Material by a Multiscale Computational Method. \emph{Advanced Functional Materials} \textbf{2023}, 2301093\relax
\mciteBstWouldAddEndPuncttrue
\mciteSetBstMidEndSepPunct{\mcitedefaultmidpunct}
{\mcitedefaultendpunct}{\mcitedefaultseppunct}\relax
\EndOfBibitem
\bibitem[Zerulla \latin{et~al.}(2024)Zerulla, Díaz, Holzer, Rockstuhl, Fernandez-Corbaton, and Krstić]{https://doi.org/10.1002/adom.202400150}
Zerulla,~B.; Díaz,~A.~L.; Holzer,~C.; Rockstuhl,~C.; Fernandez-Corbaton,~I.; Krstić,~M. Multi-Scale Modeling of Surface Second-Harmonic Generation in Centrosymmetric Molecular Crystalline Materials: How Thick is the Surface? \emph{Advanced Optical Materials} \textbf{2024}, \emph{12}, 2400150\relax
\mciteBstWouldAddEndPuncttrue
\mciteSetBstMidEndSepPunct{\mcitedefaultmidpunct}
{\mcitedefaultendpunct}{\mcitedefaultseppunct}\relax
\EndOfBibitem
\bibitem[Carletti \latin{et~al.}(2021)Carletti, Gandolfi, Rocco, Tognazzi, de~Ceglia, Vincenti, and De~Angelis]{carletti2021reconfigurable}
Carletti,~L.; Gandolfi,~M.; Rocco,~D.; Tognazzi,~A.; de~Ceglia,~D.; Vincenti,~M.~A.; De~Angelis,~C. Reconfigurable nonlinear response of dielectric and semiconductor metasurfaces. \emph{Nanophotonics} \textbf{2021}, \emph{10}, 4209--4221\relax
\mciteBstWouldAddEndPuncttrue
\mciteSetBstMidEndSepPunct{\mcitedefaultmidpunct}
{\mcitedefaultendpunct}{\mcitedefaultseppunct}\relax
\EndOfBibitem
\bibitem[Carletti \latin{et~al.}(2024)Carletti, Rocco, Vincenti, de~Ceglia, and De~Angelis]{carletti2024intrinsic}
Carletti,~L.; Rocco,~D.; Vincenti,~M.~A.; de~Ceglia,~D.; De~Angelis,~C. Intrinsic nonlinear geometric phase in SHG from zincblende crystal symmetry media. \emph{Nanophotonics} \textbf{2024}, \emph{13}, 3321--3326\relax
\mciteBstWouldAddEndPuncttrue
\mciteSetBstMidEndSepPunct{\mcitedefaultmidpunct}
{\mcitedefaultendpunct}{\mcitedefaultseppunct}\relax
\EndOfBibitem
\bibitem[Makarov \latin{et~al.}(2017)Makarov, Petrov, Zywietz, Milichko, Zuev, Lopanitsyna, Kuksin, Mukhin, Zograf, Ubyivovk, \latin{et~al.} others]{makarov2017efficient}
Makarov,~S.~V.; Petrov,~M.~I.; Zywietz,~U.; Milichko,~V.; Zuev,~D.; Lopanitsyna,~N.; Kuksin,~A.; Mukhin,~I.; Zograf,~G.; Ubyivovk,~E. \latin{et~al.}  Efficient second-harmonic generation in nanocrystalline silicon nanoparticles. \emph{Nano letters} \textbf{2017}, \emph{17}, 3047--3053\relax
\mciteBstWouldAddEndPuncttrue
\mciteSetBstMidEndSepPunct{\mcitedefaultmidpunct}
{\mcitedefaultendpunct}{\mcitedefaultseppunct}\relax
\EndOfBibitem
\bibitem[Smirnova and Kivshar(2016)Smirnova, and Kivshar]{smirnova2016multipolar}
Smirnova,~D.; Kivshar,~Y.~S. Multipolar nonlinear nanophotonics. \emph{Optica} \textbf{2016}, \emph{3}, 1241--1255\relax
\mciteBstWouldAddEndPuncttrue
\mciteSetBstMidEndSepPunct{\mcitedefaultmidpunct}
{\mcitedefaultendpunct}{\mcitedefaultseppunct}\relax
\EndOfBibitem
\bibitem[Zheng \latin{et~al.}(2023)Zheng, Rocco, Ren, Sergaeva, Zhang, Whaley, Ying, de~Ceglia, De-Angelis, Rahmani, \latin{et~al.} others]{zheng2023advances}
Zheng,~Z.; Rocco,~D.; Ren,~H.; Sergaeva,~O.; Zhang,~Y.; Whaley,~K.~B.; Ying,~C.; de~Ceglia,~D.; De-Angelis,~C.; Rahmani,~M. \latin{et~al.}  Advances in nonlinear metasurfaces for imaging, quantum, and sensing applications. \emph{Nanophotonics} \textbf{2023}, \emph{12}, 4255--4281\relax
\mciteBstWouldAddEndPuncttrue
\mciteSetBstMidEndSepPunct{\mcitedefaultmidpunct}
{\mcitedefaultendpunct}{\mcitedefaultseppunct}\relax
\EndOfBibitem
\bibitem[Rahmani \latin{et~al.}(2018)Rahmani, Leo, Brener, Zayats, Maier, De~Angelis, Tan, Gili, Karouta, Oulton, \latin{et~al.} others]{rahmani2018nonlinear}
Rahmani,~M.; Leo,~G.; Brener,~I.; Zayats,~A.~V.; Maier,~S.~A.; De~Angelis,~C.; Tan,~H.; Gili,~V.~F.; Karouta,~F.; Oulton,~R. \latin{et~al.}  Nonlinear frequency conversion in optical nanoantennas and metasurfaces: materials evolution and fabrication. \emph{Opto-Electronic Advances} \textbf{2018}, \emph{1}, 180021--1\relax
\mciteBstWouldAddEndPuncttrue
\mciteSetBstMidEndSepPunct{\mcitedefaultmidpunct}
{\mcitedefaultendpunct}{\mcitedefaultseppunct}\relax
\EndOfBibitem
\bibitem[TUR(2022)]{TURBOMOLE2022}
{{TURBOMOLE}} 7.7. 2022; \url{https://www.turbomole.org}\relax
\mciteBstWouldAddEndPuncttrue
\mciteSetBstMidEndSepPunct{\mcitedefaultmidpunct}
{\mcitedefaultendpunct}{\mcitedefaultseppunct}\relax
\EndOfBibitem
\bibitem[Franzke \latin{et~al.}(2023)Franzke, Holzer, Andersen, Begušić, Bruder, Coriani, Della~Sala, Fabiano, Fedotov, Fürst, Gillhuber, Grotjahn, Kaupp, Kehry, Krstić, Mack, Majumdar, Nguyen, Parker, Pauly, Pausch, Perlt, Phun, Rajabi, Rappoport, Samal, Schrader, Sharma, Tapavicza, Treß, Voora, Wodyński, Yu, Zerulla, Furche, Hättig, Sierka, Tew, and Weigend]{TM_TODAY_TOMORROW}
Franzke,~Y.~J.; Holzer,~C.; Andersen,~J.~H.; Begušić,~T.; Bruder,~F.; Coriani,~S.; Della~Sala,~F.; Fabiano,~E.; Fedotov,~D.~A.; Fürst,~S. \latin{et~al.}  TURBOMOLE: Today and Tomorrow. \emph{Journal of Chemical Theory and Computation} \textbf{2023}, \emph{19}, 6859--6890\relax
\mciteBstWouldAddEndPuncttrue
\mciteSetBstMidEndSepPunct{\mcitedefaultmidpunct}
{\mcitedefaultendpunct}{\mcitedefaultseppunct}\relax
\EndOfBibitem
\bibitem[Klamt and Sch{\"u}{\"u}rmann(1993)Klamt, and Sch{\"u}{\"u}rmann]{klamtCOSMONewApproach1993}
Klamt,~A.; Sch{\"u}{\"u}rmann,~G. {{COSMO}}: A New Approach to Dielectric Screening in Solvents with Explicit Expressions for the Screening Energy and Its Gradient. \emph{J. Chem. Soc., Perkin Trans. 2} \textbf{1993}, 799--805\relax
\mciteBstWouldAddEndPuncttrue
\mciteSetBstMidEndSepPunct{\mcitedefaultmidpunct}
{\mcitedefaultendpunct}{\mcitedefaultseppunct}\relax
\EndOfBibitem
\bibitem[Söptei \latin{et~al.}(2015)Söptei, Mihály, Szigyártó, Wacha, Németh, Bertóti, May, Baranyai, Sajó, and Bóta]{SOPTEI20158}
Söptei,~B.; Mihály,~J.; Szigyártó,~I.~C.; Wacha,~A.; Németh,~C.; Bertóti,~I.; May,~Z.; Baranyai,~P.; Sajó,~I.~E.; Bóta,~A. The supramolecular chemistry of gold and l-cysteine: Formation of photoluminescent, orange-emitting assemblies with multilayer structure. \emph{Colloids and Surfaces A: Physicochemical and Engineering Aspects} \textbf{2015}, \emph{470}, 8--14\relax
\mciteBstWouldAddEndPuncttrue
\mciteSetBstMidEndSepPunct{\mcitedefaultmidpunct}
{\mcitedefaultendpunct}{\mcitedefaultseppunct}\relax
\EndOfBibitem
\bibitem[Fakhouri \latin{et~al.}(2019)Fakhouri, Peri{\ifmmode\acute{c}\else\'{c}\fi}, \latin{et~al.} others]{Fakhouri2019}
Fakhouri,~H.; Peri{\ifmmode\acute{c}\else\'{c}\fi},~M.; others {Sub-100 nanometer silver doped gold{\textendash}cysteine supramolecular assemblies with enhanced nonlinear optical properties}. \emph{Phys. Chem. Chem. Phys.} \textbf{2019}, \emph{21}, 12091--12099\relax
\mciteBstWouldAddEndPuncttrue
\mciteSetBstMidEndSepPunct{\mcitedefaultmidpunct}
{\mcitedefaultendpunct}{\mcitedefaultseppunct}\relax
\EndOfBibitem
\bibitem[Ni \latin{et~al.}(2024)Ni, Vivod, Avaro, Qi, Zahn, Wang, and C{\"o}lfen]{Ni2024}
Ni,~B.; Vivod,~D.; Avaro,~J.; Qi,~H.; Zahn,~D.; Wang,~X.; C{\"o}lfen,~H. Reversible chirality inversion of an AuAgx-cysteine coordination polymer by pH change. \emph{Nature Communications} \textbf{2024}, \emph{15}, 2042\relax
\mciteBstWouldAddEndPuncttrue
\mciteSetBstMidEndSepPunct{\mcitedefaultmidpunct}
{\mcitedefaultendpunct}{\mcitedefaultseppunct}\relax
\EndOfBibitem
\bibitem[Waterman(1965)]{Waterman}
Waterman,~P. Matrix formulation of electromagnetic scattering. \emph{Proceedings of the IEEE} \textbf{1965}, \emph{53}, 805--812\relax
\mciteBstWouldAddEndPuncttrue
\mciteSetBstMidEndSepPunct{\mcitedefaultmidpunct}
{\mcitedefaultendpunct}{\mcitedefaultseppunct}\relax
\EndOfBibitem
\bibitem[Tsang \latin{et~al.}(2000)Tsang, Kong, Ding, and Kong]{Tsang2000-BasicTheoryofElect}
Tsang,~L.; Kong,~J.~A.; Ding,~K.-H.; Kong,~J.~A. \emph{{Scattering of Electromagnetic Waves: Theories and Applications}}; John Wiley {\&} Sons, Ltd: Chichester, England, UK, 2000; pp 53--106\relax
\mciteBstWouldAddEndPuncttrue
\mciteSetBstMidEndSepPunct{\mcitedefaultmidpunct}
{\mcitedefaultendpunct}{\mcitedefaultseppunct}\relax
\EndOfBibitem
\bibitem[Rahimzadegan()]{Rahimzadegan}
Rahimzadegan,~A. A Comprehensive Multipolar Theory for Periodic Metasurfaces. \relax
\mciteBstWouldAddEndPunctfalse
\mciteSetBstMidEndSepPunct{\mcitedefaultmidpunct}
{}{\mcitedefaultseppunct}\relax
\EndOfBibitem
\bibitem[Beutel \latin{et~al.}(2021)Beutel, Groner, \latin{et~al.} others]{Beutel2021Jun}
Beutel,~D.; Groner,~A.; others {Efficient simulation of biperiodic, layered structures based on the T-matrix method}. \emph{J. Opt. Soc. Am. B, JOSAB} \textbf{2021}, \emph{38}, 1782--1791\relax
\mciteBstWouldAddEndPuncttrue
\mciteSetBstMidEndSepPunct{\mcitedefaultmidpunct}
{\mcitedefaultendpunct}{\mcitedefaultseppunct}\relax
\EndOfBibitem
\bibitem[Not()]{Note-1}
Ref.~\citenum {Zerulla2024Feb}, Eq.~(7)\relax
\mciteBstWouldAddEndPuncttrue
\mciteSetBstMidEndSepPunct{\mcitedefaultmidpunct}
{\mcitedefaultendpunct}{\mcitedefaultseppunct}\relax
\EndOfBibitem
\bibitem[Beutel \latin{et~al.}(2024)Beutel, Fernandez-Corbaton, \latin{et~al.} others]{Beutel2024Apr}
Beutel,~D.; Fernandez-Corbaton,~I.; others {treams {\textendash} a T-matrix-based scattering code for nanophotonics}. \emph{Comput. Phys. Commun.} \textbf{2024}, \emph{297}, 109076\relax
\mciteBstWouldAddEndPuncttrue
\mciteSetBstMidEndSepPunct{\mcitedefaultmidpunct}
{\mcitedefaultendpunct}{\mcitedefaultseppunct}\relax
\EndOfBibitem
\bibitem[Antoine and Bona{\ifmmode\check{c}\else\v{c}\fi}i{\ifmmode\acute{c}\else\'{c}\fi}-Kouteck{\ifmmode\acute{y}\else\'{y}\fi}()Antoine, and Bona{\ifmmode\check{c}\else\v{c}\fi}i{\ifmmode\acute{c}\else\'{c}\fi}-Kouteck{\ifmmode\acute{y}\else\'{y}\fi}]{Antoine}
Antoine,~R.; Bona{\ifmmode\check{c}\else\v{c}\fi}i{\ifmmode\acute{c}\else\'{c}\fi}-Kouteck{\ifmmode\acute{y}\else\'{y}\fi},~V. \emph{{Liganded silver and gold quantum clusters. Towards a new class of nonlinear optical nanomaterials}}; Springer International Publishing: Cham, Switzerland\relax
\mciteBstWouldAddEndPuncttrue
\mciteSetBstMidEndSepPunct{\mcitedefaultmidpunct}
{\mcitedefaultendpunct}{\mcitedefaultseppunct}\relax
\EndOfBibitem
\end{mcitethebibliography}


\providecommand{\latin}[1]{#1}
\makeatletter
\providecommand{\doi}
  {\begingroup\let\do\@makeother\dospecials
  \catcode`\{=1 \catcode`\}=2 \doi@aux}
\providecommand{\doi@aux}[1]{\endgroup\texttt{#1}}
\makeatother
\providecommand*\mcitethebibliography{\thebibliography}
\csname @ifundefined\endcsname{endmcitethebibliography}  {\let\endmcitethebibliography\endthebibliography}{}
\begin{mcitethebibliography}{28}
\providecommand*\natexlab[1]{#1}
\providecommand*\mciteSetBstSublistMode[1]{}
\providecommand*\mciteSetBstMaxWidthForm[2]{}
\providecommand*\mciteBstWouldAddEndPuncttrue
  {\def\EndOfBibitem{\unskip.}}
\providecommand*\mciteBstWouldAddEndPunctfalse
  {\let\EndOfBibitem\relax}
\providecommand*\mciteSetBstMidEndSepPunct[3]{}
\providecommand*\mciteSetBstSublistLabelBeginEnd[3]{}
\providecommand*\EndOfBibitem{}
\mciteSetBstSublistMode{f}
\mciteSetBstMaxWidthForm{subitem}{(\alph{mcitesubitemcount})}
\mciteSetBstSublistLabelBeginEnd
  {\mcitemaxwidthsubitemform\space}
  {\relax}
  {\relax}

\bibitem[Kühne \latin{et~al.}(2020)Kühne, Iannuzzi, Del~Ben, Rybkin, Seewald, Stein, Laino, Khaliullin, Schütt, Schiffmann, Golze, Wilhelm, Chulkov, Bani-Hashemian, Weber, Borštnik, Taillefumier, Jakobovits, Lazzaro, Pabst, Müller, Schade, Guidon, Andermatt, Holmberg, Schenter, Hehn, Bussy, Belleflamme, Tabacchi, Glöß, Lass, Bethune, Mundy, Plessl, Watkins, VandeVondele, Krack, and Hutter]{CP2K}
Kühne,~T.~D.; Iannuzzi,~M.; Del~Ben,~M.; Rybkin,~V.~V.; Seewald,~P.; Stein,~F.; Laino,~T.; Khaliullin,~R.~Z.; Schütt,~O.; Schiffmann,~F.; Golze,~D.; Wilhelm,~J.; Chulkov,~S.; Bani-Hashemian,~M.~H.; Weber,~V.; Borštnik,~U.; Taillefumier,~M.; Jakobovits,~A.~S.; Lazzaro,~A.; Pabst,~H.; Müller,~T.; Schade,~R.; Guidon,~M.; Andermatt,~S.; Holmberg,~N.; Schenter,~G.~K.; Hehn,~A.; Bussy,~A.; Belleflamme,~F.; Tabacchi,~G.; Glöß,~A.; Lass,~M.; Bethune,~I.; Mundy,~C.~J.; Plessl,~C.; Watkins,~M.; VandeVondele,~J.; Krack,~M.; Hutter,~J. CP2K: An electronic structure and molecular dynamics software package - Quickstep: Efficient and accurate electronic structure calculations. \emph{J. Chem. Phys.} \textbf{2020}, \emph{152}, 194103\relax
\mciteBstWouldAddEndPuncttrue
\mciteSetBstMidEndSepPunct{\mcitedefaultmidpunct}
{\mcitedefaultendpunct}{\mcitedefaultseppunct}\relax
\EndOfBibitem
\bibitem[Perdew \latin{et~al.}(1996)Perdew, Burke, and Ernzerhof]{PBE_a}
Perdew,~J.~P.; Burke,~K.; Ernzerhof,~M. Generalized Gradient Approximation Made Simple. \emph{Phys. Rev. Lett.} \textbf{1996}, \emph{77}, 3865--3868\relax
\mciteBstWouldAddEndPuncttrue
\mciteSetBstMidEndSepPunct{\mcitedefaultmidpunct}
{\mcitedefaultendpunct}{\mcitedefaultseppunct}\relax
\EndOfBibitem
\bibitem[Perdew \latin{et~al.}(1997)Perdew, Burke, and Ernzerhof]{PBE_b}
Perdew,~J.~P.; Burke,~K.; Ernzerhof,~M. Generalized Gradient Approximation Made Simple [Phys. Rev. Lett. 77, 3865 (1996)]. \emph{Phys. Rev. Lett.} \textbf{1997}, \emph{78}, 1396--1396\relax
\mciteBstWouldAddEndPuncttrue
\mciteSetBstMidEndSepPunct{\mcitedefaultmidpunct}
{\mcitedefaultendpunct}{\mcitedefaultseppunct}\relax
\EndOfBibitem
\bibitem[VandeVondele and Hutter(2007)VandeVondele, and Hutter]{MOLOPT}
VandeVondele,~J.; Hutter,~J. Gaussian basis sets for accurate calculations on molecular systems in gas and condensed phases. \emph{J. Chem. Phys.} \textbf{2007}, \emph{127}, 114105\relax
\mciteBstWouldAddEndPuncttrue
\mciteSetBstMidEndSepPunct{\mcitedefaultmidpunct}
{\mcitedefaultendpunct}{\mcitedefaultseppunct}\relax
\EndOfBibitem
\bibitem[Goedecker \latin{et~al.}(1996)Goedecker, Teter, and Hutter]{GTH-PP1}
Goedecker,~S.; Teter,~M.; Hutter,~J. Separable dual-space Gaussian pseudopotentials. \emph{Phys. Rev. B} \textbf{1996}, \emph{54}, 1703--1710\relax
\mciteBstWouldAddEndPuncttrue
\mciteSetBstMidEndSepPunct{\mcitedefaultmidpunct}
{\mcitedefaultendpunct}{\mcitedefaultseppunct}\relax
\EndOfBibitem
\bibitem[Hartwigsen \latin{et~al.}(1998)Hartwigsen, Goedecker, and Hutter]{GTH-PP2}
Hartwigsen,~C.; Goedecker,~S.; Hutter,~J. Relativistic separable dual-space Gaussian pseudopotentials from H to Rn. \emph{Phys. Rev. B} \textbf{1998}, \emph{58}, 3641--3662\relax
\mciteBstWouldAddEndPuncttrue
\mciteSetBstMidEndSepPunct{\mcitedefaultmidpunct}
{\mcitedefaultendpunct}{\mcitedefaultseppunct}\relax
\EndOfBibitem
\bibitem[Krack(2005)]{GTH-Pseudopotentials}
Krack,~M. Pseudopotentials for H to Kr optimized for gradient-corrected exchange-correlation functionals. \emph{Theoretical Chemistry Accounts} \textbf{2005}, \emph{114}, 145--152\relax
\mciteBstWouldAddEndPuncttrue
\mciteSetBstMidEndSepPunct{\mcitedefaultmidpunct}
{\mcitedefaultendpunct}{\mcitedefaultseppunct}\relax
\EndOfBibitem
\bibitem[Caldeweyher \latin{et~al.}(2019)Caldeweyher, Ehlert, Hansen, Neugebauer, Spicher, Bannwarth, and Grimme]{caldeweyherGenerallyApplicableAtomiccharge2019}
Caldeweyher,~E.; Ehlert,~S.; Hansen,~A.; Neugebauer,~H.; Spicher,~S.; Bannwarth,~C.; Grimme,~S. A Generally Applicable Atomic-Charge Dependent {{London}} Dispersion Correction. \emph{The Journal of Chemical Physics} \textbf{2019}, \emph{150}, 154122\relax
\mciteBstWouldAddEndPuncttrue
\mciteSetBstMidEndSepPunct{\mcitedefaultmidpunct}
{\mcitedefaultendpunct}{\mcitedefaultseppunct}\relax
\EndOfBibitem
\bibitem[Caldeweyher \latin{et~al.}(2017)Caldeweyher, Bannwarth, and Grimme]{caldeweyherExtensionD3Dispersion2017}
Caldeweyher,~E.; Bannwarth,~C.; Grimme,~S. Extension of the {{D3}} Dispersion Coefficient Model. \emph{The Journal of Chemical Physics} \textbf{2017}, \emph{147}, 034112\relax
\mciteBstWouldAddEndPuncttrue
\mciteSetBstMidEndSepPunct{\mcitedefaultmidpunct}
{\mcitedefaultendpunct}{\mcitedefaultseppunct}\relax
\EndOfBibitem
\bibitem[TUR(2022)]{TURBOMOLE2022}
{{TURBOMOLE}} 7.7. 2022; \url{https://www.turbomole.org}\relax
\mciteBstWouldAddEndPuncttrue
\mciteSetBstMidEndSepPunct{\mcitedefaultmidpunct}
{\mcitedefaultendpunct}{\mcitedefaultseppunct}\relax
\EndOfBibitem
\bibitem[Franzke \latin{et~al.}(2023)Franzke, Holzer, Andersen, Begušić, Bruder, Coriani, Della~Sala, Fabiano, Fedotov, Fürst, Gillhuber, Grotjahn, Kaupp, Kehry, Krstić, Mack, Majumdar, Nguyen, Parker, Pauly, Pausch, Perlt, Phun, Rajabi, Rappoport, Samal, Schrader, Sharma, Tapavicza, Treß, Voora, Wodyński, Yu, Zerulla, Furche, Hättig, Sierka, Tew, and Weigend]{TM_TODAY_TOMORROW}
Franzke,~Y.~J.; Holzer,~C.; Andersen,~J.~H.; Begušić,~T.; Bruder,~F.; Coriani,~S.; Della~Sala,~F.; Fabiano,~E.; Fedotov,~D.~A.; Fürst,~S.; Gillhuber,~S.; Grotjahn,~R.; Kaupp,~M.; Kehry,~M.; Krstić,~M.; Mack,~F.; Majumdar,~S.; Nguyen,~B.~D.; Parker,~S.~M.; Pauly,~F.; Pausch,~A.; Perlt,~E.; Phun,~G.~S.; Rajabi,~A.; Rappoport,~D.; Samal,~B.; Schrader,~T.; Sharma,~M.; Tapavicza,~E.; Treß,~R.~S.; Voora,~V.; Wodyński,~A.; Yu,~J.~M.; Zerulla,~B.; Furche,~F.; Hättig,~C.; Sierka,~M.; Tew,~D.~P.; Weigend,~F. TURBOMOLE: Today and Tomorrow. \emph{Journal of Chemical Theory and Computation} \textbf{2023}, \emph{19}, 6859--6890\relax
\mciteBstWouldAddEndPuncttrue
\mciteSetBstMidEndSepPunct{\mcitedefaultmidpunct}
{\mcitedefaultendpunct}{\mcitedefaultseppunct}\relax
\EndOfBibitem
\bibitem[Klamt and Sch{\"u}{\"u}rmann(1993)Klamt, and Sch{\"u}{\"u}rmann]{klamtCOSMONewApproach1993}
Klamt,~A.; Sch{\"u}{\"u}rmann,~G. {{COSMO}}: A New Approach to Dielectric Screening in Solvents with Explicit Expressions for the Screening Energy and Its Gradient. \emph{J. Chem. Soc., Perkin Trans. 2} \textbf{1993}, 799--805\relax
\mciteBstWouldAddEndPuncttrue
\mciteSetBstMidEndSepPunct{\mcitedefaultmidpunct}
{\mcitedefaultendpunct}{\mcitedefaultseppunct}\relax
\EndOfBibitem
\bibitem[Chai and Head-Gordon(2008)Chai, and Head-Gordon]{B810189B}
Chai,~J.-D.; Head-Gordon,~M. Long-range corrected hybrid density functionals with damped atom–atom dispersion corrections. \emph{Phys. Chem. Chem. Phys.} \textbf{2008}, \emph{10}, 6615--6620\relax
\mciteBstWouldAddEndPuncttrue
\mciteSetBstMidEndSepPunct{\mcitedefaultmidpunct}
{\mcitedefaultendpunct}{\mcitedefaultseppunct}\relax
\EndOfBibitem
\bibitem[Weigend and Ahlrichs(2005)Weigend, and Ahlrichs]{weigendBalancedBasisSets2005}
Weigend,~F.; Ahlrichs,~R. Balanced Basis Sets of Split Valence, Triple Zeta Valence and Quadruple Zeta Valence Quality for {{H}} to {{Rn}}: {{Design}} and Assessment of Accuracy. \emph{Phys. Chem. Chem. Phys.} \textbf{2005}, \emph{7}, 3297--3305\relax
\mciteBstWouldAddEndPuncttrue
\mciteSetBstMidEndSepPunct{\mcitedefaultmidpunct}
{\mcitedefaultendpunct}{\mcitedefaultseppunct}\relax
\EndOfBibitem
\bibitem[Eichkorn \latin{et~al.}(1995)Eichkorn, Treutler, {\"O}hm, H{\"a}ser, and Ahlrichs]{eichkornAuxiliaryBasisSets1995}
Eichkorn,~K.; Treutler,~O.; {\"O}hm,~H.; H{\"a}ser,~M.; Ahlrichs,~R. Auxiliary Basis Sets to Approximate {{Coulomb}} Potentials. \emph{Chemical Physics Letters} \textbf{1995}, \emph{242}, 652--660\relax
\mciteBstWouldAddEndPuncttrue
\mciteSetBstMidEndSepPunct{\mcitedefaultmidpunct}
{\mcitedefaultendpunct}{\mcitedefaultseppunct}\relax
\EndOfBibitem
\bibitem[Eichkorn \latin{et~al.}(1997)Eichkorn, Weigend, Treutler, and Ahlrichs]{eichkornAuxiliaryBasisSets1997}
Eichkorn,~K.; Weigend,~F.; Treutler,~O.; Ahlrichs,~R. Auxiliary Basis Sets for Main Row Atoms and Transition Metals and Their Use to Approximate {{Coulomb}} Potentials. \emph{Theor Chem Acta} \textbf{1997}, \emph{97}, 119--124\relax
\mciteBstWouldAddEndPuncttrue
\mciteSetBstMidEndSepPunct{\mcitedefaultmidpunct}
{\mcitedefaultendpunct}{\mcitedefaultseppunct}\relax
\EndOfBibitem
\bibitem[Weigend(2006)]{weigendAccurateCoulombfittingBasis2006}
Weigend,~F. Accurate {{Coulomb-fitting}} Basis Sets for {{H}} to {{Rn}}. \emph{Phys. Chem. Chem. Phys.} \textbf{2006}, \emph{8}, 1057--1065\relax
\mciteBstWouldAddEndPuncttrue
\mciteSetBstMidEndSepPunct{\mcitedefaultmidpunct}
{\mcitedefaultendpunct}{\mcitedefaultseppunct}\relax
\EndOfBibitem
\bibitem[Andrae \latin{et~al.}(1990)Andrae, H{\"a}u{\ss}ermann, Dolg, Stoll, and Preu{\ss}]{Andrae1990}
Andrae,~D.; H{\"a}u{\ss}ermann,~U.; Dolg,~M.; Stoll,~H.; Preu{\ss},~H. Energy-adjustedab initio pseudopotentials for the second and third row transition elements. \emph{Theoretica chimica acta} \textbf{1990}, \emph{77}, 123--141\relax
\mciteBstWouldAddEndPuncttrue
\mciteSetBstMidEndSepPunct{\mcitedefaultmidpunct}
{\mcitedefaultendpunct}{\mcitedefaultseppunct}\relax
\EndOfBibitem
\bibitem[Ahlrichs(2004)]{ri}
Ahlrichs,~R. Efficient evaluation of three-center two-electron integrals over Gaussian functions. \emph{Phys. Chem. Chem. Phys.} \textbf{2004}, \emph{6}, 5119--5121\relax
\mciteBstWouldAddEndPuncttrue
\mciteSetBstMidEndSepPunct{\mcitedefaultmidpunct}
{\mcitedefaultendpunct}{\mcitedefaultseppunct}\relax
\EndOfBibitem
\bibitem[Sierka \latin{et~al.}(2003)Sierka, Hogekamp, and Ahlrichs]{sierkaFastEvaluationCoulomb2003}
Sierka,~M.; Hogekamp,~A.; Ahlrichs,~R. Fast Evaluation of the {{Coulomb}} Potential for Electron Densities Using Multipole Accelerated Resolution of Identity Approximation. \emph{The Journal of Chemical Physics} \textbf{2003}, \emph{118}, 9136--9148\relax
\mciteBstWouldAddEndPuncttrue
\mciteSetBstMidEndSepPunct{\mcitedefaultmidpunct}
{\mcitedefaultendpunct}{\mcitedefaultseppunct}\relax
\EndOfBibitem
\bibitem[Holzer(2020)]{Holzer2020senex}
Holzer,~C. {An improved seminumerical Coulomb and exchange algorithm for properties and excited states in modern density functional theory}. \emph{The Journal of Chemical Physics} \textbf{2020}, \emph{153}, 184115\relax
\mciteBstWouldAddEndPuncttrue
\mciteSetBstMidEndSepPunct{\mcitedefaultmidpunct}
{\mcitedefaultendpunct}{\mcitedefaultseppunct}\relax
\EndOfBibitem
\bibitem[Drake \latin{et~al.}(1985)Drake, Lesiecki, and Camaioni]{DRAKE1985530}
Drake,~J.; Lesiecki,~M.~L.; Camaioni,~D.~M. Photophysics and cis-trans isomerization of DCM. \emph{Chemical Physics Letters} \textbf{1985}, \emph{113}, 530--534\relax
\mciteBstWouldAddEndPuncttrue
\mciteSetBstMidEndSepPunct{\mcitedefaultmidpunct}
{\mcitedefaultendpunct}{\mcitedefaultseppunct}\relax
\EndOfBibitem
\bibitem[Waterman(1965)]{waterman1965matrix}
Waterman,~P.~C. Matrix formulation of electromagnetic scattering. \emph{Proceedings of the IEEE} \textbf{1965}, \emph{53}, 805--812\relax
\mciteBstWouldAddEndPuncttrue
\mciteSetBstMidEndSepPunct{\mcitedefaultmidpunct}
{\mcitedefaultendpunct}{\mcitedefaultseppunct}\relax
\EndOfBibitem
\bibitem[Beutel \latin{et~al.}(2021)Beutel, Groner, \latin{et~al.} others]{Beutel2021Jun}
Beutel,~D.; Groner,~A.; others {Efficient simulation of biperiodic, layered structures based on the T-matrix method}. \emph{J. Opt. Soc. Am. B, JOSAB} \textbf{2021}, \emph{38}, 1782--1791\relax
\mciteBstWouldAddEndPuncttrue
\mciteSetBstMidEndSepPunct{\mcitedefaultmidpunct}
{\mcitedefaultendpunct}{\mcitedefaultseppunct}\relax
\EndOfBibitem
\bibitem[Beutel \latin{et~al.}(2024)Beutel, Fernandez-Corbaton, \latin{et~al.} others]{Beutel2024Apr}
Beutel,~D.; Fernandez-Corbaton,~I.; others {treams {\textendash} a T-matrix-based scattering code for nanophotonics}. \emph{Comput. Phys. Commun.} \textbf{2024}, \emph{297}, 109076\relax
\mciteBstWouldAddEndPuncttrue
\mciteSetBstMidEndSepPunct{\mcitedefaultmidpunct}
{\mcitedefaultendpunct}{\mcitedefaultseppunct}\relax
\EndOfBibitem
\bibitem[Zerulla \latin{et~al.}(2022)Zerulla, Krsti{\ifmmode\acute{c}\else\'{c}\fi}, \latin{et~al.} others]{Zerulla2022May}
Zerulla,~B.; Krsti{\ifmmode\acute{c}\else\'{c}\fi},~M.; others {A Multi-Scale Approach for Modeling the Optical Response of Molecular Materials Inside Cavities}. \emph{Adv. Mater.} \textbf{2022}, \emph{34}, 2200350\relax
\mciteBstWouldAddEndPuncttrue
\mciteSetBstMidEndSepPunct{\mcitedefaultmidpunct}
{\mcitedefaultendpunct}{\mcitedefaultseppunct}\relax
\EndOfBibitem
\bibitem[Fakhouri \latin{et~al.}(2019)Fakhouri, Peri{\ifmmode\acute{c}\else\'{c}\fi}, \latin{et~al.} others]{Fakhouri2019}
Fakhouri,~H.; Peri{\ifmmode\acute{c}\else\'{c}\fi},~M.; others {Sub-100 nanometer silver doped gold{\textendash}cysteine supramolecular assemblies with enhanced nonlinear optical properties}. \emph{Phys. Chem. Chem. Phys.} \textbf{2019}, \emph{21}, 12091--12099\relax
\mciteBstWouldAddEndPuncttrue
\mciteSetBstMidEndSepPunct{\mcitedefaultmidpunct}
{\mcitedefaultendpunct}{\mcitedefaultseppunct}\relax
\EndOfBibitem
\end{mcitethebibliography}

\end{document}


\section{The details of the quantum chemistry calculations}

The geometrical properties of the layered hybrid gold-cysteine nanomaterial have been obtained by periodic density functional theory calculations using CP2K electronic structure program.\cite{CP2K} The periodic cell was composed of 16 Au atoms connected in 4 strands by 16 Sulfur-Cysteine complexes as depicted in Fig S1 (a)-(b). Additionally, 34 H\textsubscript{2}O were explicitly added to the periodic cell to include the effects of the solvent as it is the case in the realistic experimental studies. The GGA PBE density functional\cite{PBE_a, PBE_b} in combination with DZVP-MOLOPT-SR-GTH basis set\cite{MOLOPT} and GTH-PBE potentials\cite{GTH-PP1, GTH-PP2, GTH-Pseudopotentials} for all atoms was used to minimize the energy of the system using gradient-based algorithms. Grimme's D3 dispersion correction\cite{ caldeweyherGenerallyApplicableAtomiccharge2019} with Becke-Johnson damping\cite{caldeweyherExtensionD3Dispersion2017} was also employed to account for long-range London dispersion interactions. The energy cut-off for the plane waves was 600 Rydbergs with a relative cut-off at 60 Rydbergs. Simultaneously, we screened for the unit cell vectors while performing optimization of the positions of all atoms within the periodic unit cell. The angles between the cell vectors were set to 90° and kept fixed during optimizations. The optimized unit cell vectors were: $\vec{\emph{a}}$ = 17.126856 \r{A} and $\vec{\emph{b}}$ = 14.832264 \r{A}. The vector $\vec{\emph{c}}$ was set to be 13.0 \r{A} as previously determined in the experiments and was not changed during the calculations.  

\begin{figure*}[htbp!]
\includegraphics[width=1.0\textwidth]{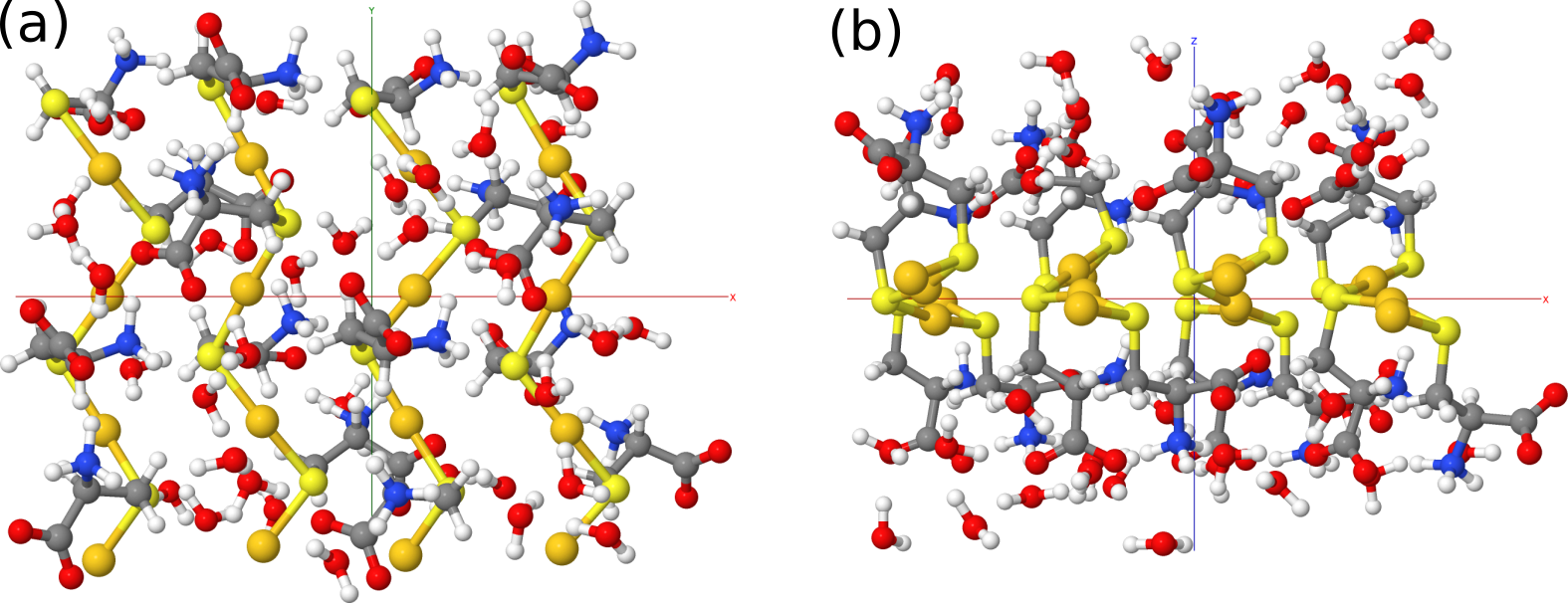}
\caption{ (a) A top-down view of optimized periodic cell of Au-Cys nanoparticle using CP2K program with explicit water molecules. (b) A side view of the same optimized periodic cell from (a) with explicit water molecules.} \label{structures_with_H2O} 
\end{figure*}

\begin{figure*}[htbp!]
\includegraphics[width=0.9\textwidth]{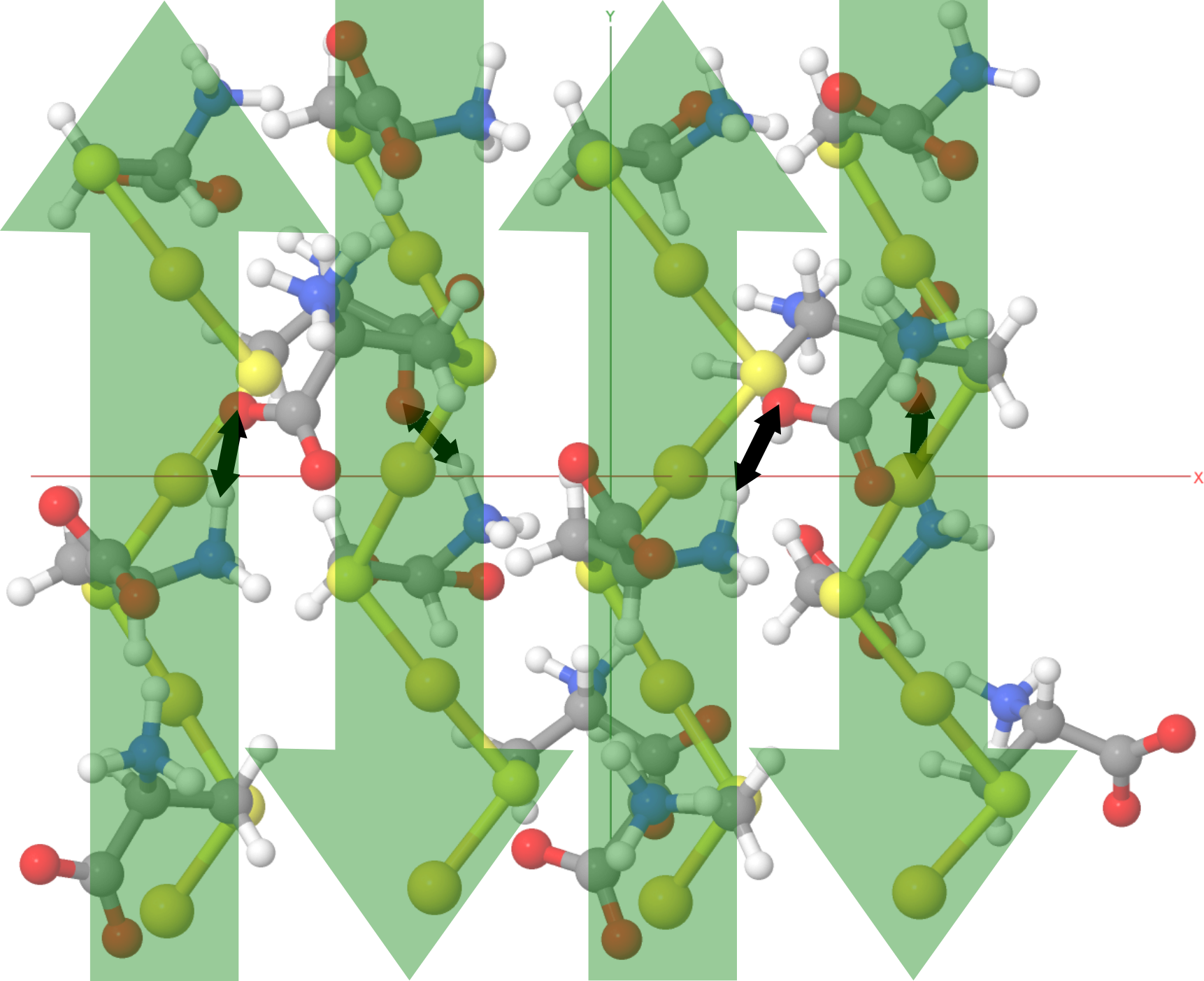}
\caption{A structure of periodic cell of Au-cysteine nanoparticle single-layer in analogy with $\beta$-sheet secondary structure of the proteins where individual $\beta$-strands are noncovalently coordinated to stabilize the structure.} \label{structures_with_H2O} 
\end{figure*}

To calculate the optical properties of this gold-cysteine nanomaterial, we switch to finite-size molecular models compatible to calculations in the development version of TURBOMOLE 7.7 electronic structure program\cite{TURBOMOLE2022, TM_TODAY_TOMORROW}. The finite-size model was selected to be a periodic cell of the nanomaterial optimized in CP2K without explicit water molecules, Fig 1. (a)-(b). We perform calculations of the linear polarizabilities with models consisting of implicit H\textsubscript{2}O surrounding based on Conductor-like Screening Model (COSMO)\cite{klamtCOSMONewApproach1993} for water. The nonlinear complex dynamic first hyperpolarizability rank-3-tensors were also calculated with the COSMO implicit solvation model containing 224 atoms in the quantum region to reduce computational costs. The damping parameter was set to 0.15 eV for half-width at half-maximum (HWHM) in calculations of polarizabilities and first hyperpolarizabilities. We chose long-rang corrected wb97x-d density functional\cite{B810189B} in combination with Karlsruhe def2-TZVP basis set\cite{weigendBalancedBasisSets2005, eichkornAuxiliaryBasisSets1995, eichkornAuxiliaryBasisSets1997, weigendAccurateCoulombfittingBasis2006} and matching 19-electron effective core-potential (ECP) for Au atoms\cite{Andrae1990} in excited-state calculations in TURBOMOLE. To further speed up the calculation we employed resolution-of-identity (rij) approach\cite{ri} in combination with multipole-accelerated-resolution-of-identity (marij)\cite{sierkaFastEvaluationCoulomb2003} and semi-numerical integration for exchange (esenex)\cite{Holzer2020senex}.

\begin{figure*}[htbp!]
\centering
\subfloat{
\hspace{0.0cm}
\includegraphics[width=0.47\textwidth]{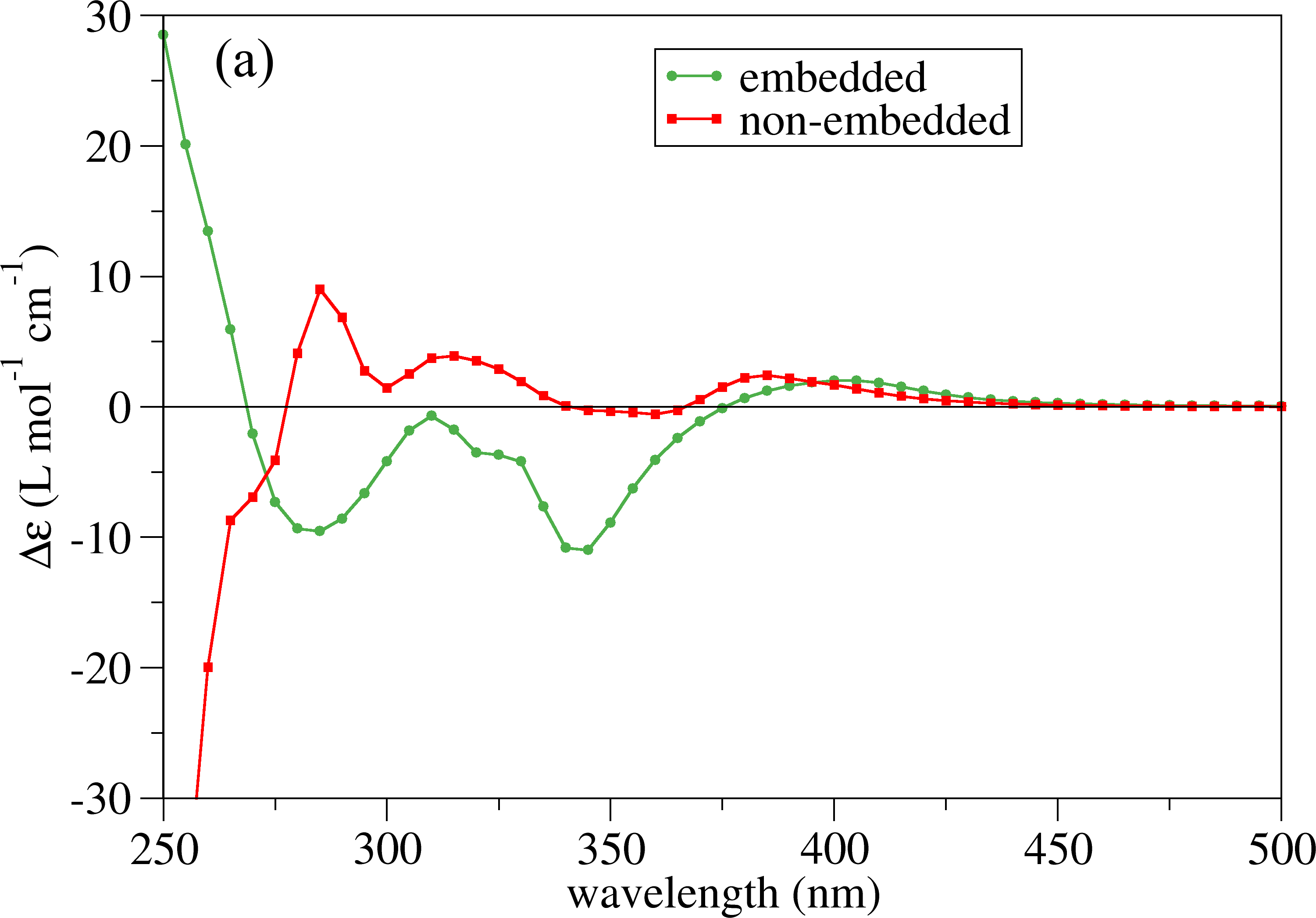}
}\hspace{0.0cm}
\subfloat{
\includegraphics[width=0.47\textwidth]{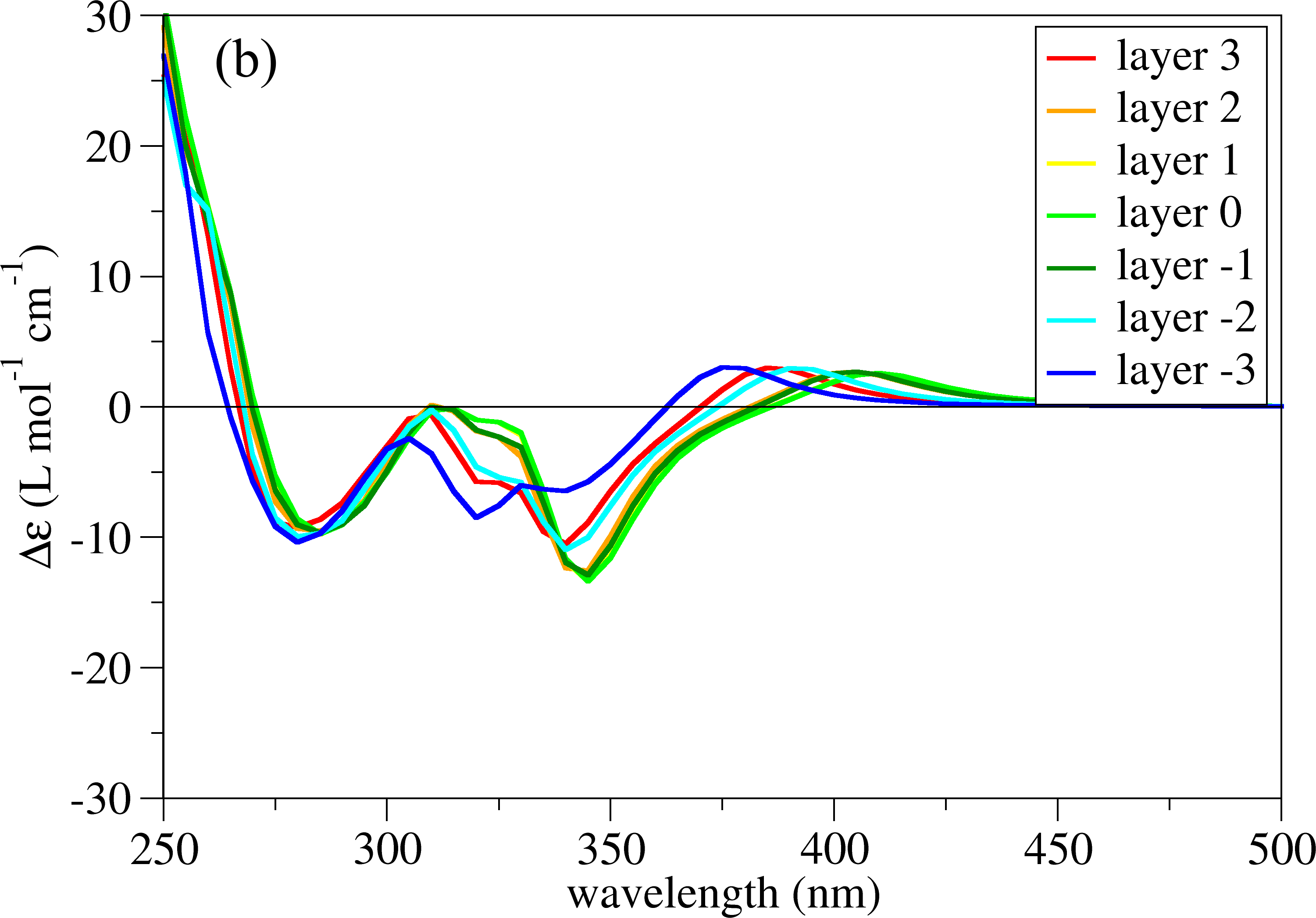}
}
\caption{ (a) A comparison of differential absorption for left and right circularly polarized light for non-embedded molecular model based on one periodic cell with average of all 7 layers of embedded models ("-3" to "3") presented individually in subfigure (b).} \label{ECD_TDDFT} 
\end{figure*}

\section{Transmission electron microscopy (TEM) and dynamic light scattering (DLS) measurements of Gold-cysteine nanoparticles}

Morphological studies were performed using transmission electron microscopy (TEM, JEOL-JEM-F2100) at an operating voltage of 200 kV. TEM samples were prepared by drop-casting solutions onto carbon-coated gold grids.

\begin{figure*}[htbp!]
\centering
\subfloat{
\hspace{0.0cm}
\includegraphics[width=0.7\textwidth]{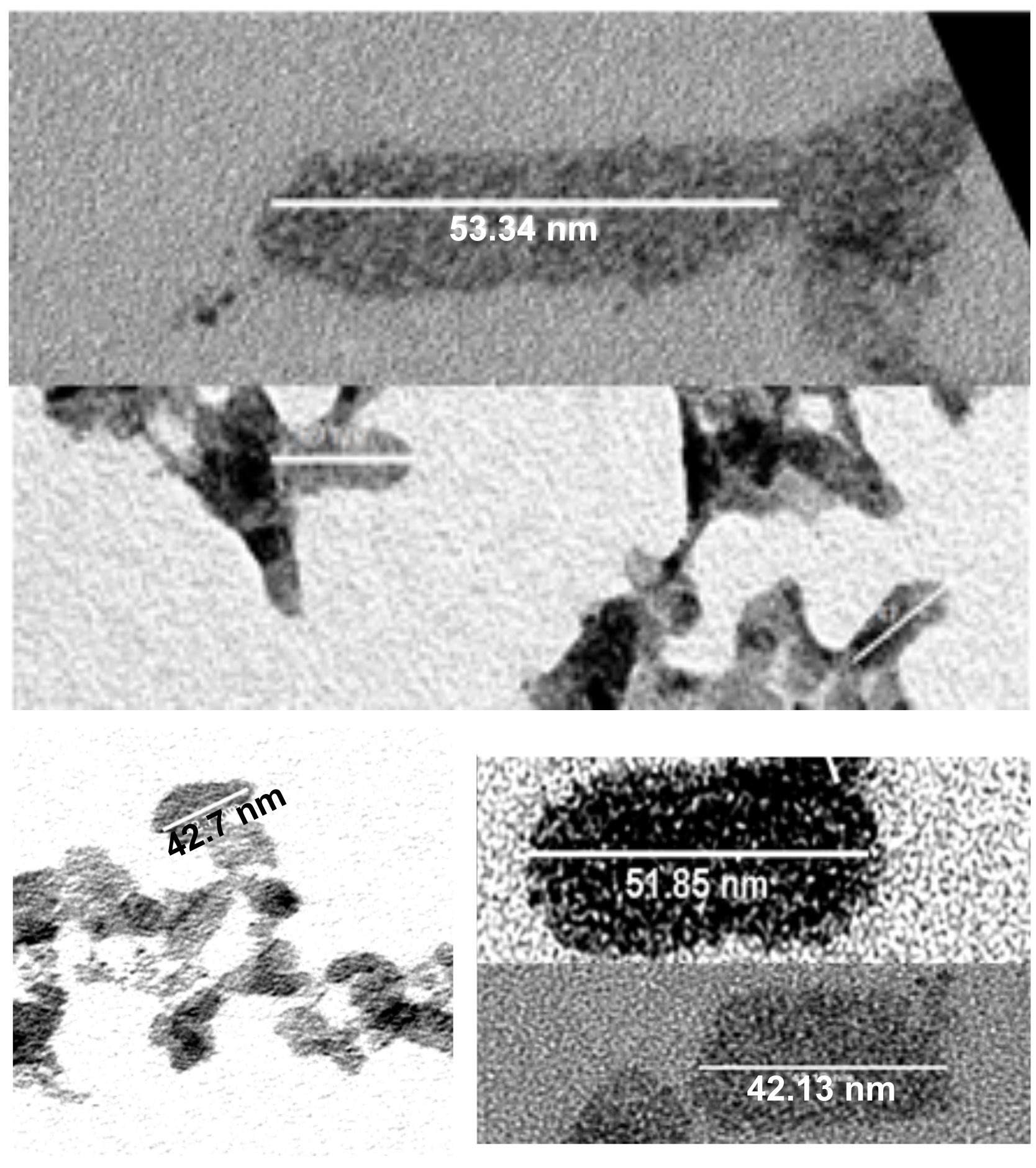}
}\hspace{0.0cm}
\caption{TEM images of deposited Gold-Cysteine beta-sheet supramolecular nanoparticles showing flat, more rectangular-like shapes.} \label{TEM} 
\end{figure*}

The hydrodynamic radius of gold-cysteine supramolecular nanoparticles was measured by dynamic light scattering (DLS) using a Malvern Zetasizer Nano ZS. 

\begin{figure*}[htbp!]
\centering
\subfloat{
\hspace{0.0cm}
\includegraphics[width=0.65\textwidth]{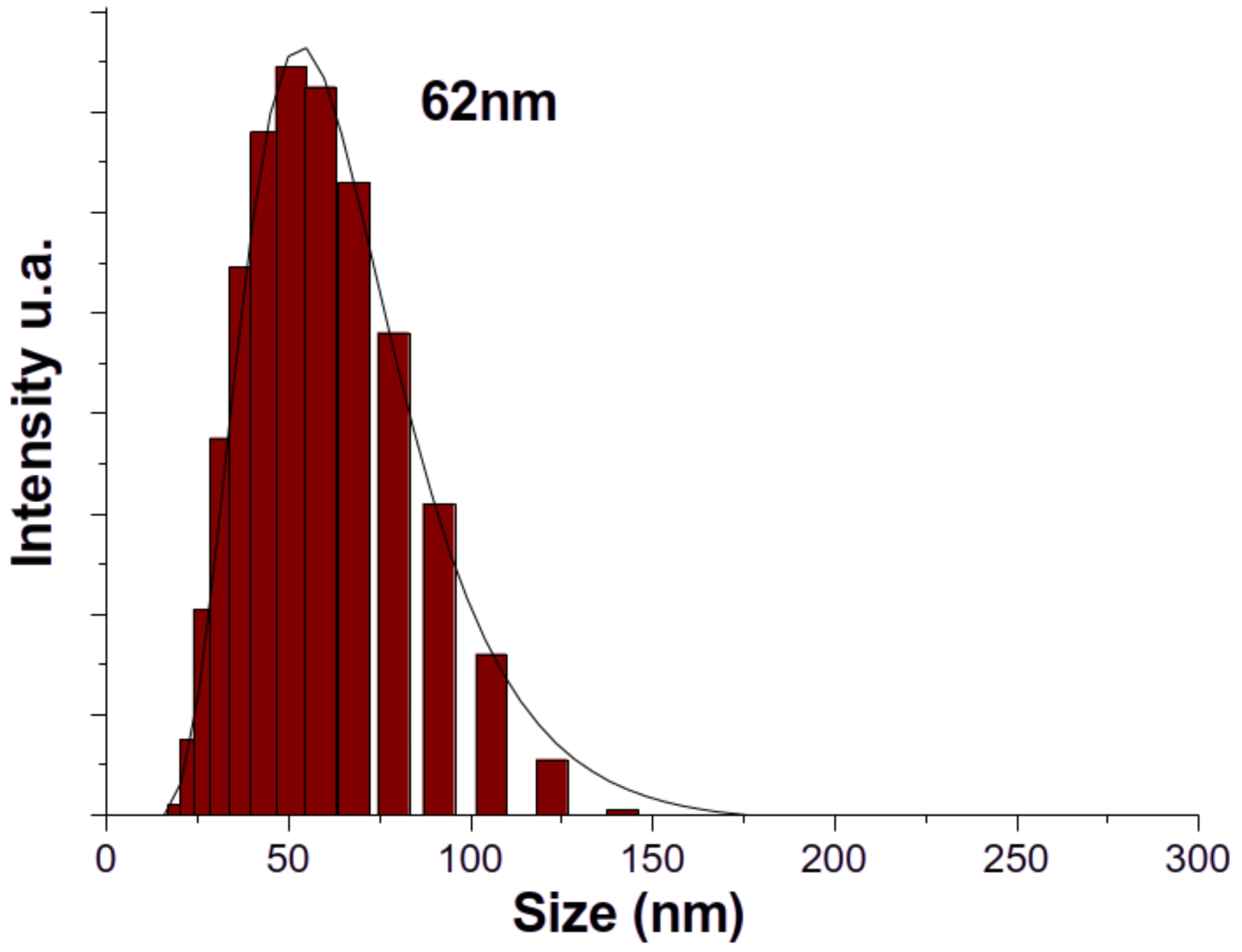}
}\hspace{0.0cm}
\caption{A dynamic light scattering (DLS) measurements of hydrodynamic radius of Gold-cysteine nanoparticles showing average size of $\sim$62 nm.} \label{DLS} 
\end{figure*}

\section{Experimental linear optical spectra of gold-L-cysteine supramolecular nanoparticles}

\begin{figure*}[h!]
\centering
\subfloat{
\hspace{0.0cm}
\includegraphics[width=0.7\textwidth]{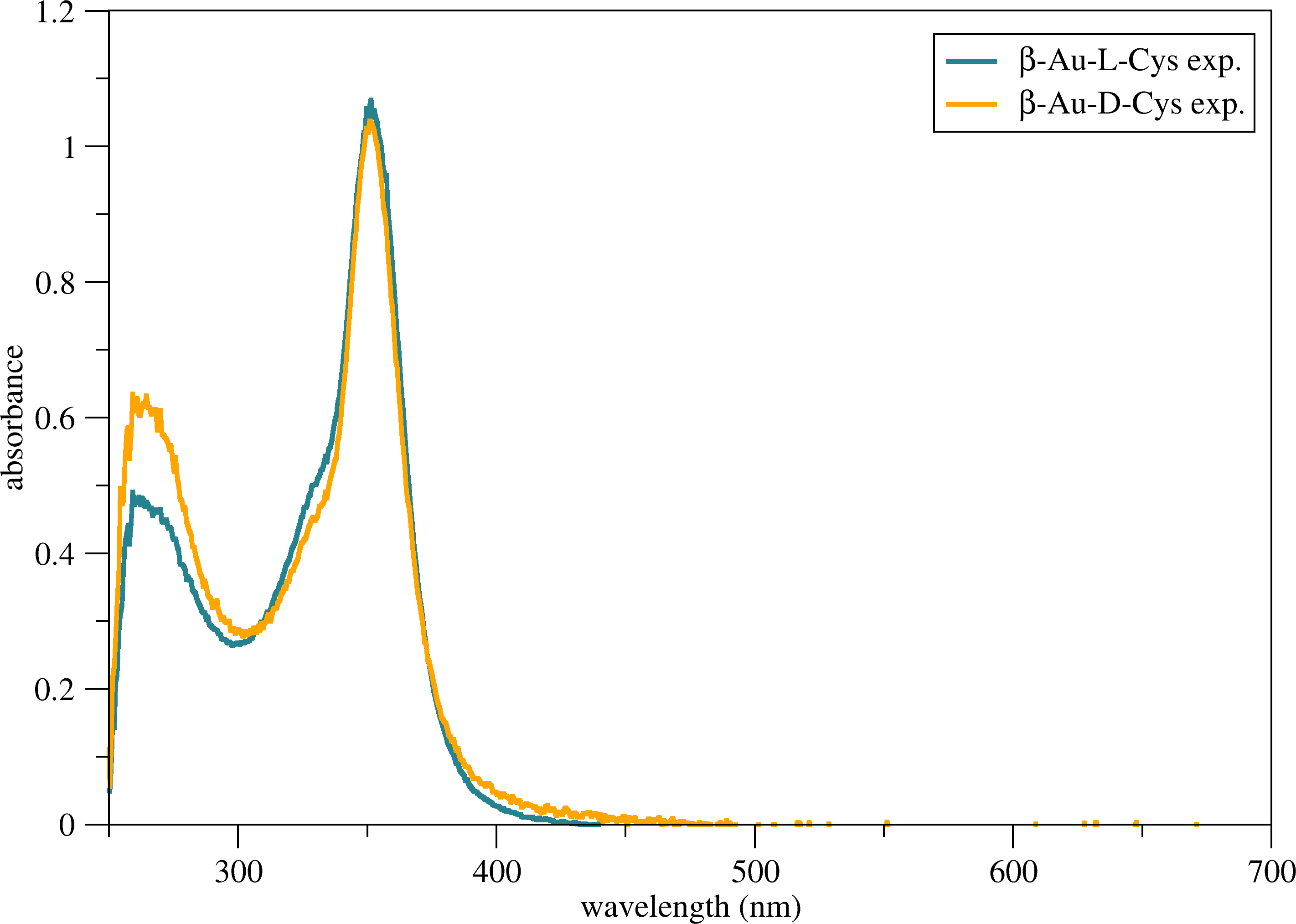}
}\hspace{0.0cm}
\caption{Experimentally obtained spectrogram of both enantiomers of gold-cysteine supramolecular nanoparticles shows two maxima at ~350 nm and ~275 nm.} \label{ABS_exp} 
\end{figure*}

\begin{figure*}[h!]
\centering
\subfloat{
\includegraphics[width=0.7\textwidth]{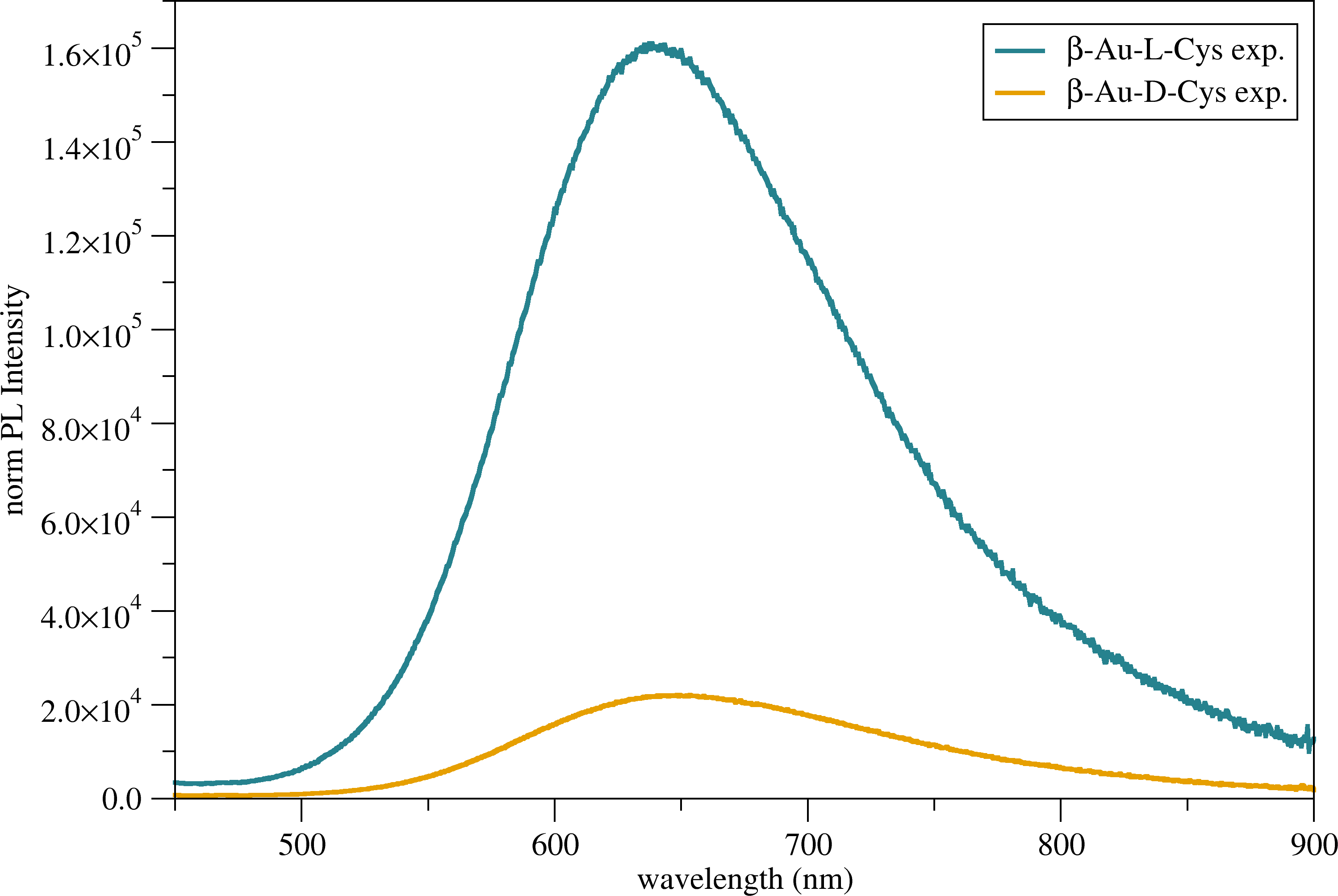}
 }\hspace{0.1cm}
 \vspace{0.5cm}
\subfloat{
\includegraphics[width=0.69\textwidth]{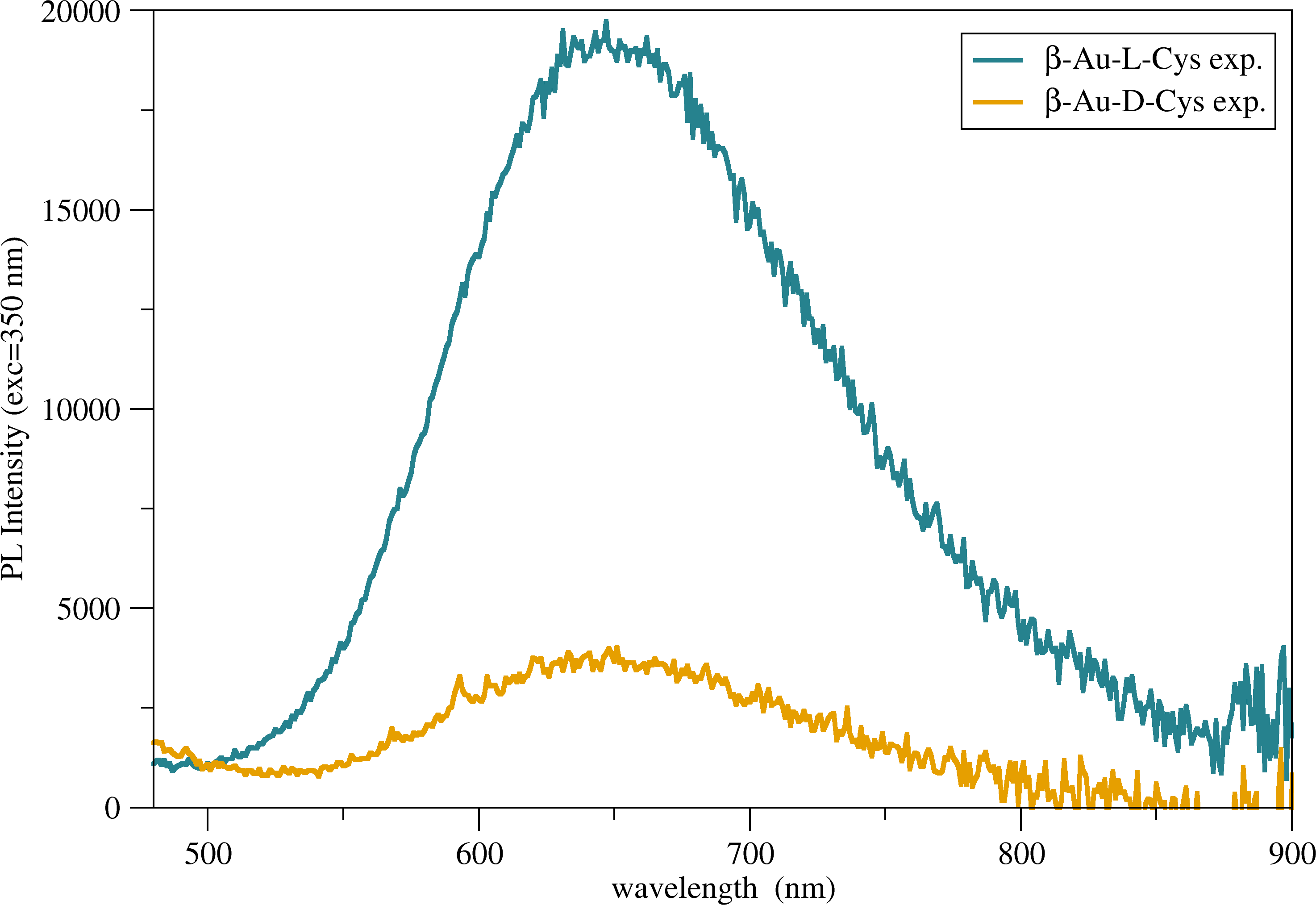}
}
\caption{\textbf{(a)} Normalized photoluminescence intensity of both enantiomer species of Au-cysteine nanoparticles. \textbf{(b)} Photoluminescence intensity of left and right enantiomer of gold-cysteine beta-sheet nanoparticles upon excitation with incident light of 350 nm wavelength.}
\label{fig:emi_exp}
\end{figure*}

\setlength{\extrarowheight}{1pt}
\fontsize{10}{13}\selectfont
\setlength{\tabcolsep}{0.2em}
\begin{table}[h!]
\centering
\caption{Quantum yield of gold-cysteine beta-sheet nanoparticles. A DCM dye in methanol was used as a baseline.\cite{DRAKE1985530}}
\begin{tabular}{|c|c|}
\hline\hline
Sample & $\Phi$ (exc = 350 nm)   \\
\hline\hline
$\beta$-Au-L-Cys & 0.32\%   \\
$\beta$-Au-D-Cys & 0.40\%   \\
\hline\hline
\end{tabular}
\end{table}

\begin{figure*}[h!]
\centering
\vspace{0.5cm}
\subfloat{
\includegraphics[width=0.69\textwidth]{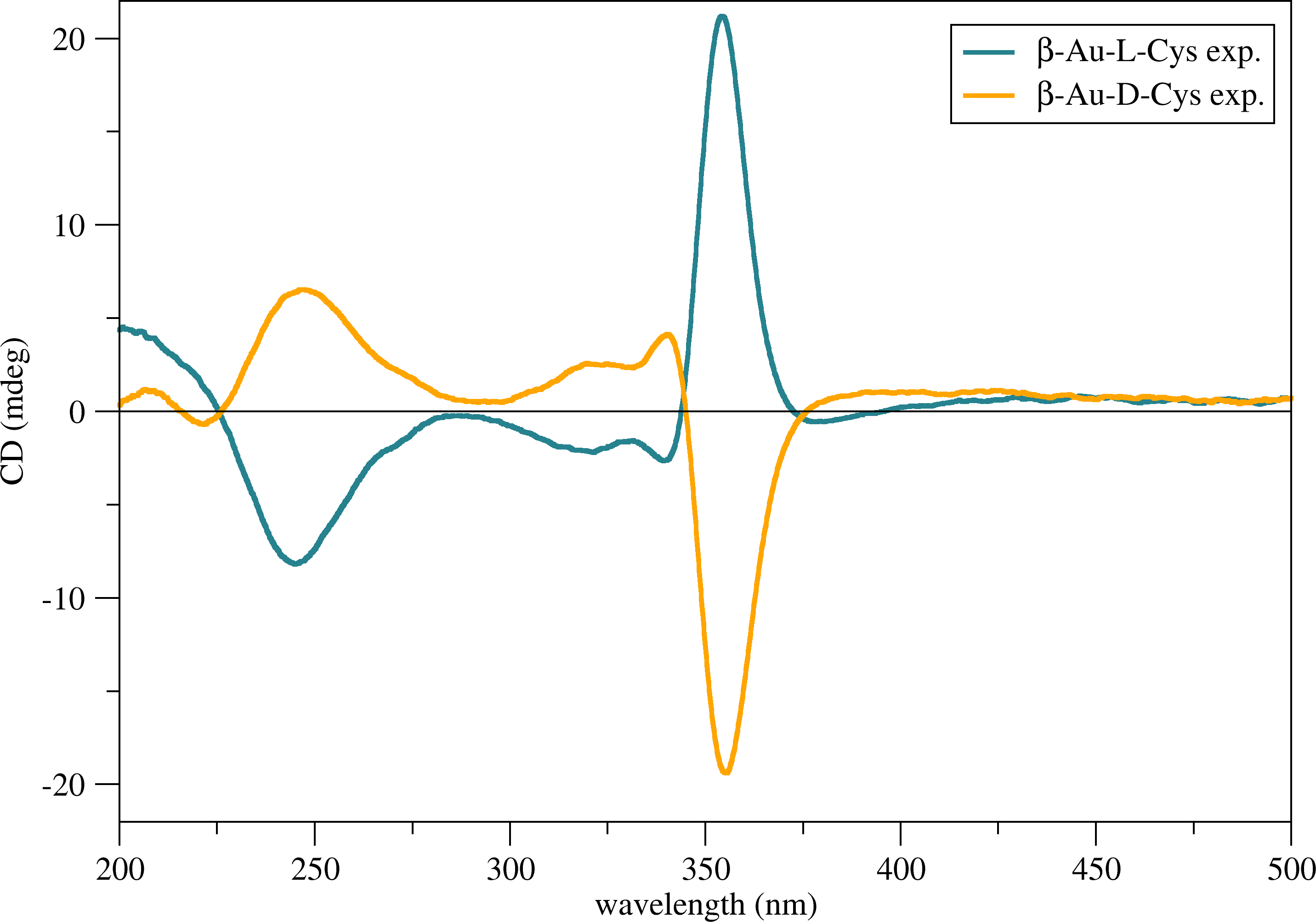}
 }\hspace{0.1cm}
\subfloat{
\includegraphics[width=0.7\textwidth]{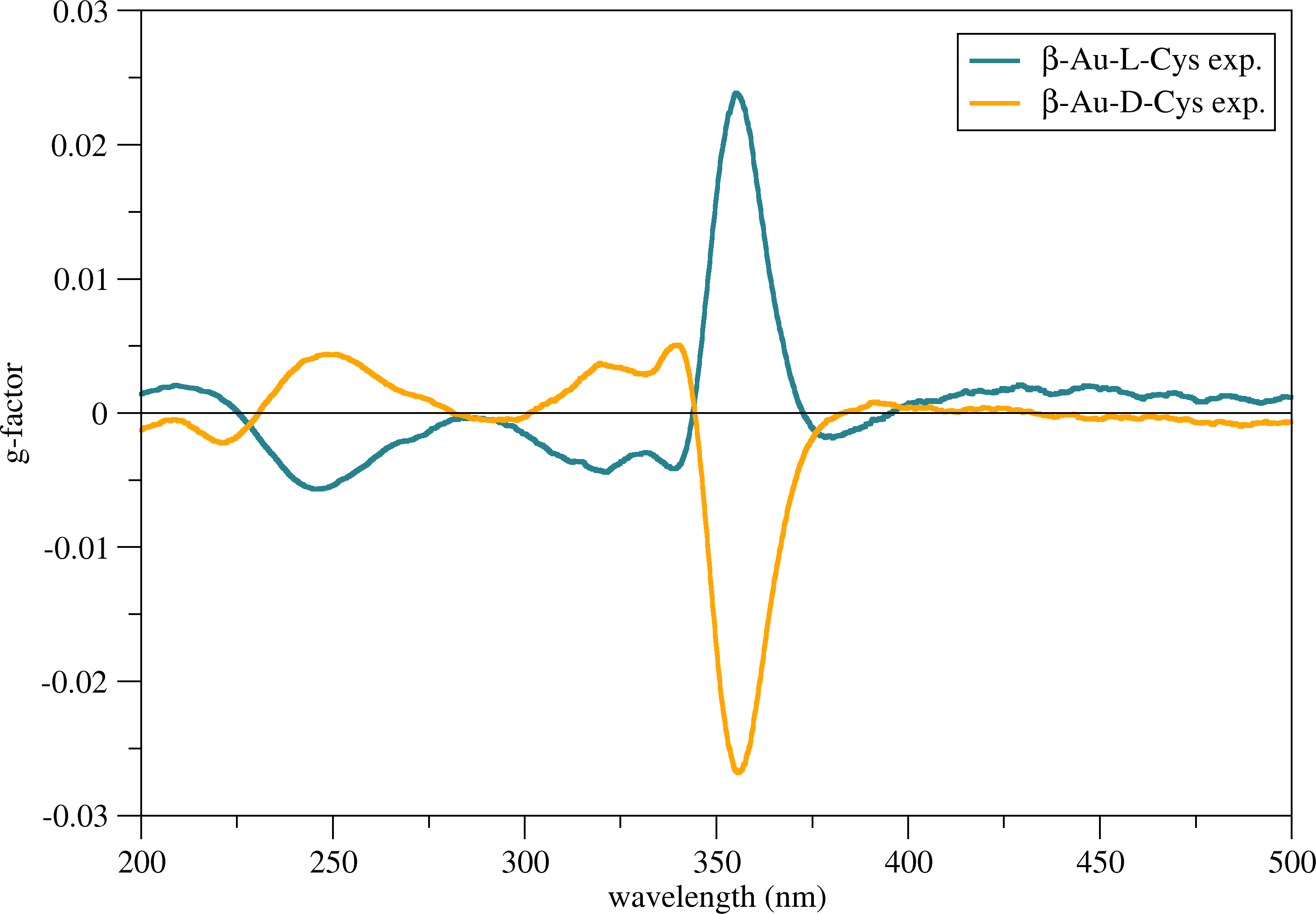}
}
\caption{\textbf{(a)} Linear electronic circular dichroism (CD) spectra of both enantiomers measured in milli-degrees (mdeg). \textbf{(b)} Experimentally determined g-factor of both enantiomers of Au-cysteine nanoparticles.}
\label{fig:ecd_exp}
\end{figure*}

\newpage
\section{T-matrix formalism}
In this section, we define a T-matrix of a single scatterer \cite{waterman1965matrix}. The total linear electric field outside from an isolated scatterer can be expressed as follows,
\begin{equation}\label{field}
\begin{array}{r}
\mathbf{E}(\boldsymbol{r})=\sum_{l=1}^{\infty} \sum_{m=-l}^l\left[a_{l m, \mathrm{~N}} \mathbf{N}_{l m}^{(1)}\left(k\boldsymbol{r}\right)+a_{l m, \mathrm{M}} \mathbf{M}_{l m}^{(1)}\left(k \boldsymbol{r}\right)+c_{l m, \mathrm{~N}} \mathbf{N}_{l m}^{(3)}\left(k\boldsymbol{r}\right)+c_{l m, \mathrm{M}} \mathbf{M}_{l m}^{(3)}\left(k\boldsymbol{r}\right)\right]
\end{array},
\end{equation}
where $\mathbf{N}_{l m}^{(1)}\left(k\boldsymbol{r}\right)$ and $\mathbf{M}_{l m}^{(1)}\left(k \boldsymbol{r}\right)$
are the regular, and $\mathbf{N}_{l m}^{(3)}\left(k\boldsymbol{r}\right)$ and $\mathbf{M}_{l m}^{(3)}\left(k \boldsymbol{r}\right)$ are singular
vector spherical waves. The latter are defined as \cite{Beutel2021Jun, Beutel2024Apr},

\begin{equation} \label{eq5}
\mathbf{M}_{\ell m}\left(k \boldsymbol{r}\right) =\nabla \times\left(\boldsymbol{r} \psi_{ \ell m}\left(k r\right)\right),
\end{equation}
\begin{equation}\label{eq6}
\mathbf{N}_{\ell m }\left(k \boldsymbol{r}\right)=\frac{\nabla \times \mathbf{M}_{\ell m}\left(k r\right)}{k},
\end{equation}
where
\begin{equation}\label{eq7}
\psi_{\ell m}\left(kr\right)=B_{lm}P_{\ell}^{m}(\cos \theta) z_{\ell}^{(n)}(k r) \mathrm{e}^{\mathrm{i} m \varphi}. \\\end{equation}
Here, $B_{lm}$ are the normalization constants
\cite{Beutel2024Apr},
\begin{equation}
B_{\ell m} = \sqrt{\frac{(2\ell + 1)(\ell - m)!}{4\pi \ell (\ell + 1)(\ell + m)!}},
\end{equation}
functions $P_l^m(\cos \theta)$ are the associated Legendre polynomials, $z_l^{(n)}(k r)$ are spherical Bessel ($n = 1$) or
Hankel ($n = 3$) functions.

The T-matrix of a single scatterer connects the scattered field coefficients $c_{l m, \mathrm{~N/M}}$ with the coefficients of the incident field $a_{l m, \mathrm{~N/ M}}$ in Eq. \ref{field},
\begin{equation}
    \boldsymbol{c} = T\boldsymbol{a}.
\end{equation}

The connection between the T-matrix and dipolar polarizability tensors is explained in detail in Ref. \cite{Zerulla2022May}.

\section{Linear scattering properties of the gold-cysteine film}
\begin{figure*}[htbp!]
\includegraphics[width=0.9\textwidth]{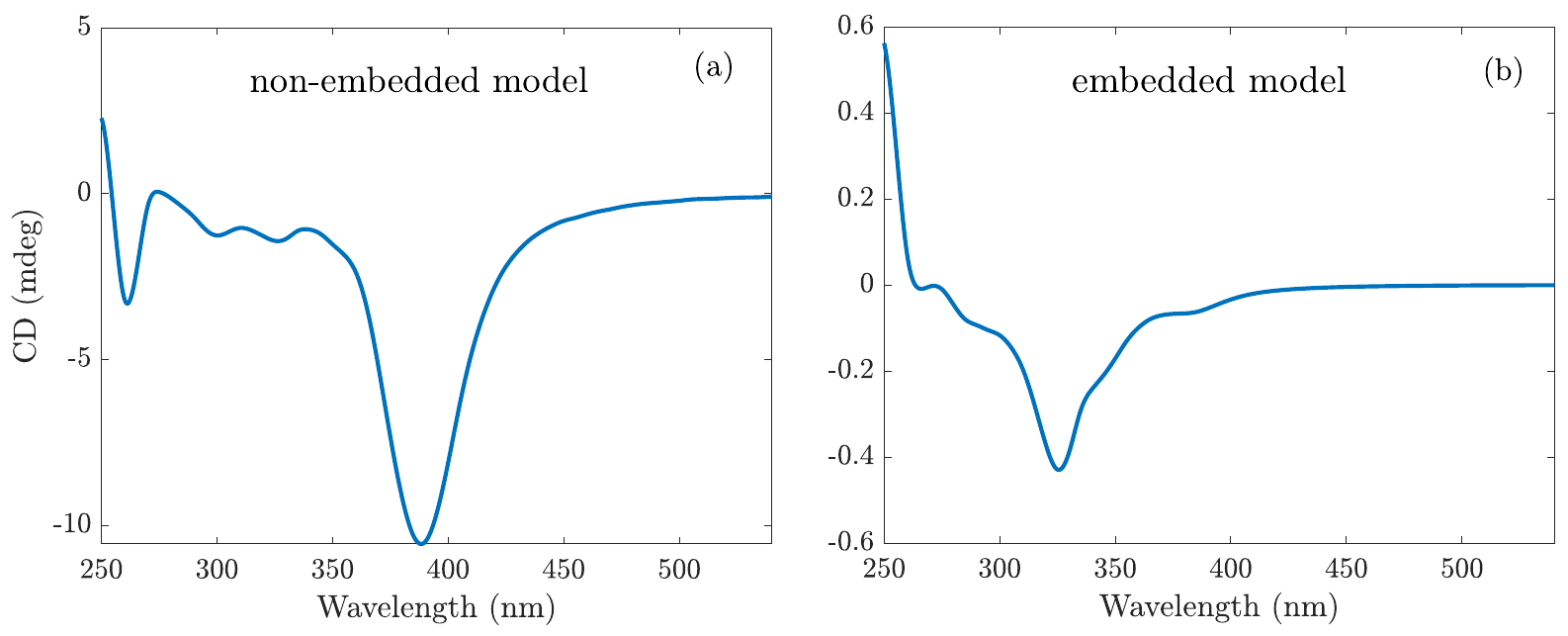}
\caption{ Circular dichroism spectra of a thin film of gold-cysteine nanoparticles with 9.1 nm width (7 layers) for (a) "bulk" and (b) 7-layer models.} \label{abs_acd_cd_50nm} 
\end{figure*}

This section presents the additional linear properties of thin films composed of gold-cysteine molecules, specifically circular dichroism (CD). Fig. \ref{abs_acd_cd_50nm}(a) illustrates the calculations for a $9.1$ nm film, where each unit cell corresponds to the same T-matrix (the "bulk model"). Here, we see again that the most pronounced resonance for the ``bulk'' model experiences a blue shift compared to the experimental results \cite{Fakhouri2019}. On the other hand, in Figures \ref{abs_acd_cd_50nm}(b), the 7-layer model shows better correspondence with the experiment. The formula used for the calculation of the CD is as follows,
\begin{equation}\label{CD}
    \text{CD} = \arctan{\frac{t^--t^+}{t^- + t^+}},
\end{equation}
where $t^{+/-}$ is the transmission coefficient of left-, or right-circular polarized wave.

\section{Averaging over the angle of incidence}
\begin{figure*}[htbp!]
\includegraphics[width=1.0\textwidth]{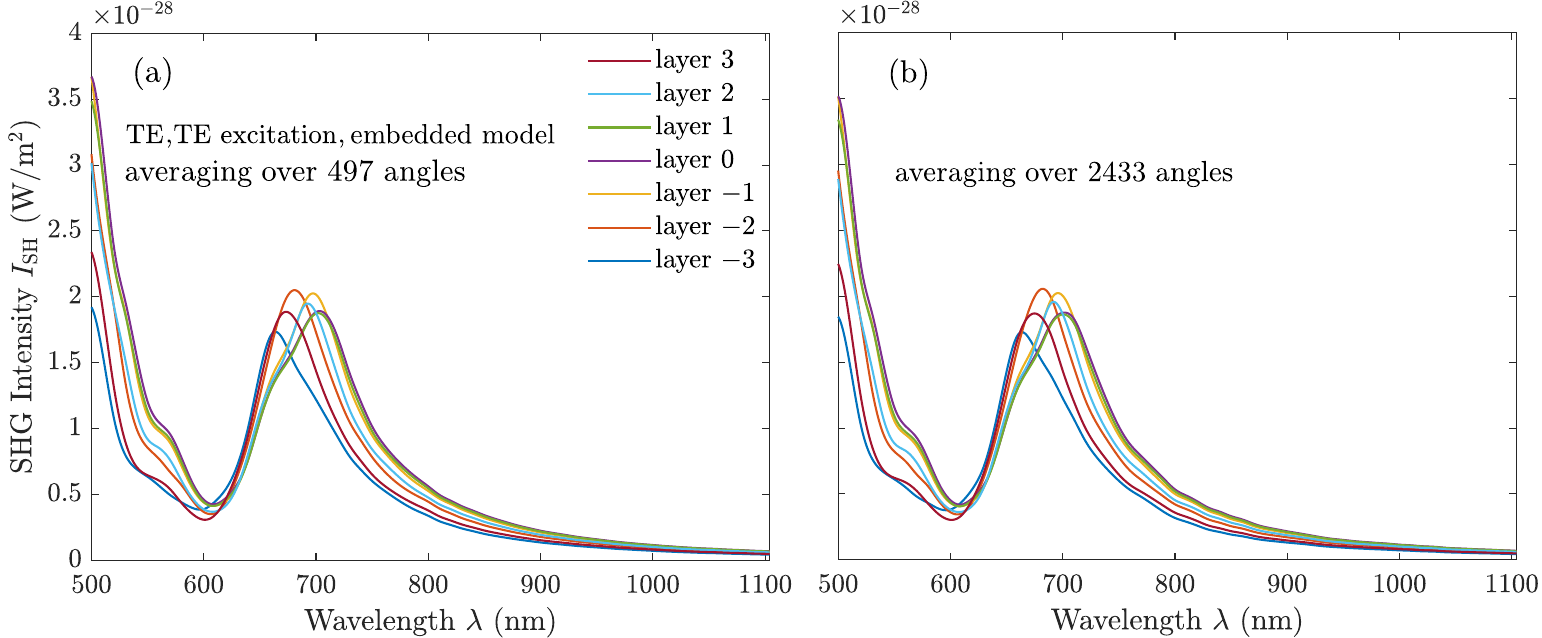}
\caption{SH intensity contributions from each of the seven layers within the 7-layer model under TE, TE excitation averaged over (a) 497, (b) 2433 angles of incidence in the lower-half-space.} \label{averaging} 
\end{figure*}
To better approximate our simulation results with the experimental conditions, we averaged the results over the angle of incidence in the lower half-space. Figure \ref{averaging} compares the outcomes of averaging over 497 angles versus 2433 angles. The results are nearly identical, with only minor differences, indicating that 497 angles are sufficient to achieve convergence.

\bibliography{suppl}